\documentclass[11pt]{article}
\usepackage[utf8]{inputenc}
\usepackage[T1]{fontenc}
\usepackage[normalem]{ulem}
\usepackage{verbatim}
\usepackage{amsmath,amssymb,amsthm,amsfonts,mathrsfs,dsfont,braket,array,float}
\usepackage{color,graphicx,cite}
\usepackage{slashed}
\usepackage{enumitem}

\usepackage{epsfig}
\usepackage{mathabx}
\usepackage[mathscr]{eucal}
\usepackage{stmaryrd}
\usepackage{esint}
\usepackage{physics}
\usepackage{setspace}
\usepackage{braket}
\usepackage{float}
\usepackage{url}
\usepackage{mathtools}
\usepackage{bbold}
\usepackage{slashed}
\usepackage{tikz}
\usepackage{graphicx}
\usepackage{epstopdf}
\usepackage{subfigure}
\usepackage[labelfont=bf]{caption}
\usepackage{pgfplots}
\usepackage{fancybox}
\usepackage{eurosym}
\usepackage{tcolorbox}
\usepackage{tensor}
\allowdisplaybreaks

\usepackage[margin = 2.2cm]{geometry}
\usepackage[ragged]{footmisc}
\usepackage[bookmarks=true,bookmarksnumbered=false,
hyperindex=true,bookmarksopen=true,hyperfigures=true,
colorlinks=true,linkcolor=darkgray,citecolor=custom,urlcolor=custom,breaklinks]{hyperref}
\definecolor{webgreen}{rgb}{0, 0.5, 0}
\definecolor{webblue}{rgb}{0.06, 0.2, .65}
\definecolor{webblue2}{rgb}{0, 0.33, .71}
\definecolor{webred}{rgb}{0.9, 0.1, 0}
\definecolor{darkgreen}{rgb}{0,0.4,0.2}
\definecolor{custom}{rgb}{0.05,0.31,0.55}
\definecolor{darkgray}{rgb}{0.4,0.4,0.4}

\def\be{\begin{equation}}
	\def\ee{\end{equation}}

\def\bfm{\mathbf{m}}

\numberwithin{equation}{section}

\def\tbeta{\tilde{\beta}}

\usepackage{formatting}

\usepackage{shortcuts}

\begin{document} 
\begin{titlepage}
\begin{center}
\phantom{ }
\vspace{1.5cm}

{\bf \Large{Microscopic origin of the entropy of black holes in general relativity}}
\vskip 0.5cm
Vijay Balasubramanian${}^{\dagger}$, Albion Lawrence${}^{*}$, Javier M. Mag\'an${}^{\ddagger}$, Martin Sasieta${}^{\mathsection}$
\vskip 0.05in
\small{${}^{\dagger}$ ${}^{\ddagger}$  \textit{David Rittenhouse Laboratory, University of Pennsylvania}}
\vskip -.4cm
\small{\textit{ 209 S.33rd Street, Philadelphia, PA 19104, USA}}
\vskip -.10cm
\small{  ${}^{\dagger}$ \textit{Santa Fe Institute}}
\vskip -.4cm
\small{\textit{ 1399 Hyde Park Road, Santa Fe, NM 87501, USA}}
\vskip -.10cm
\small{ ${}^{\dagger}$ ${}^{\ddagger}$ \textit{Theoretische Natuurkunde, Vrije Universiteit Brussel}}
\vskip -.4cm
\small{\textit{Pleinlaan 2,  B-1050 Brussels, Belgium}}
\vskip -.10cm
\small{ ${}^{*}$ ${}^{\mathsection}$ \textit{Martin Fisher School of Physics, Brandeis University}}
\vskip -.4cm
\small{\textit{Waltham, Massachusetts 02453, USA}}
\vskip -.10cm
\small{  ${}^{\ddagger}$ \textit{Instituto Balseiro, Centro At\'omico Bariloche}}
\vskip -.4cm
\small{\textit{ 8400-S.C. de Bariloche, R\'io Negro, Argentina}}

\begin{abstract}
We construct an infinite family of  microstates  with geometric interiors for eternal black holes in general relativity with a negative cosmological constant in any dimension.  Wormholes in the Euclidean path integral for gravity cause these states to have small, but non-zero, quantum mechanical overlaps that have a universal form.  The overlaps have a dramatic consequence:  the microstates span a Hilbert space of log dimension equal to the Bekenstein-Hawking entropy. The semiclassical microstates we construct contain Einstein-Rosen bridges of arbitrary size behind their horizons. Our results imply that all these bridges can be interpreted as quantum superpositions of wormholes of size at most exponential in the entropy. 
\end{abstract}
\end{center}

\small{\vspace{2 cm}\noindent ${}^{\dagger}$vijay@physics.upenn.edu\\
${}^{*}$albion@brandeis.edu\\
${}^{\ddagger}$magan@sas.upenn.edu\\
${}^{\mathsection}$martinsasieta@brandeis.edu
}

\end{titlepage}

\setcounter{tocdepth}{2}

{\parskip = .4\baselineskip \tableofcontents}
\newpage


\section{Introduction}\label{SecI}

In this article, we consider one old problem and one new problem concerning black holes. As we will describe, these two problems turn out to be precisely related.

The old problem concerns the statistical origin and finiteness of the Bekenstein-Hawking black hole entropy \cite{PhysRevD.7.2333,Hawking:1975vcx}
\be
S_{BH}=\frac{A}{4G}\;.
\label{exp} 
\ee
where $A$ is the area of the black hole horizon, and we have chosen units to set $\hbar =1$.  One way to derive this equation is from an analysis of the partition function in Euclidean quantum gravity \cite{Hawking1979}. With different identifications of entropy $S$ and area $A$, it also appears more generally in the context of quantum gravity with a negative cosmological constant  \cite{Maldacena:1997re,Witten:1998qj,Gubser:1998bc} through the Ryu-Takayanagi conjecture \cite{Ryu:2006bv,Lewkowycz:2013nqa}.
Equation~\ref{exp} suggests that the Hilbert space describing the quantum dynamics of black holes is finite dimensional, spanned by $e^{S_{BH}}$ orthogonal states, usually dubbed the  ``black hole microstates''. The problem arises when one tries to find these microstates, thus providing a statistical account of black hole entropy.

Many routes towards solving this problem have been explored. On the one hand, we have approaches in string theory such as precise microstate counting 
\cite{Strominger:1996sh},
and the fuzzball approach \cite{https://doi.org/10.48550/arxiv.2204.13113}. The problem with these celebrated results is that they apply to restricted scenarios, and  do not provide a satisfactory understanding of the universality of~(\ref{exp}). On the other hand we have approximate approaches, such as the interpretation of black hole entropy as entanglement entropy \cite{Bombelli:1986rw,Srednicki:1993im}, or relatedly the association of black hole entropy with the entropy of thermally excited quantum fields in the vicinity of the horizon \cite{tHooft:1984kcu}. These approaches naively give an infinite answer, due to the infinity of quantum field theory modes in the ultraviolet. For sufficiently low-spin fields, at least, such divergences can be absorbed into a renormalization of Newton's constant, so that the denominator in Eq. \ref{exp} is understood as the observed low-energy Newton's constant (see \cite{Jacobson:2012ek,Cooperman:2013iqr}\ for discussion). However, puzzles remain, such as the interpretation of the contribution of the bare Newton's constant to Eq.~\ref{exp}, the validity of the approach in a full quantum theory of  gravity, and the relationship of this interpretation to the state-counting interpretation when the latter is available.\footnote{In the case of the AdS/CFT correspondence, \cite{Chandrasekaran:2022eqq} has shown an equivalence between the Bekenstein-Hawking  entropy in AdS and an entanglement entropy of a Type II algebra defined in the dual CFT.  Here we will follow a different logic, and work directly with the underlying discrete Hilbert space of the black hole.}

The new problem concerns a conjecture about the behavior of the volume of the Einstein-Rosen bridge behind the horizon of an eternal black hole \cite{Susskind:2014rva,Stanford:2014jda}. The conjecture suggests that the volume of such Einstein-Rosen bridges is related to the ``quantum complexity'' \cite{Aaronson:2016vto} of the underlying state. In classical gravity, the volume of Einstein-Rosen bridge grows linearly with respect to the asymptotic time.  Thus the conjecture predicts that the complexity, i.e., the number of gates of the minimal  circuit simulating the time evolution of the quantum state, will increase linearly with time.  On the other hand, it is known that the complexity of any circuit is bounded by the dimension of the Hilbert space on which the circuit acts (see Chapter 7 of \cite{Aaronson:2016vto} and the discussion in \cite{Balasubramanian:2019wgd}).  Thus, the conjecture also suggests that the volume of the Einstein-Rosen bridge must saturate at exponential time, and hence at a value   exponentially large in the black hole entropy, namely at $ \mathcal{O}(e^{S_{BH}})$. However, in semiclassical gravity, the Einstein-Rosen bridge can grow indefinitely and no saturation is observed.\footnote{The authors of \cite{Iliesiu:2021ari} put forward an interesting approach to the interior volume of black holes in 2d gravity, where they defined a path integral definition of ``volume'' that saturates at exponentially large times. Here we will follow a different approach that applies in any dimension.}

We will show that there is a relation between these two problems:  the expected finiteness of black hole entropy, and the conjectured finiteness of Einstein-Rosen bridges.  Indeed, we will provide a simple and universal solution to both problems in gravitational models in any dimension. To do so we will only assume that general relativity, the well-established low-energy theory of gravity, has some ultraviolet completion with a well-defined semiclassical path integral approximation.

The first step in our analysis is to recognize that the problem at hand is not to find a {\it specific} family of $e^{S_{BH}}$ microstates. Indeed, below  we will explicitly construct  many different families of black hole microstates, each with infinitely many members.\footnote{In this work we define ``black hole microstate'' to be any gravitational state with same exterior geometry as the black hole. } These will already, at the time-symmetric point, include geometries with Einstein-Rosen bridges of arbitrarily large volume. This connects our puzzle about time evolution of the interior Einstein-Rosen bridge to the puzzle regarding the apparently excessive number of microstates. However, we will show that the dimension of the Hilbert space spanned by any of these infinite families is actually finite, and given by the exponential of the Bekenstein-Hawking entropy~(\ref{exp}).  This enormous reduction in the Hilbert space dimension as compared to the naive expectation originates from the  existence of wormhole saddlepoints of the gravitational path integral, discussed recently in the context of heavy operator statistics in the AdS/CFT correspondence \cite{Martin}. We demonstrate that these wormholes give rise to tiny, but universal, contributions to the quantum overlap of the candidate black hole microstates.   Thus, states in the infinitely many families that we construct are not orthogonal, and a computation establishes that they span a Hilbert space of dimension that precisely equals the exponential of the Bekenstein-Hawking entropy.\footnote{Our results recall the idea that semiclassical gravity contains an enormously many null states \cite{Marolf:2020xie,Balasubramanian:2020jhl} and the proposal that the AdS/CFT encoding map is highly non-isometric because many classically distinct states are mapped to non-orthogonal quantum states \cite{Balasubramanian:2022fiy,Akers:2022qdl,Kar:2022qkf}.  Several authors have also pointed out that a finite dimensional Hilbert space, like a microcanonical sector of the CFT dual to AdS, can host a large number of approximately orthogonal states. In fact the number of states with a fixed small overlap is exponentially large in the Hilbert space dimension, see, e.g.,  \cite{Chakravarty:2020wdm,chao2017overlapping}.  In our case the overlaps and the Hilbert space dimension will be determined by the Euclidean gravity path integral.}  The same overlaps will allow us to interpret Einstein-Rosen bridges of any volume as superpositions of wormholes of volume bounded by an exponential in the entropy.

Our results will generalize and clarify recent insights from low dimensional gravity, in particular  concerning the spectral form factor in black hole physics \cite{Cotler:2016fpe,Saad:2018bqo,Saad:2019lba}, and the derivation of the Page curve via replica wormholes (see \cite{Mertens:2022irh} for a review). In particular our construction will not use artifices like  ``end-of-the-world branes'' with unknown degrees of freedom, and will be applicable
to general relativity in general dimensions.

Five sections follow. In Sec.~\ref{SecII} we construct infinite families of geometric solutions to general relativity with a negative cosmological constant that have different interior geometries behind the horizons of a fixed exterior eternal black hole.  The two  asymptotic regions of this  black hole can have different masses.
The interior regions include shells of dust moving  within a long Einstein-Rosen bridge connecting the two asymptotic regions.  Our states are semiclassically well-defined and can be constructed via Lorentzian continuation of an Euclidean geometry.
In the boundary dual conformal field theory this construction has a well-defined interpretation in terms of the state-operator correspondence. Thus these distinct interior geometries  necessarily contribute to the dimension of the black hole Hilbert space. However, they naively lead to a black hole entropy far exceeding the Bekenstein-Hawking formula. The apparent overcounting resembles an old puzzle of Wheeler, concerning ``bag of gold'' geometries \cite{Wheeler,Almheiri:2020cfm}. Thus, our  construction  sharpens a basic question: how do we count the independent degrees of freedom contributing to the entropy of a black hole.

In Sec.~\ref{SecIV} we show why the states we constructed in Sec.~\ref{SecII} are not orthogonal: they overlap quantum mechanically because of the effects of the wormholes found in \cite{Martin}. These wormholes contribute to the Euclidean path integral, exist for all dimensions and for all our states, and their action is straightforward to compute using general relativity. For most states in every family, the contribution from the wormhole is a universal functional of the partition function of the theory. We also compute higher moments of the overlaps, finding universal answers as well. Schematically, in the limit of heavy shells of dust with different masses, these wormholes contribute to the statistics of the overlaps so that, universally, the n-th moment of the overlap is
   \be\label{UniAmp2}
    \overline{\bra{\Psi_{\mathbf{m}}} \ket{\Psi_{\mathbf{m}'}}...\bra{\Psi_{\mathbf{m}'...'}} \ket{\Psi_{\mathbf{m}}}} \, \simeq \, \dfrac{Z(n\beta)^{2}}{Z(\beta)^{2n}}\;,
    \ee
where the overline means we are performing the computation using the gravitational path integral, $Z(\beta)$ is the Euclidean partition function of the gravitational theory, and the square comes from the consideration of eternal black holes, that are naturally associated to twice the entropy.

In Sec.~\ref{SecV} we describe how such exponentially small, but non-vanishing overlaps determine a Hilbert space of finite dimension given by the exponential of black hole entropy. Physically, the excess microstates do not add to the entropy since they do not generate new orthogonal Hilbert space directions. We show this by demonstrating an explicit transition in the overlap matrix, the so-called Gram matrix of the black hole microstates, as we increase the number of microstates under consideration. The transition happens when the Gram matrix goes from being positive definite to positive semidefinite, implying that the added vectors are not linearly independent. This transition does not depend on the specific set of microstates that we start with. Once the transition occurs,  we do not generate new orthogonal Hilbert space directions. Our arguments provide a way of computing the entropy for general black holes, using the large, overcomplete basis of microstates that we construct.

In Sec.~\ref{SecVI} we describe the implications of our construction for the interior volumes of black holes.
First, we compute the volumes of the Einstein-Rosen bridges associated with the families of black hole microstates constructed in Sec.~\ref{SecII}. We will see that these volumes are simple functions of the proper masses (scaling dimensions in the AdS/CFT context) of the operators used to construct the states in the gravitational description. Since there is no  limit on the energy or scaling dimension of such operators, there is no limit on the interior volumes even at a fixed time. This sharpens the problem regarding the conjecture that relates wormhole volume and quantum complexity, since all these states are geometrical and semiclassical.  We will also compute the maximal transversal size of our Einstein-Rosen bridges, namely the sizes of their  ``pythons'' \cite{Brown:2019rox}.  We will then show that any state which has an Einstein-Rosen bridge whose size is super-exponential in the entropy can  be rewritten as a superposition of states with at most exponentially large wormholes. Thus, there is  a transition beyond which the notion of ``volume'' requires refinement. If we define volume by a minimization over the possible linear definitions, this notion saturates at an exponentially large value. We use recent advances on spread complexity, the tridiagonalization of random matrices, and the associated Krylov basis \cite{Balasubramanian:2022tpr,Balasubramanian:2022dnj} to construct a unitary toy model of these effects. 

In Sec.~\ref{SecVII} we end with conclusions and open problems.

{\it Reader's guide:} to understand the basic points of the paper, the reader can skip Sec. \ref{Sec:2_3}, Sec. \ref{Sec:3_4}, Sec. \ref{Sec:4_3} and Sec. \ref{SecVI} on a first read. 

\section{Infinite geometric families of black hole microstates}\label{SecII}

The no-hair theorem states that all stationary black hole solutions of the Einstein equations are completely characterized by a few independent externally observable classical parameters, such as mass, electric charge, and angular momentum. This is sometimes incorrectly understood as an obstruction to finding black hole microstates. In fact, the same black hole can develop from many initial states that differ in their late-time interiors -- we can regard these different interiors as examples of ``microstates''.  Likewise, an operation creating an excitation of a quantum field behind the horizon that does not modify the mass can also be regarded as making a microstate.  
In the holographic AdS/CFT context, such operations in the interior of the black hole should be encoded in the exact dual CFT description.\footnote{See \cite{Kim:2020cds,Balasubramanian:2022fiy,Akers:2022qdl,Kar:2022qkf} for recent progress in this direction.} This idea also appears in more general terms in the old notion of ``black hole complementarity'' \cite{Susskind:1993if}.

Here, we will build infinite families of geometric microstates of a large eternal AdS black hole, by exploring families of states with exteriors given by the black hole geometry, but with distinct interiors arising from  backreaction caused by shells of dust.\footnote{Black hole like solutions in general relativity which differ only in the interior geometry have appeared in e.g. \cite{Frolov:1998wf,Hsu:2008yi,Fu:2019oyc}.} We will prepare these states via the Euclidean path integral, generalizing the construction in \cite{Kourkoulou:2017zaj,Goel:2018ubv} for SYK/JT gravity to a generic higher dimensional AdS/CFT setup. Our states also generalize the infalling shell states considered in \cite{Chandra:2022fwi} to various numbers of  shells behind the horizon of a spacetime with two asymptotic boundaries and interior backreaction. The dual CFT interpretation of our construction shows that these are pure states of  a quantum theory of gravity with two asymptotically AdS boundaries.

We examine a small subset of the possible configurations that can be built using our construction.   However, as we will argue later, the additional states would not increase the dimension of the spanned Hilbert space. In fact, our construction already provides an infinite number of candidate microstates, which naively overcount the Bekenstein-Hawking entropy.
One of our main messages is that it does not really matter how  we count states, or which states we count. The only important thing is to count them carefully.  Namely, we have to determine the actual dimension of the spanned Hilbert space. The main advantage of the families of states that we present below  is that they can be described geometrically and permit explicit calculations.

\subsection{State preparation and thin shell operators}

In detail, we consider states in a theory of gravity with two asymptotic AdS boundaries with topology $\mathbf{S}^{d-1}\times \mathbf{R}$, where $\mathbf{R}$ represents time, and   asymptotic AdS curvature radius $\ell$. Quantum gravity with such boundary conditions is equivalent to the physics of two copies of a Conformal Field Theory (CFT), dubbed CFT$_L$ and CFT$_R$, living at the left/right boundaries and with Hamiltonians $H_L=H_R=H$. The spatial component of each of these boundaries is  $\mathbf{S}^{d-1}$. The total Hilbert space is just the tensor product $\mathcal{H}_L^{CFT} \,\otimes\, \mathcal{H}_R^{CFT}$. The microcanonical Hilbert space dimension at a given energy in this theory is thus twice the microcanonical dimension of a single sided black hole. We define the energy basis as
\be 
H_L\ket{n,m}=E_n\ket{n,m}\,\,\,\,\,\,\,\,\,\,\,\,\,\,\,\,\,H_R\ket{n,m}=E_m\ket{n,m}\;.
\ee
\begin{figure}[h]
		\centering
		\includegraphics[width=.8\textwidth]{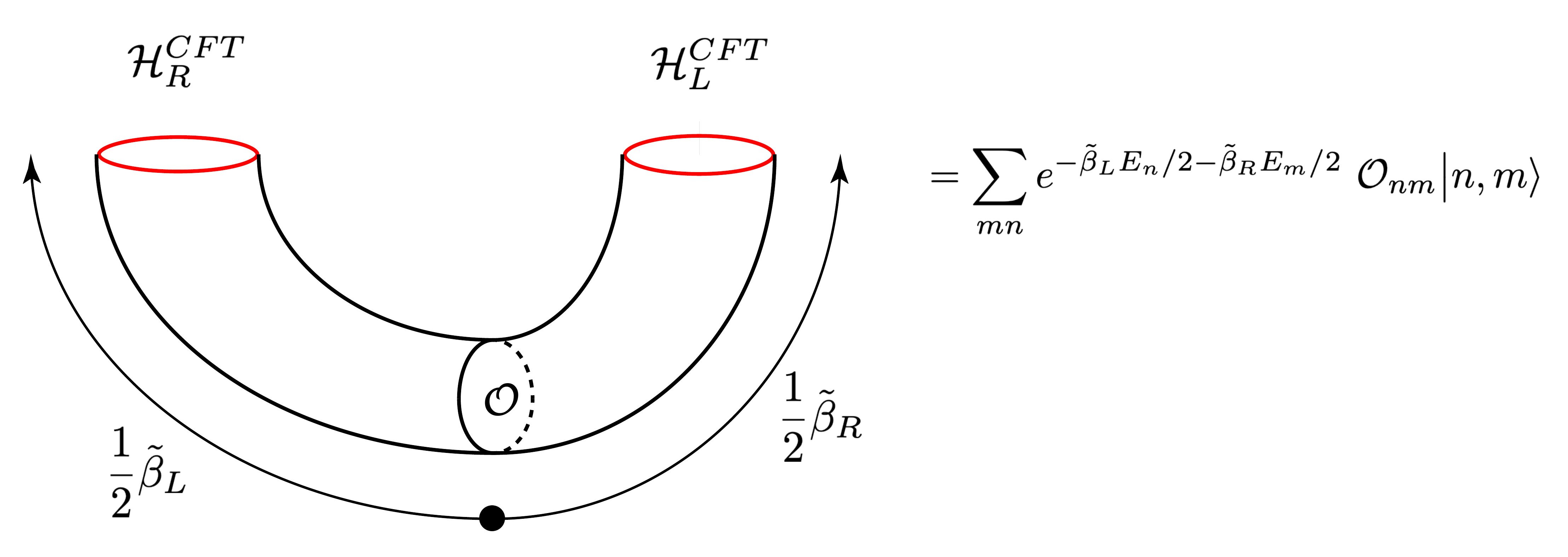}
		\caption{Schematic form of the Euclidean path integral in the boundary field theory, which is two copies of a holographic CFT with Hilbert space ${\cal H}^{CFT}$. The cross section of the tube represents the spatial direction, and the length is Euclidean time. The operator ${\cal O}$ is inserted in the circle at the bottom, and the tube corresponds to propagation in Euclidean time via the  Hamiltonian of the underlying CFT, by ${\tilde \beta}_{L}$ (${\tilde \beta}_{R}$) along to the left (right) of the operator insertion.} 
		\label{fig::Euclid}
	\end{figure}
Quantum states in the doubled Hilbert space can be prepared by a Euclidean path integral for a single CFT on a finite cylinder, the argument of the wavefunctional for each factor given by the value of the fields at the ends of the cylinder  (Fig.~\ref{fig::Euclid}). We insert an operator ${\cal O}$ in the single CFT at the spatial slice denoted by the bottom circle in Fig.~\ref{fig::Euclid}. The cylinder extends by Euclidean time ${\tilde \beta}_L/2$ to the left and  ${\tilde \beta}_R/2$ to the right. The resulting state can be described by starting with the operator
\be 
\label{CJop} \rho_{\tilde{\beta}_{L}/2}\,\mathcal{O}\,\rho_{\tilde{\beta}_{R}/2}\ ,
\ee
where $\rho_{\beta}=e^{-\beta\,H}$ is the unnormalized thermal density matrix,
defined in a single copy of the CFT on $S^d\times\mathbf{R}$. This is identified with a state in a doubled Hilbert space via the usual isomorphism. We can write this schematically as
\be
\ket{\Psi}\, =  \,\ket{ \rho_{\tilde{\beta}_{L}/2}\,\mathcal{O}\,\rho_{\tilde{\beta}_{R}/2}}
 =  \dfrac{1}{\sqrt{Z_1}}\, \sum_{n,m}\,e^{-\frac{1}{2}(\tbeta_L E_n + \tbeta_R E_m)}\,\mathcal{O}_{nm}\,\ket{n,m}\;,\label{PETS2a}
\ee
where $Z_1 =  \text{Tr}(\mathcal{O}^\dagger e^{-\tbeta_L H}\mathcal{O}e^{-\tbeta_R H})$ normalizes these pure states. Note that the trace defining $Z_1$ is taken in a {\it single}\ copy of the CFT.\footnote{The reason for the notation $Z_1$ will become clearer as we move along the article.}

We will show that these states correspond to two-sided black holes with independent masses $M_{+,-}$ and inverse temperatures $\beta_{R,L}$ respectively.\footnote{We keep the notation $M_{+,-}$ since it is conventional in the analysis of thin shells in general relativity. The same will follow for the associated manifolds.} Note that the Euclidean times $\tilde{\beta}_{R,L}$ used to prepare the states through Euclidean evolution in (\ref{CJop}) are not necessarily equal to the physical inverse temperatures of the black holes. We will derive the precise relationship for a certain class of operators ${\cal O}$ below.
	
	\begin{figure}[h]
		\centering
		\includegraphics[width=.65\textwidth]{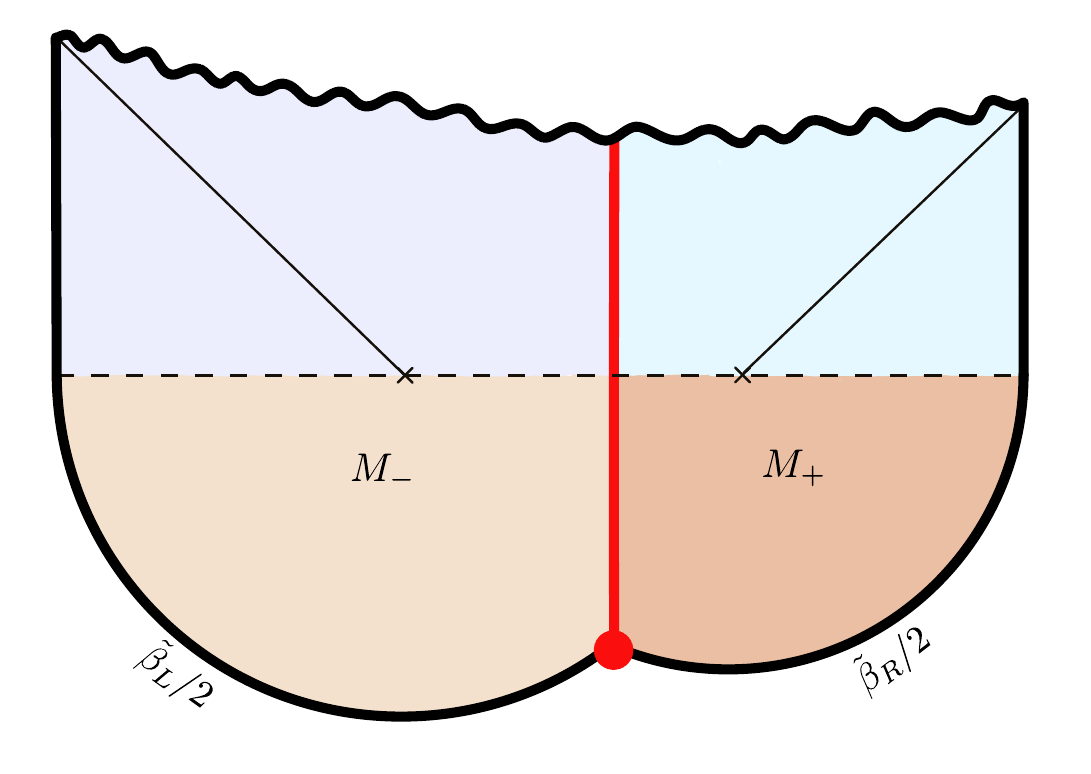}
		\caption{Preparation of the state from the Euclidean path integral. The lower part of the figure (tan colors) shows time evolutions from a past Euclidean boundary with a shell operator (red dot) inserted.  In the Euclidean section the  horizon is a point, here represented by crosses for the  the left and right black holes on either side of the shell.   At $t=0$ the Euclidean preparation geometry is matched onto a Lorentzian time evolution (blue and purple).  The black hole horizons (thin black lines) become null surfaces, while the shell (red line) continues to move into the future behind both horizons.   Spacetime is glued across the trajectory of the thin shell by Israel's junction conditions. Here, we have allowed for the possibility that the black on either side of the shell have different temperatures or masses.
  The solid black line is the spacetime boundary.  For $d$-dimensional AdS geometries this will be a cylinder  $\mathbf{S}^{d-1} \times R$ on which the dual CFT lives.}
		\label{fig::one}
	\end{figure}

We can construct another family of states via the Euclidean path integral, with Euclidean time-evolution by $\tilde{\beta}_R/2$ at the right boundary, followed by the insertion of an operator $\mathcal{O}_1$, followed by  evolution with $\tilde{\beta}_{1}$, then by another operator $\mathcal{O}_2$, and finally by a further  time-evolution of $\tilde{\beta}_L/2$ up to the left boundary. This prepares a pure state which can be schematically written as
\be 
\ket{\Psi}\,=\,\ket{ \rho_{\tilde{\beta}_L/2}\,\mathcal{O}_2\,\rho_{\tilde{\beta}_1}\,\mathcal{O}_1\,\rho_{\tilde{\beta}_R/2}}\;.
\ee
More generally, given operators $\mathcal{O}_{1},\cdots ,\mathcal{O}_n$ and Euclidean lengths $\tilde{\beta}_{1},\cdots ,\tilde{\beta}_{n-1}$ we can build
\be\label{eq:multishell}
\ket{\Psi}\,=\,\ket{ \rho_{\tilde{\beta}_L/2}\,\mathcal{O}_n\,\rho_{\tilde{\beta}_{n-1}}\,\cdots\,\rho_{\tilde{\beta}_1}\,\mathcal{O}_1\,\rho_{\tilde{\beta}_{R}/2}}\;.
\ee
By construction all these states belong to the Hilbert space $\mathcal{H}_L^{CFT} \,\otimes\, \mathcal{H}_R^{CFT}$ associated with the two asymptotic CFT's. We will construct bulk duals for states of this form that contribute to the semiclassical Hilbert space of a black hole with fixed exterior geometry, which is to say with fixed ADM masses and exterior temperatures in each asymptotic exterior region. Thus they are candidates for microstates of these black holes.

We are interested in states that have a simple semiclassical and geometrical description in the gravitational theory. To this end, we consider operators that create spherically symmetric ``dust shells''.  To do this we can can start with any scalar operator ${\cal O}_{\Delta}$ with a scaling dimension $\Delta$ of $O(1)$.  Such operators are dual to scalar fields in the gravity theory with mass $m_{\Delta}\,\ell=\sqrt{\Delta(\Delta-d)}$.  Standard constructions \cite{Balasubramanian:1998de,Banks:1998dd,Balasubramanian:1999ri,Hamilton:2006az} show that application of ${\cal O}_{\Delta}$ at a spherically symmetric distributions of points in the  CFT is dual, after appropriate holographic renormalization \cite{Balasubramanian:1999re,Henningson:1998gx,Skenderis:2002wp}, to applying the bulk scalar field to the same distribution of points at the AdS boundary. This creates a shell of matter $\mathcal{O}$ at the boundary that will propagate into the bulk.  In our construction, any operator ${\cal O}_{\Delta}$ will suffice because, in an interacting theory, we expect that the algebra generated by ${\cal O}_{\Delta}$ will contain all operators of the theory, allowing us access to any state in the theory, although in a possibly complicated way.

We also require that the operator we are constructing should be sufficiently heavy, i.e., that it creates states with masses of $O(1/G)$, in units of the AdS radius $\ell$, where $G$ is Newton's constant. With this choice,  $\mathcal{O}$ creates a spherically symmetric heavy shell of dust particles, which classically backreacts on the geometry at leading order in the $G\rightarrow 0$ expansion. To this end, we use a number of dust particles which scales parametrically as $n\sim \ell^{d-1}/G $, homogeneously distributed along the sphere.\footnote{In the AdS/CFT context this means a number scaling with the central charge of the CFT.} The final dust shell operator can be effectively described as a presureless perfect fluid localized at the worldvolume $\mathscr{W}$ of the shell. This fluid has the usual energy-momentum tensor
	\be
	T_{\mu\nu}\Big|_\mathscr{W} = \sigma\, u_\mu \,u_\nu \;,
	\ee
where $u^\mu$ is the proper fluid velocity, tangent to $\mathscr{W}$, and $\sigma$ is the surface density. The total rest mass of the dust shell is
	\be
	m =  \sigma \,V_\Omega\, r_\infty^{d-1}=n\,m_{\Delta}\,,
	\label{eq:littlem}
	\ee
where $V_\Omega = \text{Vol}(\mathbf{S}^{d-1})$ is the volume of the sphere, and  $m_{\Delta}$ is the mass of the individual operator insertions. The effective perfect fluid description of the operator works for fluid densities $\sigma$ that are sufficiently large, compared to one dust particle per unit volume, but also sufficiently small, when measured in Planck units, so that we can trust the classical description.

\subsection{Geometry of single-shell states}

Our strategy for preparing bulk microstates is to find the bulk Euclidean geometry dual to the Euclidean calculation shown in Figure \ref{fig::Euclid}. The full geometry is symmetric under Euclidean time reversal; the fields at the $\tau = 0$ spatial slice are fixed under this symmetry. So we can continue to Lorentzian signature at this slice.

\subsection*{Euclidean geometry}

We start with one-shell microstates. The operator $\mathcal{O}$  inserts a thin spherical  domain wall of dust particles. This shell then propagates in Euclidean time. Since the mass is large in Planck units, we have to account for its backreaction on the geometry.

More precisely, the worldvolume $\mathscr{W}$ of the thin shell bisects the Euclidean manifold $X$ and generates two connected components $X^\pm \subset X$, one on each side of $\mathcal{W}$. This is depicted in Fig. \ref{fig::one}. Given the spherical symmetry, the geometry of each component $X^\pm$ is that of a Euclidean black hole
	\be\label{eq:geometry}
	\text{d}s_\pm^2\, = \,  f_\pm(r)\,\text{d}\tau_\pm \,+\,\dfrac{\text{d}r^2}{f_\pm(r)}\,+\,r^2\,\text{d}\Omega_{d-1}^2\;,
	\ee
where
	\begin{gather}
	f_\pm(r)\, = \, \dfrac{r^2}{\ell^2}\,+\,1\,-\,\dfrac{16\pi G M_\pm }{(d-1) V_\Omega \,r^{d-2}}\,\hspace{.5cm}\text{for } d>2\;,\label{metric}\\[.4cm]
	f_\pm(r)\, = \, \dfrac{r^2}{\ell^2}\,\,-\,8GM_\pm\hspace{.5cm}\text{for } d=2\;,
	\end{gather}
 Here $r$ is a radial coordinate on the Euclidean disc and $\tau_\pm \sim \tau_\pm + \beta_\pm$ are angular coordinates around periodic Euclidean time. 
In what follows we will write all dimensionful parameters in units of the AdS radius $\ell$, effectively $\ell = 1$. Here $M_\pm$ is the ADM mass of the black hole in component $X_{\pm}$, with inverse temperature $\beta_\pm = 4\pi/f'(r_\pm)$, where $f_\pm'$ is the derivative of $f_\pm$ evaluated at the horizon radii $r_\pm$ where $f_\pm(r) = 0$.

The shell follows a trajectory  parametrized in terms of $T$, the synchronous proper time of the dust particles, via functions $r=R(T)$ and $\tau_{\pm}=\tau_{\pm}(T)$.  The rest mass of the shell $m$ is conserved along its trajectory. Its surface density at radius $R$ is therefore defined by
	\be\label{eq:density}
	\sigma = \dfrac{m}{V_\Omega\,R^{d-1}}\;,
	\ee
 where $V_\Omega$ is the volume of the unit transverse sphere.
In the thin-shell formalism, the shell's dynamics, described just by its trajectory, is reduced to the motion of a non-relativistic effective particle with zero total energy
	\be\label{eq:eomshell}
	\left(\dfrac{\text{d}R}{\text{d}T}\right)^2 + V_{\text{eff}}(R) = 0\;,
	\ee
subject to the effective potential
	\be\label{eq:Veff}
	V_{\text{eff}}(R) = -f_+(R) + \left(\dfrac{M_+-M_-}{m} - \dfrac{4\pi G m}{(d-1)V_\Omega R^{d-2}}\right)^2\;.
	\ee
This potential arises from Israel's junctions conditions \cite{Israel:1966rt} along the shell. Appendix~\ref{appendix:A} gives details about this gluing proces and the derivation of the potential.  Thus each single shell configuration for a black hole of fixed mass is characterized by the shell mass $m$.
 
\begin{figure}[h]
		\centering
		\includegraphics[width=.5\textwidth]{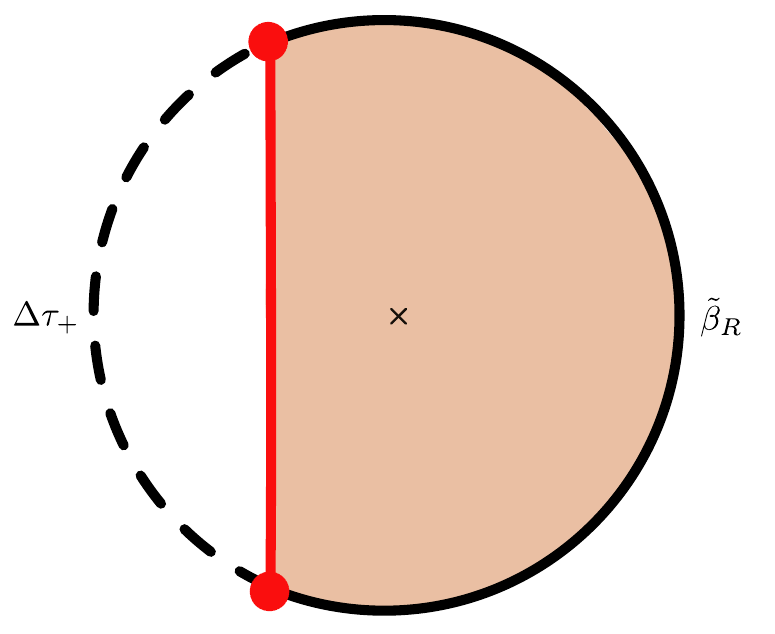}
		\caption{We show here the Euclidean geometry of the right black hole. This is a disk of circumference $\beta_R$. The shell (red line) cuts the disk, with an associated Euclidean travel time $\Delta \tau_+$. Consistency fixes the state preparation temperature to satisfy $\beta_R=\tilde{\beta}_R+\Delta\tau_+$.}
		\label{fig::three}
	\end{figure}

Qualitatively, the thin shell starts at the boundary $R = r_\infty$ and falls inwards towards the tip of the geometry (the Euclidean horizon) at $R=r_\pm$.  The trajectory is subject to a repulsive force in the Euclidean section. Thus it bounces back at a minimum radius $R = R_* \geq r_\pm $ at the axis of time-reflection symmetry of the solution, and then returns to $R = r_\infty$. In our calculations we will send $r_\infty \to \infty$. The Euclidean time elapsed by the shell $\Delta \tau_\pm$ as described in patches to on either of side of it, $X^\pm$, is 
\be\label{eq:shifttime}
	\Delta \tau_{\pm}  = 2 \int_{R_*}^{r_\infty}\dfrac{\text{d}R}{f_\pm(R)} \, \sqrt{\dfrac{f_\pm(R) + V_{\text{eff}}(R)}{- V_{\text{eff}}(R)}}\;.
\ee
Therefore, the  evolution of the one-shell states~\eqref{PETS2a} is performed for the imaginary ``preparation times"
	\begin{gather}
	\tbeta_{L}\, = \, \beta_{L} \,- \,{\Delta \tau_-}\;,\label{efftempL}\\
	\tbeta_{R}\, = \, \beta_{R} \,- \,{\Delta \tau_+}\;,\label{efftempR}
	\end{gather}
which differ from the fixed inverse temperatures $\beta_{L,R}$ of the actual black holes (see Fig. \ref{fig::three}), which also fixes their masses.\footnote{Basically, the total Euclidean time, say $\beta_L$ should equal the sum of $\Delta \tau$ and the preparation temperature $\tilde{\beta}_L$ as in Fig.~\ref{fig::three}.} 
This condition ensures that the backreaction of the shell does not change the asymptotic mass of the geometry on either side.  Notice that left and right asymptotic black hole regions can have different temperatures $\beta_L$ and $\beta_R$ because of the presence of the shell which serves as a domain wall supporting this temperature difference.  Here, the trajectory of the shell in (\ref{eq:shifttime}) is fixed by the mass (temperature) of the background black hole and the mass of the shell, and the resulted elapsed Euclidean time in the trajectory  fixes the required preparation time.  If the mass of the shell is taken to zero such solutions do not exist for different left/right black hole masses. Equation~(\ref{eq:shifttime}), together with~(\ref{efftempL}) and (\ref{efftempR}), ensure that the trajectory passes in between both horizons in the Euclidean geometry (Fig.~\ref{fig::one}) so that the Lorentzian continuation is an eternal black hole with a shell inside.\footnote{It is also possible to find classical trajectories for the shell separating $X^\pm$ where the elapsed Euclidean time (\ref{eq:shifttime}) is too large to satisfy (\ref{efftempL}) and (\ref{efftempR}).  In this case the shell will lie to the ``right'' of both Euclidean horizons indicated in (Fig.~\ref{fig::one}); i.e., in the Lorentzian continuation it will lie outside the horizon of the right-hand black hole.  We will not consider this class of gravitational states.}
	
\subsection*{Lorentzian geometry}

In the Lorentzian section (the blue and purple parts of Fig.~\ref{fig::one}), the shell follows the equation of motion $\dot{R}^2-{V}_{\text{eff}}(R) =  0$, arising from  analytic continuation of the shell's proper time $T\rightarrow iT$. The shell now emerges from the past singularity at finite proper time $T=-T_0$ in the past, reaches the radius $R=R_*$ inside the black hole at $T=0$ and dives into the future singularity at $T=T_0$.

The geometry is that of two Schwarzschild-AdS black holes, with inverse temperatures $\beta_{L,R}$, glued  along the trajectory of the shell's worldvolume $\mathcal{W}$ in the black hole interior (Fig. \ref{fig::one}). The induced metric in the time-reflection symmetric slice is that of a `python's lunch', that is, there is a bulge where  area of the transverse spatial spheres reaches a maximum, at $r=R_*$, between the two horizons.\footnote{Notice the marked differences of the present construction of black hole microstates, in comparison with the fuzzball program \cite{https://doi.org/10.48550/arxiv.2204.13113}. The fuzzball program aims to quantize a particular space of smooth supergravity solutions with no horizons. Here we are identifying an infinite set of black hole microstates with horizons and singularities. That this should not be considered as a problem is already clear in the construction of the Hartle-Hawking state itself (i.e. the state with no dust shell inserted), which has horizons and singularities, and corresponds to a perfectly well-defined state in the dual CFT, namely the TFD state \cite{Maldacena:2001kr}.}

\subsection*{Example: 2+1 dimensions}
	
For $d=2$, the shell's trajectory can be explicitly evaluated. The BTZ horizon radii are given by $r_\pm = \sqrt{8GM_\pm}$, and the inverse temperatures by $\beta_\pm     = 2\pi/r_\pm$. The effective potential \eqref{eq:Veff} in this case is quadratic
	\be\label{2+1potential}
	{V}_{\text{eff}}(R)\, = \, -(r^2-R_*^2)\;,
	\ee  
where the turning point $R_*$ is 
	\be\label{2+1radius}
	R_*\, = \, \sqrt{r_+^2\,+\,\left(\dfrac{M_+-M_-}{m}\,-\,2Gm\right)^2}\;.
	\ee
The solution for the shell's trajectory \eqref{eq:eomshell} is
	\be\label{eq:app2+1sol}
	R(T)\, = \, R_*\,\cosh T\;,
	\ee
where we have chosen the initial condition such that the shell passes through $R_*$ at proper time $T=0$. The Euclidean time elapsed by the shell \eqref{eq:shifttime} can also be computed analytically. We find
	\be\label{eq:etime2+1}
	\Delta \tau_\pm \, = \beta_\pm \,\dfrac{\arcsin(r_\pm/R_*)}{\pi}\;.
	\ee

\subsection{Multi-shell states }\label{Sec:2_3}
We can also describe  situations with  multiple shells characterized by a mass vector ${\bf m}= \{ m_1, m_2, \cdots \}$. In the Euclidean picture, we can insert as many shells as we want behind the horizon, and the above analysis applies locally to each trajectory $R_{i}(T)$.

\begin{figure}[h]
		\centering
		\includegraphics[width=.5\textwidth]{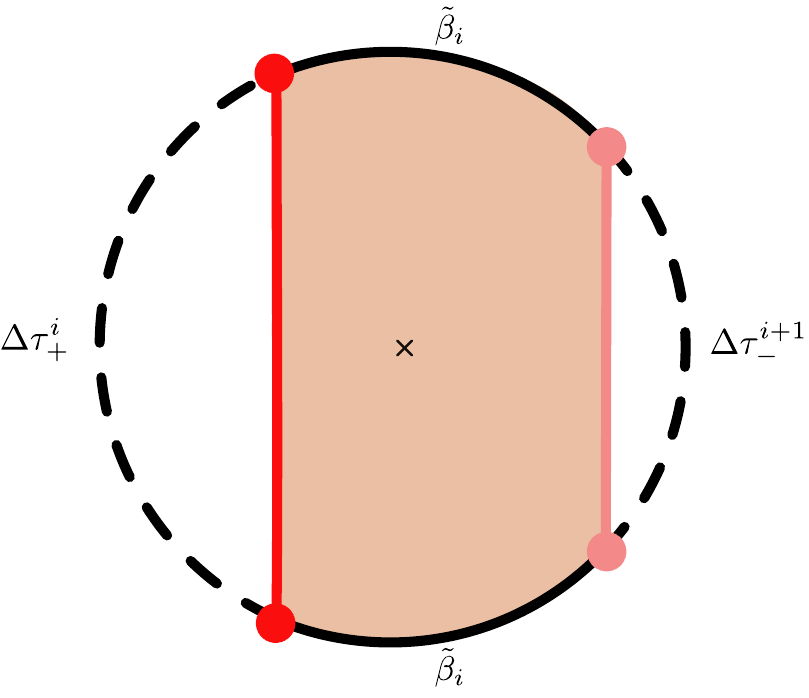}
		\caption{This is a Euclidean disk geometry (a Euclidean black hole) in between two shells (red lines). The shell travel times $\tau_+^i$, $\tau_-^{i+1}$ plus twice the preparation  temperature $\tilde{\beta}_i$ must be equal to the physical temperature $\beta_i$ of this black hole geometry between the shells.}
		\label{fig::timelapsemulti}
	\end{figure}

Such a multi-shell state will have the form \eqref{eq:multishell}.
The region between any pair of shells is locally a black hole geometry and must have (see Fig. \ref{fig::timelapsemulti})
\be\label{multishell}
\beta_i = 2\tbeta_i + \Delta \tau_+^i + \Delta \tau_-^{i+1} 
\ee
where $\Delta \tau^i_\pm$ depends on $m_i$ and on parameters $M_{i-1}$, $M_{i}$ which are ``masses'' characterizing the local geometries between the shells. From the consistency of the classical solution, the mass difference $M_i - M_{i-1}$ is constrained to be upper bounded by the $i$-th shell's gravitational self-energy at $R_*$.

\begin{figure}[h]
		\centering
		\includegraphics[width=.85\textwidth]{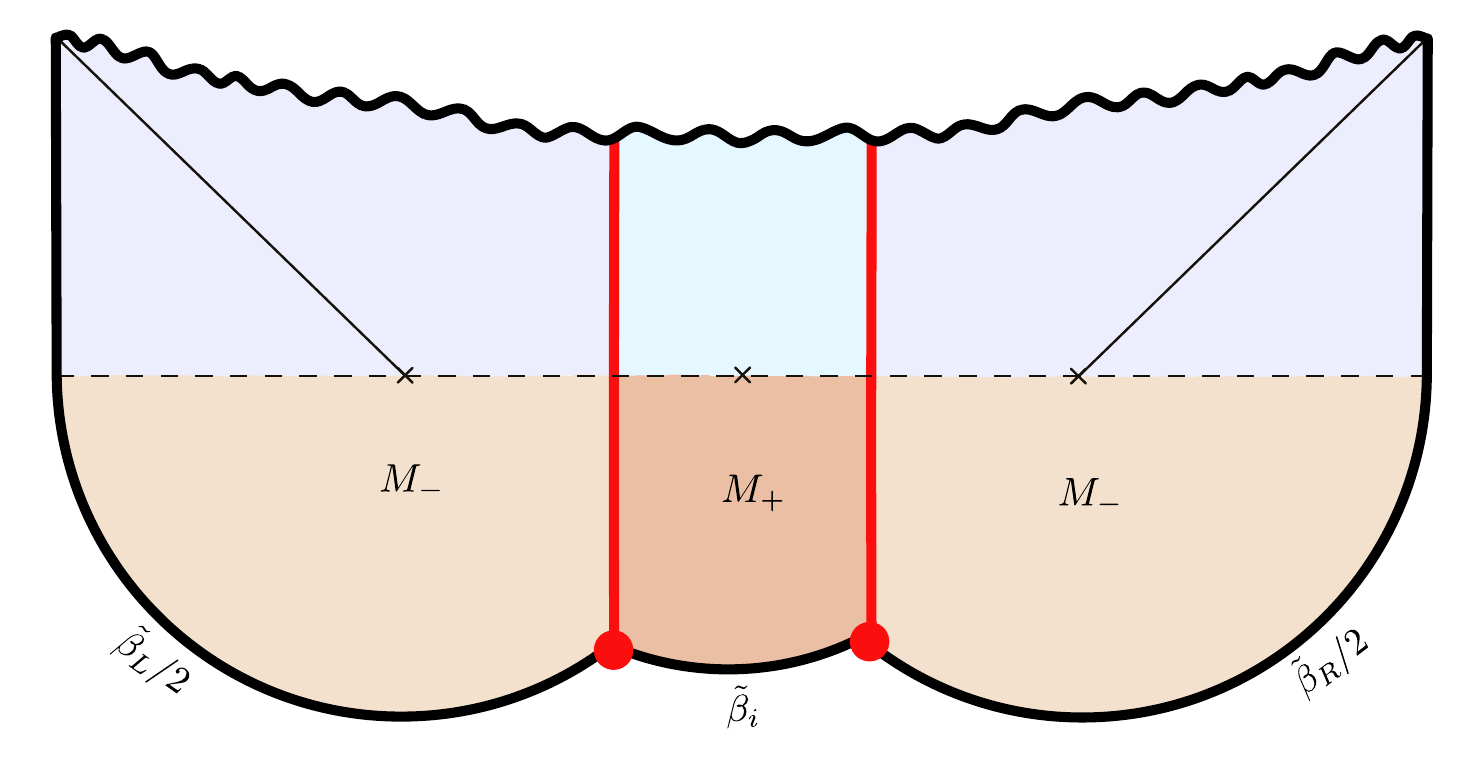}
		\caption{Example of a multi-shell state of a eternal black hole with ADM masses $M_-$ and and interior geometry with associated mass parameter $M_+$ . 
		}
		\label{fig::two}
	\end{figure}

The formulae \eqref{eq:eomshell} and \eqref{eq:Veff} go through for the dynamics of each shell, with $M_i$ and $M_{i+1}$ replacing the black hole masses, since the thin-shell formalism is a local analysis in the vicinity of the shell. The preparation of such states is shown in Fig. \ref{fig::two}.

\section{Overlaps between states from semiclassical wormholes} \label{SecIV}

Consider a pair of black hole microstates $\ket{\Psi_{\mathbf{m}}}, \ket{\Psi_{\mathbf{m}'}}$, for generic shell configurations $\mathbf{m},\mathbf{m}'$ representing classically different interior geometries in the $G\rightarrow 0$ limit. For example, for one-shell states, this means configurations whose rest mass differences scale parametrically as $ \Delta m  = O(\ell^{d-2}/G)$. The objective of this section is to compute the  overlaps and their products
\be\label{statin}
\overline{\bra{\Psi_{\mathbf{m}}} \ket{\Psi_{\mathbf{m}'}}}\;,\,\,\,\,\,\,\,\,\,\,\,\,\,\,\,\,\,\overline{\bra{\Psi_{\mathbf{m}}} \ket{\Psi_{\mathbf{m}'}}\bra{\Psi_{\mathbf{m}'}} \ket{\Psi_{\mathbf{m}''}}}\;,\,\,\,\,\,\,\,\,\,\,\,\,\,\,\,\,\,\overline{\bra{\Psi_{\mathbf{m}}} \ket{\Psi_{\mathbf{m}'}}\bra{\Psi_{\mathbf{m}'}} \ket{\Psi_{\mathbf{m}''}}\bra{\Psi_{\mathbf{m}''}} \ket{\Psi_{\mathbf{m}'''}}}\;,\,\,\,\,\,\cdots
\ee
where the overline means that we are computing these quantities within the low-energy effective theory of general relativity with negative cosmological constant, coupled to the shells.\footnote{Recall that the shell can be formed semiclassically, for example in general relativity coupled to a scalar field.} We will typically restrict to $\beta_L=\beta_R=\beta$, so that the left and right black holes have the same mass, as these are the scenarios we need while counting black hole microstates. 

At first glance, it would appear that if we compute the first inner product $\overline{\bra{\Psi_{\mathbf{m}}} \ket{\Psi_{\mathbf{m}'}}}$ in (\ref{statin}), then the remaining quantities will follow by multiplication.  However, we will see that this is not the case because of non-perturbative effects in the gravitational path integral.  To show this, we start with the combined Euclidean action
\be \label{eq:action}
I[X]=-\frac{1}{16\pi G}\int_X (\, R -2\Lambda ) + \dfrac{1}{8\pi G} \int_{\partial X} K + \int_{\mathcal{W}} \sigma + I_{\text{ct}} \;.
\ee
Here $\Lambda$ is the cosmological constant, $K$ is the extrinsic curvature of the spacetime boundary, $\sigma$ is the density of the shell, ${\cal W}$ is the worldvolume of the shell, and  $I_{\text{ct}}$ are  counterterms, localized at the asymptotic boundary $\partial X$, that remove divergences and renormalize the value of the on-shell action \cite{Balasubramanian:1999re,Skenderis:2002wp}.

Recall that inner products are computed in terms of the gravitational path integral by summing over all Euclidean geometries that are consistent with the boundary conditions determined by the preparation of the bra and ket states. In the semiclassical approximation, the path integral then reduces to a sum over the exponential of the action of the classical saddlepoints $X$ satisfying the boundary conditions. When $\mathbf{m} \neq \mathbf{m}'$, we need some way to join the different shells from the Euclidean boundary. Achieving this will require a number of bulk interactions of the order of $\Delta \mathbf{m} = |\mathbf{m}-\mathbf{m}'|$ in Planck units,  so we expect the result to be parametrically suppressed in the exponential of the mass difference $\Delta \mathbf{m}$.  Thus, we neglect such contributions by taking the mass difference to be arbitrarily large.

In this limit, the inner product is 
\be\label{eq:mean}
   \overline{\bra{\Psi_{\mathbf{m}}} \ket{\Psi_{\mathbf{m}'}}} =  \delta_{\mathbf{m},\mathbf{m}'}\;.
    \ee

Next,  $\overline{\bra{\Psi_{\mathbf{m}}} \ket{\Psi_{\mathbf{m}'}}\bra{\Psi_{\mathbf{m}'}} \ket{\Psi_{\mathbf{m}''}}}$ is computed from the  path integral with two disconnected asymptotic boundaries, which prepare the overlaps $\bra{\Psi_{\mathbf{m}}} \ket{\Psi_{\mathbf{m}'}}$ and $\bra{\Psi_{\mathbf{m}'}} \ket{\Psi_{\mathbf{m}''}}$ respectively.
This quantity receives one contribution from  a disconnected geometry -- two copies of the saddlepoint contributing to $\overline{\bra{\Psi_{\mathbf{m}}} \ket{\Psi_{\mathbf{m}'}}}$, one associated with each of the two asymptotic boundaries.  In our limit,  in which the difference in masses between different states is taken to be arbitrarily large, 
this quantity vanishes unless ${\bf m} = {\bf m}' = {\bf m}''$.   If ${\bf m} = {\bf m}''$, i.e., the first bra and the final ket have the same shells so that we are computing a ``square'' of the overlap, there is a second connected contribution -- a semiclassical wormhole  $X_2$ connecting the two boundaries that was described in \cite{Martin} in the context of operator statistics. The full semiclassical expression then reads
\be
\overline{\bra{\Psi_{\mathbf{m}}} \ket{\Psi_{\mathbf{m}'}}\bra{\Psi_{\mathbf{m}'}} \ket{\Psi_{\mathbf{m}}}} \equiv \overline{|\bra{\Psi_{\mathbf{m}}} \ket{\Psi_{\mathbf{m}'}}|^2} =\delta_{\mathbf{m},\mathbf{m}'}+\dfrac{Z_2}{Z_1 Z_1'}\,,
\label{eq:semiclOverlapSquared}
\ee
where we have defined
\be
Z_2 = e^{-I[X_2]}\hspace{1cm}Z_1 = e^{-I[X]}\hspace{1cm}Z_1' = e^{-I[X']}
\ee
as the renormalized action (\ref{eq:action}) of the wormhole $X_2$.  The normalization factors $Z_1$ and $Z_1'$ associated with each shell (see Eq.~\ref{PETS2a})  are given by the classical saddles $X$ and $X'$ respectively.\footnote{\label{f:norm} These normalizations arise from the standard procedure of filling in the bulk geometry while fixing the  boundary conditions determined by the computation of the norm of the state. The geometries $X$ and $X'$ are each given by two copies of the Euclidean part of Fig.~(\ref{fig::one}) glued on the $\tau=0$ surface.}   In fact, if ${\bf m} \neq {\bf m}''$ there will still be a connected contribution, but it will be a exponentially suppressed in the mass difference of the shells $|{\bf m} - {\bf m}''|$ is taken to be large, by the same reasoning as above.  We can continue similarly for the higher-order overlaps 
\be
\overline{\bra{\Psi_{\mathbf{m}}} \ket{\Psi_{\mathbf{m}'}}\bra{\Psi_{\mathbf{m}'}} \ket{\Psi_{\mathbf{m}''}}...\bra{\Psi_{\mathbf{m}'...'}} \ket{\Psi_{\mathbf{m}}}}|_c\,= \,\dfrac{Z_n}{Z_1\,Z_1^{'}\,\cdots\,Z_1^{'\cdots '}} \;,
\ee 
which arise from  the maximally connected $n$-boundary wormhole contributions $Z_n = e^{-I[X_n]}$.  Below we study $I[X]$ and $I[X_2]$ in detail, as well as the other $I[X_n]$, in certain limits  which we will  need in later sections. Appendix~\ref{appendix:B} provides  more details about wormholes with more than two boundaries.

\subsection{Single-shell states}

\begin{figure}[h]
		\centering
		\includegraphics[width=.6\textwidth]{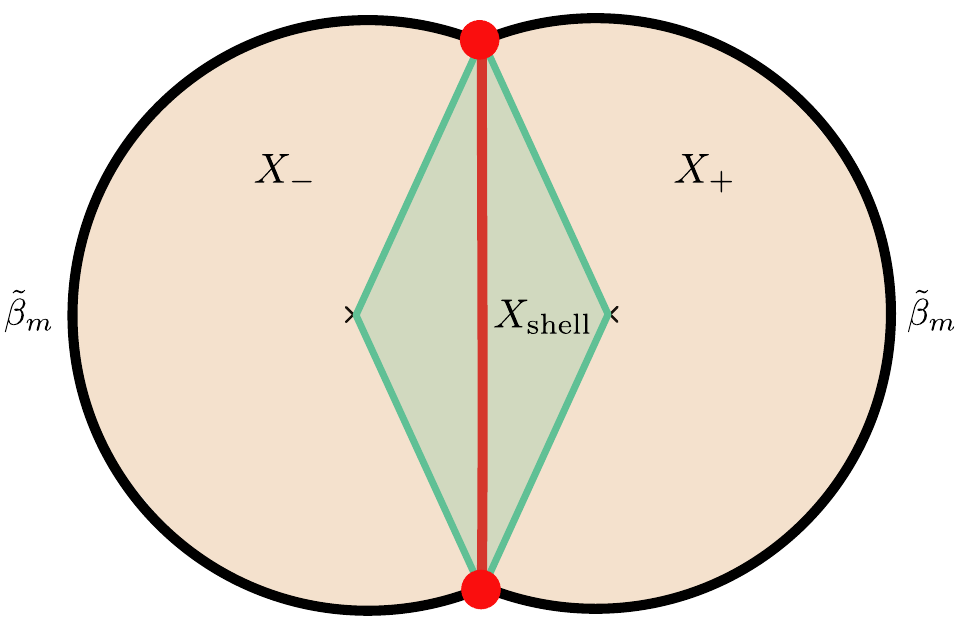}
		\caption{The different regions in the decomposition of the Euclidean gravitational action $I[X]$ for the single shell state. This figure shows the $r,\tau_{\pm}$ plane of the geometry. In this plane $\tau_{\pm}$ are angular coordinates around the horizons $r_{\pm}$, represented by the crosses. The green region $X_{\text{shell}}$  accounts for the intrinsic contribution from the shell. Its boundaries are constant Euclidean time hypersurfaces that join the horizons to the operator insertions at the boundary.}
		\label{fig:bulk2ptaction}
\end{figure}

We start with single shell states, $\bfm = m$ and $\bfm' = m'$. The normalization of these states is computed semiclassically by the space-time filling procedure described above and in footnote~\ref{f:norm}. This gives
    \be\label{eq:normf}
    {Z}_1 = e^{-I[X]}\;,
    \ee
where $X$ is the saddle-point manifold depicted in Fig. \ref{fig:bulk2ptaction}.  From Eq.~\ref{eq:shifttime} and Fig.~\ref{fig::three} with $\beta_L = \beta_R = \beta$, the effective temperature $\tilde{\beta}_m$ to prepare the state is determined by the real temperature $\beta$ of the saddle-point black hole, and  the Euclidean time $\Delta \tau$ elapsed by the shell, via the  relation
    \be\label{betapb}
    \beta = \tbeta_m + \Delta \tau\;.
    \ee
The preparation temperature is the same on both sides of the shell, so $\tilde\beta_L = \tilde\beta_R = \tilde\beta_m$.
In order to evaluate the renormalized action $I[X]$ in \eqref{eq:action}, it is convenient to divide the manifold into $X =X_- \cup X_{\text{shell}} \cup X_+$ (Fig. \ref{fig:bulk2ptaction}). The region $X_{\text{shell}}$ is bounded by the constant Euclidean time hypersurfaces that join the left/right horizon to the two operator insertions at $r_\infty$.
From additivity, the on-shell action decomposes as
\be\label{eq:renormalbulkact}
	I[X] = I[X_-] + I[X_+] + I[X_{\text{shell}}]\,.
\ee 
The first two terms just depend on the exterior black hole geometries and therefore
\be
	I[X_\pm]  = \tbeta_m \,F(\beta) =\tbeta_m\,(M-{\bf S}/\beta) \;,
\ee
where $F(\beta) = -\beta^{-1} \log Z(\beta)$ is the renormalized free energy of the respective black hole and
\be
{\bf S}\equiv \frac{A}{4G} \;,
\ee
is the Bekenstein-Hawking entropy. More explicitly the renormalized action is given by \cite{Emparan:1999pm}
\be\label{eq:renormF}
	I[X_\pm]  = \tbeta_m \,F(\beta) = \dfrac{\tbeta_m V_\Omega}{16\pi G}\left(-r_+^{d}+ r_+^{d-2} + c_d\right)\;.
\ee
The constant $c_d$ accounts for the Casimir energy of the CFT in even dimensions \cite{Balasubramanian:1999re} ($c_d = -\frac{1}{2},\frac{3}{8}, -\frac{5}{16},\ldots$ in $d=2,4,6,\ldots$). Eq.~\ref{eq:renormF} is computed by noticing that the solutions have angular symmetry around the thermal cycle.  We are computing the action of a fraction of the whole disk (i.e., regions $X_+$ and $X_-$) and so the integral over Euclidean time then goes from zero to the preparation temperature.
	
The last term $I[X_{\text{shell}}]$ arises from the presence of the shell since its value vanishes as $m\rightarrow 0$.\footnote{In the $m\rightarrow 0$ limit with fixed $\beta$, the effective temperature $\tilde{\beta}_m \rightarrow  \beta/2$, in which limit $X_{\text{shell}}$ collapses and $I[X_{\text{shell}}]\rightarrow 0$.} The Euclidean action associated with the region $X_{\text{shell}}$ has the form
\be\label{eq:Ishell}
	I[X_{\text{shell}}] = -\dfrac{1}{16\pi G}\int_{X_{\text{shell}}} \,(R-2\Lambda)\,+\,\int_{\mathcal{W}} \sigma  \;,
\ee
before counterterms are added to remove  long-distance divergences that develop in both terms as $r_\infty \rightarrow \infty$.  Here $\sigma$ is the mass density of the shell, whose radius dependence is given by \eqref{eq:density}.  The on-shell Einstein-Hilbert term gives two contributions
\be
	R-2\Lambda = -2d + \frac{16\pi G}{d-1}\delta(y)\,,
	\label{eq:R}
\ee
where $y$ is a normal coordinate to $\mathcal{W}$. The second term comes from the $\delta$ function contribution of the stress tensor of the shell, and the first term comes from the constant background curvature.  Plugging this expression into \eqref{eq:Ishell} gives
\be\label{eq:Ishell2}
	I[X_{\text{shell}}] = \dfrac{d}{8\pi G}\,\text{Vol}(X_{\text{shell}})\,+\,m\dfrac{d-2}{d-1}  L[\gamma_{\mathcal{W}}]  \;,
\ee	
where the $\text{Vol}(X_{\text{shell}})$ term arises from the bulk  integral of the background curvature and the second term takes contributions from the $\delta$ function in (\ref{eq:R}) and from the integral over the shell in (\ref{eq:Ishell}).  Here
$L[\gamma_{\mathcal{W}}]$ is the proper length of the trajectory of the shell.  Thus, the second term looks like the action of a heavy particle propagating in the $(\tau_\pm,r)$ plane. Explicitly, each term is computed from the integrals
	\begin{gather}
	L[\gamma_{\mathcal{W}}] = 2\int_{R_*}^{r_\infty} \dfrac{\text{d}R}{\sqrt{-V_{\text{eff}}(R)}}\,,\label{eq:length}\\[.4cm]
	\text{Vol}(X_{\text{shell}}) = \dfrac{4V_\Omega}{d} \int_{R_*}^{r_\infty}\dfrac{\text{d}R}{f_+(R)}\, \sqrt{\dfrac{f_+(R) + V_{\text{eff}}(R)}{- V_{\text{eff}}(R)}}\,(R^d-r_+^d)\;.\label{eq:vol}
	\end{gather}
 This volume is the same as seen from the left and right sides of the shell because we took the two asymptotic geometries to have the same mass, and hence the same horizon radius $r_+ = r_-$.
Solving these integrals requires numerical treatment when $d>2$. Summarizing, the partition function normalizing one-shell states is
\be\label{eq:norm}
	-\log Z_1 = 2\tbeta_m F(\beta) + \dfrac{d}{8\pi G}\,\text{Vol}(X_{\text{shell}})\,+\,m\dfrac{d-2}{d-1}  L[\gamma_{\mathcal{W}}] \;.
	\ee

\subsection*{Wormhole}

We now calculate the semiclassical overlap squared,  namely
\be
\overline{|\bra{\Psi_{\mathbf{m}}} \ket{\Psi_{\mathbf{m}'}}|^2} =\delta_{\mathbf{m},\mathbf{m}'}+\dfrac{Z_2}{Z_1 Z_1'}\,,
\ee
where we recall that
\be
Z_2 = e^{-I[X_2]}\;,
\ee
is the exponential of the gravitational action of the wormhole $X_2$.  The normalizations $Z_1$ and $Z_1'$ were the ones computed in the previous subsection. The wormhole $X_2$ consists of a pair of Euclidean black holes of the same mass. These are glued together along the trajectory of the two thin shells, as depicted in Fig.~\ref{fig:bulk2ptaction}. The saddle-point equations were derived in~(\ref{multishell}) and follow from an Israel junction condition analysis. Here they read
    \be
    \beta_2 = \tbeta_{m} + \tbeta_{m'} + \Delta \tau_m + \Delta \tau_{m'}\;.
    \label{eq:beta2constraint}
    \ee
In this equation, the two preparation temperatures $\tilde{\beta}_m$ and $\tilde{\beta}_{m'}$ have already been fixed by  Eq.~\ref{betapb}. Then in the wormhole solution, the equation of motion (\ref{eq:shifttime}) fixes the elapsed Euclidean time for each shell as a function of the mass of the shell and the background black hole geometry parametrized by some $\beta_2$.   The constraint (\ref{eq:beta2constraint}) then fixes $\beta_2$.
    \begin{figure}[h]
		\centering
		\includegraphics[width=.25\textwidth]{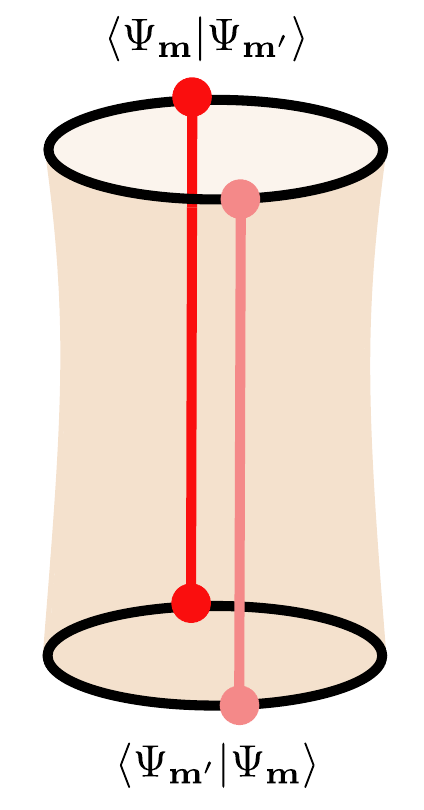}
		\hspace{2cm}
		\includegraphics[width=.56\textwidth]{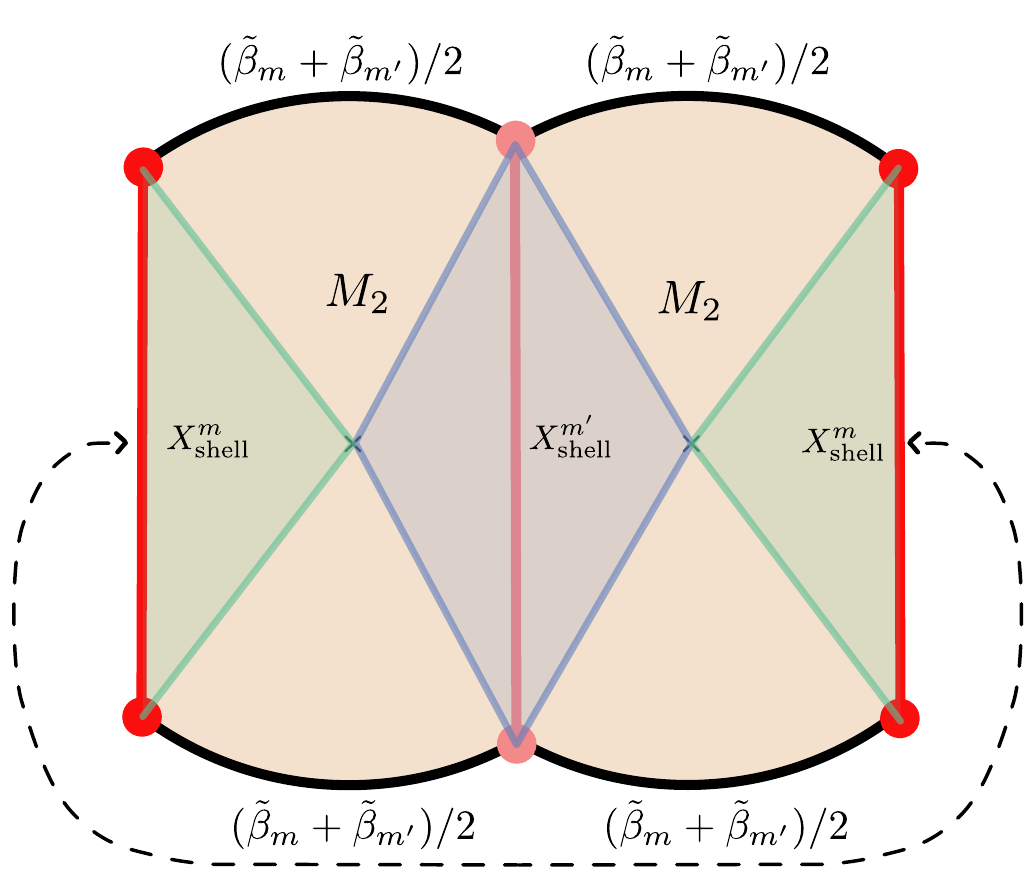}
		\caption{The wormhole $X_2$ consists of a pair of Euclidean black holes of the same mass, which are glued together along the trajectory of the two thin shells. The mass $M_2$ of the black hole that forms the wormhole is determined by the preparation temperatures and the saddle point equations.}
		\label{fig:bulk2ptaction2}
	\end{figure}
Having found this solution, the action of the wormhole after counterterm subtraction can be derived by the same logic as above, giving
\be\label{eq:whaction1}
    I[X_2] = 2(\tbeta_{m} + \tbeta_{m'}) F(\beta_2) + I[X_{\text{shell}}^{m}] + I[X_{\text{shell}}^{m'}]\;.
\ee
Including  the normalizations $Z_1$ and $Z_1'$ we obtain the semiclassical overlap squared
    \be\label{eq:Bfunction}
  \overline{|\bra{\Psi_{{m}}} \ket{\Psi_{{m}'}}|^2} = \delta_{m,m'} + \dfrac{Z_2}{Z_1 Z_1'}= \delta_{m,m'} + e^{-2\,(\tbeta_{m} + \tbeta_{m'}) \,\Delta F - \Delta I[X_{\text{shell}}^m] - \Delta I[X_{\text{shell}}^{m'}]} \;,
\ee
where $\Delta F = F(\beta_2) - F(\beta)$ and $\Delta I[X^i_{\text{shell}}] = I[X^i_{\text{shell}}]|_{\beta_2} - I[X^i_{\text{shell}}]|_{\beta}$ for $i=m,m'$.

\subsection*{2+1 dimensions}
	
For $d=2$ we can work out the details analytically. First, the saddle point equation (\ref{betapb}) relating the state preparation and physical temperatures, combined with the expression for the Euclidean time elapsed across the whole  shell trajectory (\ref{eq:etime2+1}), is solved by
	\be
\tbeta_m  = \dfrac{\beta}{\pi} \arcsin \dfrac{r_+}{R_*}\;,
	\ee
where $R_*^2 = r_+^2+ (2Gm)^2$. Thus we see that $\frac{\beta}{2}\leq \tbeta_m \leq \beta$, where the lower bound is reached for small mass $m\ll M$, while the upper bound is reached for $m\gg  M$. 
	
In this case, the second term drops out of \eqref{eq:Ishell2} and the volume of $X_{\text{shell}}$ can be analytically computed, $\text{Vol}(X_{\text{shell}}) = 4\pi G m L[\gamma_{\mathcal{W}}]$. The result equals the standard propagator of a massive particle
\be\label{eq:2daction}
	I[X_{\text{shell}}] = m L[\gamma_{\mathcal{W}}] = 2 m \,\text{cosh}^{-1}\left(\dfrac{r_\infty}{R_*}\right)\;,\hspace{.8cm}\text{for } d=2 \,,
\ee
where we have evaluated the proper length for the explicit trajectory of the particle $R(T) = R_* \cosh T$, with $R_*^2 = r_+^2+ (2Gm)^2$. 
	
To renormalize the logarithmic divergence of \eqref{eq:2daction} as $r_\infty \rightarrow \infty$, we add the counterterm $I_{\text{ct}}[X_{\text{shell}}] = -m \log r_\infty$ and then take $r_\infty \rightarrow \infty$. The final renormalized action of the shell reads
	\be\label{eq:renshellact2d}
	I_{\text{ren}}[X_{\text{shell}}]\,= \, - 2 m \log {R_*} + 2m \log 2\;, \hspace{.8cm}\text{for } d=2 \;.
	\ee
Also using \eqref{eq:renormF}, the total gravitational action of the saddle is
	\be
	-\log Z_1 = -\dfrac{\tbeta_m }{G}\left(\frac{\pi^2}{\beta^2} - \frac{1}{8}\right) -  2 m \log {R_*} + 2m \log 2 \;.
	\ee
Next, given the action of the wormhole \eqref{eq:whaction1}, we can derive an explicit analytical expression for the overlap squared:
\be
   \overline{|\bra{\Psi_{{m}}} \ket{\Psi_{{m}'}}|^2} = \delta_{m,m'} + \dfrac{Z_2}{Z_1 Z_1'} = \delta_{m,m'} +\,e^{\frac{ 2(\tbeta_{m} + \tbeta_{m'})\pi^2 }{G}\left(\frac{1}{\beta_2^2}-\frac{1}{\beta^2}\right) + 4m \log \frac{R_*(\beta_2)}{R_*(\beta)} +  4m' \log \frac{R'_*(\beta_2)}{R'_*(\beta)}}\;,
    \ee
 where again $R_*(\beta) = \sqrt{\frac{4\pi^2}{\beta^2} + (2Gm)^2}$.

\subsection{Higher moments}
We can proceed similarly to compute higher moments of the overlap.  As shown in Appendix \ref{appendix:B}, the connected part acquires the form 
    \be
\overline{\bra{\Psi_{m_1}} \ket{\Psi_{m_2}}\bra{\Psi_{m_2}} \ket{\Psi_{m_3}}...\bra{\Psi_{m_n}} \ket{\Psi_{m_1}}}|_c =\dfrac{Z_n}{Z^{(m_1)}_1...\,Z^{(m_n)}_1}\,,
\ee
where $Z_n= e^{- I[X_{n}]}$ is the contribution from the $n$-boundary wormhole (Fig. \ref{fig:nwh}) and $Z_1^{(m_i)}$ is the  normalization of the state of a shell of mass $m_i$.   The expressions for the n-boundary wormholes contributing to $Z_n$ can be obtained in a straightforward manner using the building blocks we have derived already. For details see Appendix \ref{appendix:B}.
    
\subsection{Limit of large mass and universality}

\begin{figure}[h]
		\centering
		\includegraphics[width=.9\textwidth]{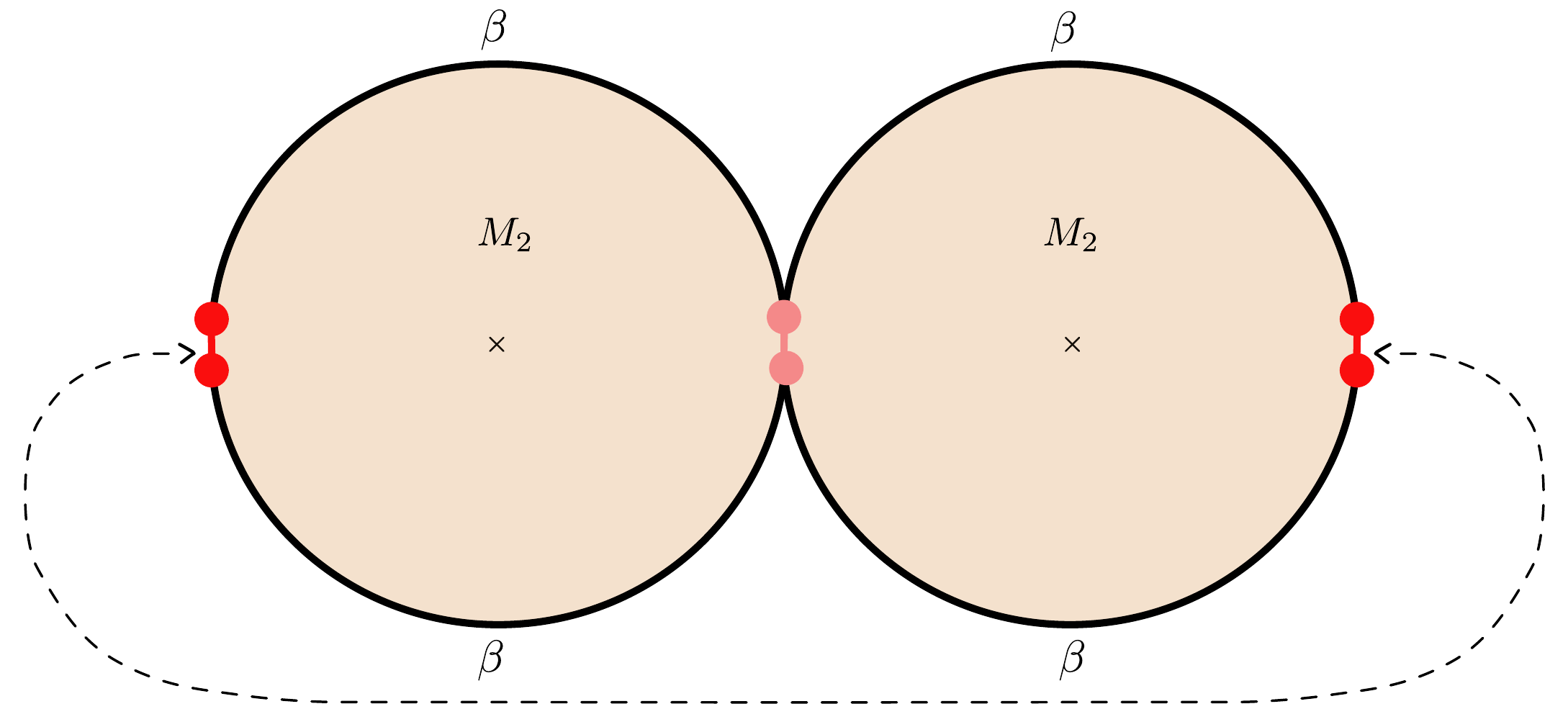}
		\caption{The wormhole $X_2$ for large masses of the shells. The preparation of the state in this case is done with Euclidean time $\tbeta_m = \tbeta_{m'} \approx \beta$. The background black hole has inverse temperature $2\beta$.}
		\label{fig:infmass}
	\end{figure}

To provide an account of black hole entropy in the next section, we will need the contribution of $n$-boundary wormholes to the $n^{{\rm th}}$ moment of the overlap.  But we will only need the large mass $m_i \gg M$ limit, where we recall that the local mass of the shell inside the black hole can actually exceed the asymptotic mass of the black hole, because of the effects of backreaction. In this limit  the Euclidean time elapsed by each shell trajectory tends to zero, $\Delta \tau_i \to 0$ (see Eq. \ref{eq:shifttime} and Eq. \ref{eq:etime2+1})  and therefore $\tbeta_{m_i} \approx \beta$ (see Eq.~\ref{betapb}). Also in this limit $ \dfrac{d}{8\pi G}\,\text{Vol}(X_{\text{shell}}) \approx \frac{m}{d-1}L[\gamma_\mathcal{W}]$ in \eqref{eq:vol} so that
    \be\label{acxsh}
    I[X_{\text{shell}}] \approx m L[\gamma_\mathcal{W}] \approx 2\,m\, \log R_*\;,
    \ee 
after including the counterterms. The proper shell action therefore becomes constant, independent of the mass of the black hole, since $R_*^{d-1} \sim G m \ell$ in this limit. The normalization of the one-shell states \eqref{eq:norm} is then given by
    \be
    {Z}_1 \sim  Z(\beta)^2\, e^{-2\,m\,\log R_*}\;,
    \ee
Similarly, the action of the wormhole $X_2$ becomes
    \be
    {Z}_2 = e^{-I[X_2]} = Z(2\beta)^{2}\, e^{-2\,m\,\log R_* -2\,m'\,\log R_*' }\;.
    \ee
This can be derived by taking the large mass limit of our previous expressions, but it is also obvious from Fig. \ref{fig:infmass}. Therefore, the overlap for $m\neq m'$ reads
    \be
    \overline{|\bra{\Psi_m}\ket{\Psi_{m'}}|^2}|_c = \dfrac{{Z}_2}{{Z}_1Z_1'} \approx \dfrac{Z(2\beta)^2}{Z(\beta)^4} ,
    \label{eq:overlap}
    \ee     
and is universal in this limit, independently of the actual masses of the shells.  

By similar reasoning, it is easy to show that the $n$-boundary wormhole introduced in Appendix \ref{appendix:B} displays a universal form in this limit
    \be
    {Z}_n  = e^{-I[X_n]} = Z(n\beta)^{2}\, e^{-2\sum_{\alpha}\,m_{\alpha}\,\log R^\alpha_*}\;,
    \ee
leading to the $n$-th moment
    \be
    \overline{\bra{\Psi_{m_1}} \ket{\Psi_{m_2}}\bra{\Psi_{m_2}} \ket{\Psi_{m_3}}...\bra{\Psi_{m_n}} \ket{\Psi_{m_1}}}|_c \,= \,\dfrac{Z_n}{Z_1^{(m_1)}...\,Z_1^{(m_n)}}\approx \dfrac{Z(n\beta)^2}{Z(\beta)^{2n}}\,.
    \ee
    
A potential concern for this analysis is that the large shell mass limit pinches the wormhole geometry (Fig.~\ref{fig:infmass}). However, as we approach this limit, there is no singularity in the Euclidean geometry because the pinching points are at the boundary of space and remaining infinitely far apart, and the Lorentzian continuation is also well-defined. One might also worry that in theory with extended states like string theory, we could get a condensation of light states in the pinching limit, requiring a modification of our analysis. However, no matter how large the mass is, the two shell operator insertions are at infinite physical distance from each other even though the Euclidean time elapsed is going to zero, because of a relative conformal factor. Thus we do not expect extended states to give rise to a problem.  This is consistent with the fact that we do not observe any singularities in the low energy theory.

\subsection*{Multi-shell states}
    
The overlaps between multi-shell states also have the same form in the limit where all of the massses $m_i \in \bfm$ are large, i.e.,  $m_i \gg M$. The Euclidean time elapsed by each shell vanishes in this regime, $\Delta \tau_i = 0$, and therefore $\tbeta_i \approx \beta_i$ as before. In this limit, the normalization of the state \eqref{eq:normf}  becomes
\be
    {Z}_1 \approx  Z(\beta)^2 e^{-2\sum_i \,m_{i}\,\log R^i_*} \prod_{i=1}^{k-1} Z(2\beta_i) \;,
\ee
where $k$ is number of shells. Each factor of $Z(2\beta_i)$ arises from the Euclidean black hole bounded by a pair of thin shells. The two factors of $Z(\beta)$ come from the exterior black holes. The last factor is the action of each shell in the limit, which was derived above~(\ref{acxsh}).
    
Similarly, the wormhole action is now
\be
    Z_2 = e^{-I[X_2]} \approx Z(2\beta)^2 e^{-2\sum_i\,m_{i}\, \log R^i_*-2\sum_j\,m_{j}\, \log R^j_*} \prod_{i=1}^{k-1} Z(2\beta_i) \prod_{j=1}^{k'-1} Z(2\beta_j)\;,
\ee
where $k'$ is the number of shells in the second state. Therefore, the overlap reduces to the universal quantity
    \be
    \overline{|\bra{\Psi_\bfm}\ket{\Psi_{\bfm'}}|^2}|_c = \dfrac{{Z}_2}{{Z}_1Z_1'} \approx \dfrac{Z(2\beta)^2}{Z(\beta)^4} \;,
    \ee     
independent  of the number of shells, and of the interior geometries which are characterized by the parameters $\beta_i$ and $\beta_i'$.

Finally, the $n$-boundary wormhole has action
    \be
    {Z}_n  = e^{-I[X_n]} \approx Z(n\beta)^{2}\,e^{2\sum_\alpha \sum_i \,m_{i\alpha}\,\log R^{i\alpha}_*}\prod_{\alpha=1}^n\prod_{i=1}^{k_\alpha-1} Z(2\beta_i) \;,
    \ee
leading to the $n$-th moment
    \be\label{Universal}
    \overline{\bra{\Psi_{\mathbf{m}_1}} \ket{\Psi_{\mathbf{m}_2}}\bra{\Psi_{\mathbf{m}_2}} \ket{\Psi_{\mathbf{m}_3}}...\bra{\Psi_{\mathbf{m}_n}} \ket{\Psi_{\mathbf{m}_1}}}|_c \,= \,\dfrac{Z_n}{Z_1^{(\mathbf{m}_1)}...\,Z_1^{(\mathbf{m}_n)}}\approx \dfrac{Z(n\beta)^2}{Z(\beta)^{2n}}\,.
    \ee
This formula is a key result of our article. It states that wormholes in the quantum gravity path integral lead to universal, non-vanishing moments of the quantum overlaps of our black hole microstates.
In Section~\ref{SecV} we will see how these overlaps lead to a bound on the dimension of the Hilbert space that equals the  Bekenstein-Hawking entropy.

\subsection{Microscopic interpretation of the overlaps}\label{Sec:3_4}

Inner products between quantum states in a Hilbert space are by definition complex numbers. As such, the product of a collection of them must factorize. However, this is not what we found for the overlaps between shell microstates computed above  via the rules prescribed by the semiclassical path integral of gravity. Indeed, the universal expression \eqref{Universal} manifestly displays non-factorization, as a consequence of connected wormhole contributions to the products of overlaps.\footnote{Recent work in 2d JT gravity especially has focused on the effect of wormholes on the factorization of gravitational amplitudes, and interpretations of this phenomenon in terms of a fundamental averaging over theories (see e.g. \cite{Saad:2019lba,Penington:2019kki,Stanford:2020wkf,PhysRevLett.125.021601,Maloney:2020nni,Cotler:2020ugk,Altland:2020ccq,Altland:2021rqn,Chandra:2022bqq}. Alternative uses of wormholes and their relation to factorization have appeared recently in \cite{Betzios:2021fnm}.} In what follows, expanding on \cite{Martin} (see also \cite{Stanford:2020wkf}), we will provide an interpretation of the semiclassical overlaps starting from a microscopic description of these states. Basically we will view the semiclassical calculation as an approximation which is only sensitive to the magnitudes and not to the erratic phases, of the real inner products between the corresponding quantum states.

We will argue that the results of the gravitational path integral are consistent with an assumption that  the underlying theory,  which in the present case is equivalent to the holographic dual CFT, satisfies something like the Eigenstate Thermalization Hypothesis (ETH) \cite{PhysRevA.43.2046,Srednicki_1994}.  ETH was originally formulated as a simple but powerful postulate that allows us to understand thermalization in isolated quantum systems. The postulate says that the matrix elements in the energy basis of a ``simple'' operator $\mathcal{O}$,  take the form
\be   \label{ETHold} 
\bra{E_n} \mathcal{O} \ket{E_m}\,=\,f(\bar{E}) \delta_{nm} \,+\,e^{-S(\bar{E})/2)}\, g(\bar{E},\omega)^{1/2} \,R_{nm}\;.
\ee
where the $|E_n\rangle$ are energy eigenstates. In this expression we  defined
\be 
\bar{E}\equiv \frac{E_n+E_m}{2}\;,\,\,\,\,\,\,\,\,\,\,\,\,\,\,\,\omega=E_m-E_n\;.
\ee
The functions $f(\bar{E}), g(\bar{E},\omega)$ are smooth functions of their arguments in the thermodynamic limit, and encode  information about the microcanonical one-point and two-point function of the operator, respectively. The coefficients $R_{nm}$ are erratic complex numbers of $\mathcal{O}(1)$ magnitude. The ETH then asserts that the $R_{nm}$ entries can be viewed as independent random variables with zero mean an unit variance.\footnote{This hypothesis is developed in the context of CFTs in \cite{Belin:2020hea}, and a reformulation in terms of multi-matrix models has appeared in \cite{Jafferis:2022uhu}.} The domain of applicability of ETH is not fully understood, but it is generally expected to apply to quantum chaotic theories, such as the ones expected to describe black holes \cite{Maldacena:2015waa,Cotler:2016fpe}.

Following \cite{Martin}, the thin fine-grained operator should be regarded ``simple'': applied near the spacetime boundary it just creates creates a gas of particles in the low-energy effective theory.\footnote{From the perspective of the dual theory, even if the shell operator is not spatially small, i.e. it is fully delocalized, it can still be considered as simple in the internal large-$N$ space of the CFT. The operator corresponds to a product of approximately local single-trace operators on each spatial site, and it displays ``Lyapunov growth" for different holographic measures of operator size.  The Euclidean time evolution transforms this simple asymptotic operator into a ``complex'' operator describing a shell behind the horizon of the Lorentzian black hole.}    As such, and in view of the chaotic Hamiltonian expected for the black hole dynamics, the shell operator should admit an ETH form (\ref{ETHold}). Under this assumption, \cite{Martin} showed that the semiclassical calculation of the product of correlation functions of the thin shell operator follows from the expected random character of the $R_{nm}$ coefficients as follows.  There are many shells of the same mass that differ in tiny microscopic, perhaps Planck-scaled details, in, e.g., the precise positioning of the dust particles.  These states should have an ETH description with coefficients of the same magnitude, but randomly varying erratic phases. The semiclassical path integral should be interpreted as a coarse-grained quantity that does not have access to the precise microscopic phases and effectively averages over them, only preserving information about the magnitudes of amplitudes in which the phases cancel. In this identification, matching the semiclassical gravity results for the thin shells requires that  $f(\bar{E})$ vanish, while the envelope function takes the form
\be\label{eq:gsol}
	 g(\bar{E},\omega) \, = \,e^{ S(\bar{E}) - \alpha_- S(\bar{E}-\omega) - \alpha_+ S(\bar{E}+\omega) -I_{\textrm{shell}}(\bar{E},\omega)}\;,
\ee
The coefficients $\alpha_\pm \equiv \Delta \tau_\pm /\beta_\pm$ are specified in terms of the shell's  Euclidean travel times \eqref{eq:shifttime}. The function $I_{\textrm{shell}}(\bar{E},\omega)$ is the value of the  shell's on-shell action (see \cite{Martin}).

The connection with the shell states presented here follows in a straightforward manner. Our states were defined as
\be
\ket{\Psi}\,=  \,\ket{ \rho_{\tilde{\beta}_{L}/2}\,\mathcal{O}\,\rho_{\tilde{\beta}_{R}/2}}
 =  \dfrac{1}{\sqrt{Z_1}}\, \sum_{n,m}\,e^{-\frac{1}{2}(\tbeta_L E_n + \tbeta_R E_m)}\,\mathcal{O}_{nm}\,\ket{n,m}\;
\ee
where ${\cal O}$ is a shell operator and 
\be
Z_1 =  \text{Tr}(\mathcal{O}^\dagger e^{-\tbeta_L H}\mathcal{O}e^{-\tbeta_R H})\;,
\ee
normalizes these pure states. Then, as described above, if we define the smooth function
    \be
    f(E_n,E_m) \equiv S(\bar{E})- \log g(\bar{E},\omega) =  \alpha_- S(\bar{E}-\omega) + \alpha_+ S(\bar{E}+\omega) +I_{\textrm{shell}}(\bar{E},\omega)\;,
    \ee
we can reproduce the gravitational results by assuming an ETH form
    \be\label{PETS2}
	\ket{\Psi} =  \dfrac{1}{\sqrt{Z_1}} \sum_{n,m}e^{-\frac{1}{2}(\tbeta_L E_m + \tbeta_R E_n-f(E_n,E_m))}R_{mn}\ket{m,n} \;,
	\ee
where the normalization becomes
\be
	Z_1 =   \text{Tr}(\mathcal{O}^\dagger e^{-\tbeta_L H}\mathcal{O}e^{-\tbeta_R H}) \approx   \sum_{n,k} e^{-\tbeta_R E_n-\tbeta_L E_k - f(E_n,E_k)} \;.
\ee

These states look approximately random in the energy basis, in each microcanonical band. Indeed, projecting onto a microcanonical band of energies $[E,E+\Delta E]$, and normalizing the state, one obtains
\be\label{PETS}
	\ket{\Psi}_E \simeq  \dfrac{1}{\Omega_E} \sum_{n,m}R_{mn}\ket{m,n} \;.
	\ee
Thus, we can regard the state preparation procedure, together with the projection on an energy window, as a procedure to generate an infinite number of random states in the band.\footnote{This randomness recalls an argument in the AdS/CFT context that typical microstates of black holes can be created by operators that are random polynomials in the CFT fields \cite{Balasubramanian:2005kk,Balasubramanian:2005mg}, and that the consequence of this randomness will be a universality of correlators computed in such states.} 
The infinite number of states  arises for us by varying the mass of the shell operator ${\cal O}$ used as the starting point, or by using multiple shell states.
In the holographic CFT, this can be achieved by increasing the scaling dimension of the operator.

We now come back to the moments of inner products,
 \be\label{gravcalc}
    \overline{\bra{\Psi_{\mathbf{m}_1}} \ket{\Psi_{\mathbf{m}_2}}\bra{\Psi_{\mathbf{m}_2}} \ket{\Psi_{\mathbf{m}_3}}...\bra{\Psi_{\mathbf{m}_k}} \ket{\Psi_{\mathbf{m}_1}}}\;,
    \ee
In this expression, the overline means we are computing these products using the gravity path integral. By writing every state in terms of the shell operators used to prepare them, and assuming ETH for the thin shell, we can use the results in \cite{Martin} to precisely interpret the overline in terms of an ETH average of the form
\be    
\overline{\mathcal{O}_{n_1m_2}\,\mathcal{O}_{n_2m_3}\,\cdots\,\mathcal{O}_{n_km_1}}\;.
\ee
In this way we arrive at a simple interpretation of the non-factorization of the semiclassical inner products of our microstates: the semiclassical path integral only computes a coarse-grained average over the microstates that are consistent with the macroscopic semiclassical description appearing in the saddlepoint.  The fine-grained phase of the overlap $\bra{\Psi_{\mathbf{m}_1}} \ket{\Psi_{\mathbf{m}_2}}$ depends erratically on the ETH coefficients of the operators $\mathcal{O}_1$ and $\mathcal{O}_2$, and averages out to zero semiclassically,  $\overline{\bra{\Psi_{\mathbf{m}_1}} \ket{\Psi_{\mathbf{m}_2}}}=0$.\footnote{Classically different shells, of mass difference scaling parametrically with $\ell^{d-1}/G$, are expected to have uncorrelated ETH coefficients. This is consistent with the semiclassical calculation.} On the contrary, the magnitude of the overlap is smooth quantity with some non-vanishing average value which is captured by the wormhole contribution \eqref{gravcalc} for $k=2$. This follows and further supports the ideas of \cite{Saad:2019lba,Saad:2019pqd,Belin:2020hea,Pollack:2020gfa, Schlenker:2022dyo, Chandra:2022bqq} connecting quantum chaos and semiclassical gravitational physics.

To end the discussion on the interpretation of the gravitational overlaps, we want to make clear that, although ETH provides the most natural and physical interpretation of our results, it is important to notice we do not need such an assumption in what follows, and in particular, in the derivation of the Bekenstein-Hawking degeneracy from the gravitational overlaps. The Bekenstein-Hawking entropy will follow from simple algebraic arguments, with the only input of the universal gravitational overlaps computed above.

\subsection*{Connection with the spectral form factor in chaotic theories}

An interesting physical interpretation of our result (\ref{Universal}) arises by considering the dynamics of chaotic quantum theories as follows. Given a quantum mechanical system with Hamiltonian $H$ and discrete spectrum, consider the so-called spectral form factor (SFF), defined by
\be 
\textrm{SFF}(t)=\frac{Z_{\beta-i\,t}\,Z^*_{\beta+it}}{Z_{\beta}^2}\;,
\ee
where $Z_\beta=\sum_i\,e^{-\beta\,E_i}$ is the partition function of the  Hamiltonian with eigenvalues $E_i$. The SFF is well known in the context of matrix models, see \cite{Guhr:1997ve}, and it has been recently studied in relation to black hole dynamics, see \cite{Cotler:2016fpe}. Consider also the TFD state, defined as follows
\be 
\vert\psi_{\beta}\rangle \equiv\frac{1}{\sqrt{Z_{\beta}}}\sum_n e^{-\frac{\beta E_n}{2}}\vert n,n\rangle\;,
\label{TFDdef}
\ee
in the tensor product of the original Hilbert space with itself. Unitary evolution with a single, say the left, Hamiltonian gives
\be 
\vert\psi_{\beta} (t)\rangle=e^{-iH_L t}\vert\psi_{\beta}\rangle=\vert\psi_{\beta+2it}\rangle\;.
\ee
The survival probability of this evolution, namely the probability that the evolved state is found in the original thermofield double, equals the spectral form factor \cite{Papadodimas:2015xma,Verlinde:2021jwu,delCampo:2017bzr,Balasubramanian:2022tpr,Stanford:2022fdt}
\be 
P(t) =\vert\langle\psi_{\beta+2it}\vert\psi_{\beta}\rangle\vert^2=\textrm{SFF}(t)\;,
\label{eq:TFDsurvival}
\ee
The time average of the spectral form factor is
\be         
\lim\limits_{T\rightarrow \infty}\,\frac{1}{T}\,\int\limits_{0}^{T}\,dt\, \textrm{SFF}(t)
=
\lim\limits_{T\rightarrow \infty} \,\frac{1}{T}\,\int\limits_{0}^{T}\,dt\,\frac{1}{Z(\beta)^2}\,
\sum\limits_{m,n}\,e^{-\beta(E_m+E_n)+i(E_m-E_n)t}\;.
\ee
If the theory is chaotic, we expect level-repulsion to remove all degeneracies in the spectrum.  Hence, the time average gives
\be
\lim\limits_{T\rightarrow \infty}\,\frac{1}{T}\,\int\limits_{0}^{T}\,dt\, \textrm{SFF}(t)=\,\dfrac{Z(2\beta)}{Z(\beta)^{2}}\;. 
\label{eq:TimeAverageSFF}
\ee
Now recall that the SFF is precisely the survival probability for the thermofield double state (\ref{eq:TFDsurvival}), i.e., the probability for the state to return to itself.  In a chaotic theory we expect an ergodic exploration of the Hilbert space.  So the time average in (\ref{eq:TimeAverageSFF}) is computing the inner product between the initial state and a typical, essentially random, state in the Hilbert space.  Notice that (\ref{Universal}) for  $n=2$ is precisely the square of (\ref{eq:TimeAverageSFF}).  The square appears because we are considering an eternal black hole, and so the Hilbert space is doubled.  But this detail aside, we see that the gravitational path integral for the square of the overlap is giving precisely the same result as the expected time-averaged survival amplitude in a chaotic theory, suggesting that the shell states we are computing are essentially random relative to each other.

The non-vanishing long time average of the spectral form factor is a proxy for the discreteness of the  underlying Hamiltonian and its dimension. Finding this long time-average by means of a gravitational computation has been a key goal in the context of quantum black holes (see \cite{Cotler:2016fpe} and references therein). This goal was been recently accomplished in the context of 2d JT gravity and its cousins \cite{Saad:2022kfe,Blommaert:2022lbh}, but here we have been able to find such typical inner products for gravity in general dimensions.

\subsection*{Reduced density matrices and quantum black hole hair}
	
Consider the reduced density matrices $\rho_{R,L} = \text{Tr}_{L,R} \ket{\Psi}\bra{\Psi}$ of each of the two boundary subsystems of our eternal black holes. In terms of the thin-shell operator making our microstates, their expression is
\begin{gather}
	\rho_L = \dfrac{1}{{Z}_1}\, e^{-\frac{\tbeta_L}{2} H}\,\mathcal{O}\,e^{-\tbeta_R H}\,\mathcal{O}^\dagger \,e^{-\frac{\tbeta_L}{2} H}\;,\\[.4cm]
	\rho_R = \dfrac{1}{{Z}_1}\,e^{-\frac{\tbeta_R}{2} H}\,\mathcal{O}\,e^{-\tbeta_L H}\,\mathcal{O}^\dagger \,e^{-\frac{\tbeta_R}{2} H}\;.
\end{gather}
This can be be verified by constructing the density matrix representation of the state in (\ref{CJop}), and tracing over the left or right Hilbert spaces.
These reduced density matrices are not exactly thermal. Since  the classical geometries outside the horizon are  equal to the black hole geometry, we are led to say that these shell states display ``quantum hair''. To analyze the effect of this quantum hair, we can use the gravitational path integral, interpreted in terms of the ETH language above, with the explicit representation of our states in the energy basis (\ref{PETS2}), to say that
 \be 
\overline{R_{nk}R_{mk}^*} = \delta_{nm} \;.
 \ee
It then follows that the semiclassical density matrices are diagonal in the energy basis (because they effectively average over the ETH phases). The reduced coarse-grained state at time $t=0$ is then given by 
	\be
	\overline{\rho_R}_{nm} = p_n \,\delta_{nm}\;,
	\ee
	where the probabilities are
	\be
	p_n = \dfrac{\sum_k e^{-\tbeta_R E_n -\tbeta_L E_k- f(E_n,E_k)}}{\sum_{n,k} e^{-\tbeta_R E_n-\tbeta_L E_k - f(E_n,E_k)}}\;.
	\ee
These probabilities are close to the thermal values. To see this, first notice that the expectation value for the left/right energy is given in terms of the canonical ensemble at temperature $\beta_{L,R}$
	\begin{gather}
	 M_\pm  =\, \bra{\Psi}H_{R,L}\ket{\Psi}\, =\,\text{Tr} (\rho_{\beta_{R,L}}\,H_{R,L}) \;.
	\end{gather}
The first equality simply expresses the fact that the ADM mass measured in the left or the right external geometries must equal the expected value of the Hamiltonian in the state by the general reasoning of \cite{Balasubramanian:1999re}.  The second equality simply states that we could have obtained the same result by taking the expectation value of the left or right Hamiltonian in a exactly thermal density matrix.  In other words, although our density matrix is not exactly thermal, it gives the same expectation value for the energy. Also, we can consider the large mass limit $m\gg E$. In this case we have $\tbeta_L = \beta_L$ and therefore
\be
p_n = e^{-\beta_R(E_n-E_R) - S(E_R)} = \dfrac{e^{-\beta_R E_n}}{Z(\beta_R)}\;,
\ee
meaning that the reduced density matrix looks increasingly thermal. At small $m$ the expression is more complicated but not very illuminating for the present purposes. This structure of reduced density matrices suggest that there will be  modes of the thermal radiation on both exterior geometries which will be sensitive to the fact that the reduced state is not thermal.  Effectively, these modes detect the presence of the interior shell through its quantum hair.   One might expect that the wavelength of these modes is determined by the distance of the shell to the horizon. When the shell is very far away (its mass is large),  these modes become extremely low energy compared to the temperature, and are suppressed by the corresponding Boltzmann factors, making them undetectable in practice. In a certain sense, the shell can be interpreted as a sort of soft firewall deep inside the black hole.

\section{Counting the microstates of black holes}\label{SecV}

In Sec.~\ref{SecII} we  constructed several infinite families of black hole microstates with geometric descriptions, whose existence sharpens the problem of black hole microstate counting.  These states also challenge the holographic principle \cite{tHooft:1993dmi,Susskind:1994vu,Bousso:1999xy}, since it appears that we can fit arbitrarily many inside a black hole of a given mass.
However, as we  showed in Sec.~\ref{SecIV}, these states are not orthogonal because of non-perturbative effects in quantum gravity.  In fact, they have universal overlaps with each other (see Eq.~\ref{Universal}). Because of these overlaps, the dimension of the Hilbert space spanned by these states is smaller than the number of states.  Below, we will calculate this dimension.

\subsection{One-parameter family of states}
	
First, we consider a convenient one-parameter family of states: the one-shell states with equal black hole temperatures at both sides. Equivalently, these states have the same ADM mass on both sides. Therefore, these are microstates of an eternal, finite temperature black hole.  For a given temperature $\beta$, our states are labeled by the rest mass $m$ of the shell. In the context of AdS/CFT, this mass is related to the scaling dimension of the field by
\be 
m_n=\, n\,m_{\Delta}=n\,\ell^{-1}\,\sqrt{\Delta(\Delta-d)}\;,
\ee
where $n\sim \ell^{d-1}/G \gg 1$ is the number of operator insertions. Every shell backreacts on the interior geometry as described previously, leaving the outside geometry unchanged.	

Given a field of mass $m_{\Delta}$, we can thus create a discrete family of states labeled by $n$. In terms of the CFT, the family is defined as
	\be
	\ket{\Psi_n} = \dfrac{1}{\sqrt{Z_1}}\sum_{i,j} e^{-\frac{1}{2}(\tilde{\beta}^{m_n}_L E_i + \tilde{\beta}^{m_n}_R E_j)}\,\mathcal{O}^{m_n}_{ij}\ket{i,j}\;,
	\ee
where $Z_1$ normalizes the state and where we need the condition
\be 
\tilde{\beta}^{m_n}_L=\tilde{\beta}^{m_n}_R=\beta-\Delta\,\tau\;
\ee
on the Euclidean lengths used to prepare the state
for the left/right physical black hole temperatures to be the same, i.e.,  $\beta_L=\beta_R=\beta$.
Here $\Delta\tau$ was defined in~(\ref{eq:shifttime}) as the Euclidean time required by the shell trajectory. Note that, as described in Sec.~\ref{SecIV}, we can choose the initial size of the shell for any $n$ so that all these states are within the regime of validity of our approximations. Specifically, in the Euclidean geometry the minimum shell radius is $R_*$ and so in our limit, it can be verified that the shell density in Planck units never exceeds a maximum  of $\mathcal{O}(\ell_P/\ell)\ll 1$.

Defined in this way, the difference in wormhole lengths between subsequent shell states is less than Planckian, and one cannot consider them to be geometrically different at a semiclassical level. This also implies that the inner product between subsequent states is  not going to be exponentially suppressed; indeed the universal result found earlier for geometrically different states does not apply.\footnote{In modern jargon, going from one shell to another by adding one particle, effectively constitutes a motion in the code subspace of the original shell. Overlaps in the code subspace need not be exponentially suppressed in $G$. Within those subspaces we can neglect the dynamical effects of gravity such as backreaction in the semiclassical limit.} For our purposes it is  convenient to choose an infinite family in which every member is geometrically different from the others at scales bigger than the Planck length. To achieve this it is enough that subsequent states have wormholes with lengths differing by order $m=n\,m_{\Delta}$ where we take $n \sim O(1/G_N)$. We thus consider microstates with masses
\be 
m_p=p\,m\,\,\,\,\,\,\,\,\,p=1,2,\cdots\;.
\ee
This choice also ensures that as $p$ grows, the overlap between different states is controlled by the universal answer derived before.

\subsection{The microstate Gram matrix and the Hilbert space dimension}

The infinite family of different, geometrical, and semiclassical states $\ket{\Psi_{p}}$ with masses $m_p=p\,m$ naively overcounts the Bekenstein-Hawking entropy. We now show this is not actually the case. \footnote{A similar situation was encountered in \cite{Hsin:2020mfa} when trying to argue against the existence of global symmetries in two-dimensional models of quantum gravity describing near extremal black holes. In that context, wormholes provide the necessary global symmetry violating amplitudes to ensure that the number of linearly independent states remains below the Bekenstein-Hawking entropy of the black hole.} The key question is not how many semiclassical/geometrical states we have at our disposal, but what is the dimension of the Hilbert space they span. To find this dimension we examine the overlap matrix, i.e., the Gram matrix $G$ of the microstates, whose entries are defined as
\be 
G_{pq}\equiv \braket{\Psi_p}{\Psi_q}\;.    
\ee
We will be considering this matrix $G$ for $p,q=1,\cdots \,\Omega$, with $\Omega$ 
 finite. We then vary $\Omega=1,\cdots,\infty$. This generates a sequence of $\Omega\times\Omega$ Gram matrices $G$ of microstate overlaps.

Gram matrices are Hermitian and positive semidefinite by construction. We can see this by noticing that they can always be written as
\be 
G=B^{\dagger}\,B\;,
\ee
with $B$ being the matrix that has the vectors $\ket{\Psi_p}$ as its columns. $B$ is a $\Omega\times\tilde{\Omega}$ matrix, where $\tilde{\Omega}$ is the dimension of the Hilbert space where the states $\ket{\Psi_{p}}$ live. Because it is positive semidefinite, its  eigenvalues  are positive or zero. Also, from the definition of the Gram matrix, it is clear that the microstate vectors $\ket{\Psi_{p}}$ will be linearly independent if and only if the Gram matrix is positive {\it definite}, namely if it has no zero eigenvalues. More generally, the rank of the Gram matrix will count the number of linearly independent vectors in the family $\vert\psi_p\rangle$.

Therefore, for each $\Omega$ and associated Gram matrix $G$, we need to compute the number of zero eigenvalues of $G$. To this end we first compute the resolvent of the matrix $G$. This is defined as
\be \label{resol}
R_{ij}(\lambda)\,\equiv\,\left( \frac{1}{\lambda \mathds{1} -G}\right)_{ij} \,=\,\frac{1}{\lambda}\,\delta_{ij}+\sum\limits_{n=1}^{\infty}\,\frac{1}{\lambda^{n+1}}\,(G^n)_{ij}\;.
\ee
The density of eigenvalues of the Gram matrix then follows from the discontinuity across the real axis of the trace of the resolvent $R(\lambda) = \sum\limits_{i=1}^\Omega R_{ii}(\lambda)$.  To be precise, the density of eigenvalues is 
\be\label{eq:densitydisc}
D(\lambda)=\frac{1}{2\,\pi\,i}\left(\,R(\lambda-i\epsilon)-R(\lambda+i\epsilon)\,\right)\;.
\ee

Ref. \cite{Penington:2019kki} (see also \cite{CVITANOVIC198149,https://doi.org/10.48550/arxiv.0911.0087,Hsin:2020mfa}) showed how these types of resolvent matrices give rise to a Schwinger-Dyson equation. One starts by writing the definition of the resolvent in a diagrammatic expansion.
\begin{figure}[h]
		\centering
		\includegraphics[width = \textwidth]{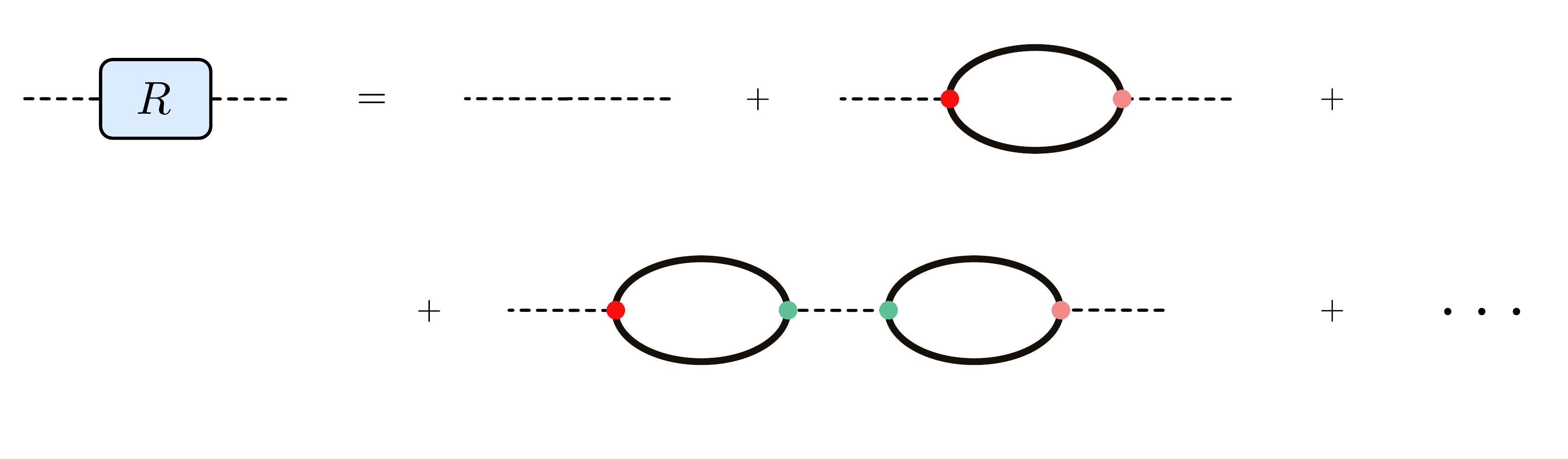}
		\caption{Diagrammatic expansion of the resolvent matrix. The external dashed lines represent the $(i,j)$ indices. The internal dashed lines represent summation over indices. The blobs represent the inner products $G_{ij}=\langle \Psi_i\vert \Psi_j\rangle$, and the color dots are the operator insertions.}
		\label{fig:sd}
	\end{figure}
In this  version of Eq.~(\ref{resol}), the blobs are just the inner products $G_{ij}=\langle \Psi_i\vert \Psi_j\rangle$ that we have been discussing all along. They are just drawn horizontally instead of vertically as they were in our gravitational path integrals. The colored dots correspond to the shell insertions. To get Eq.~(\ref{resol}) from this diagram, we note that the dashed lines are understood as ``free propagators'', and they are assigned factors of $1/\lambda$.

\begin{figure}[h]
		\centering
		\includegraphics[width = \textwidth]{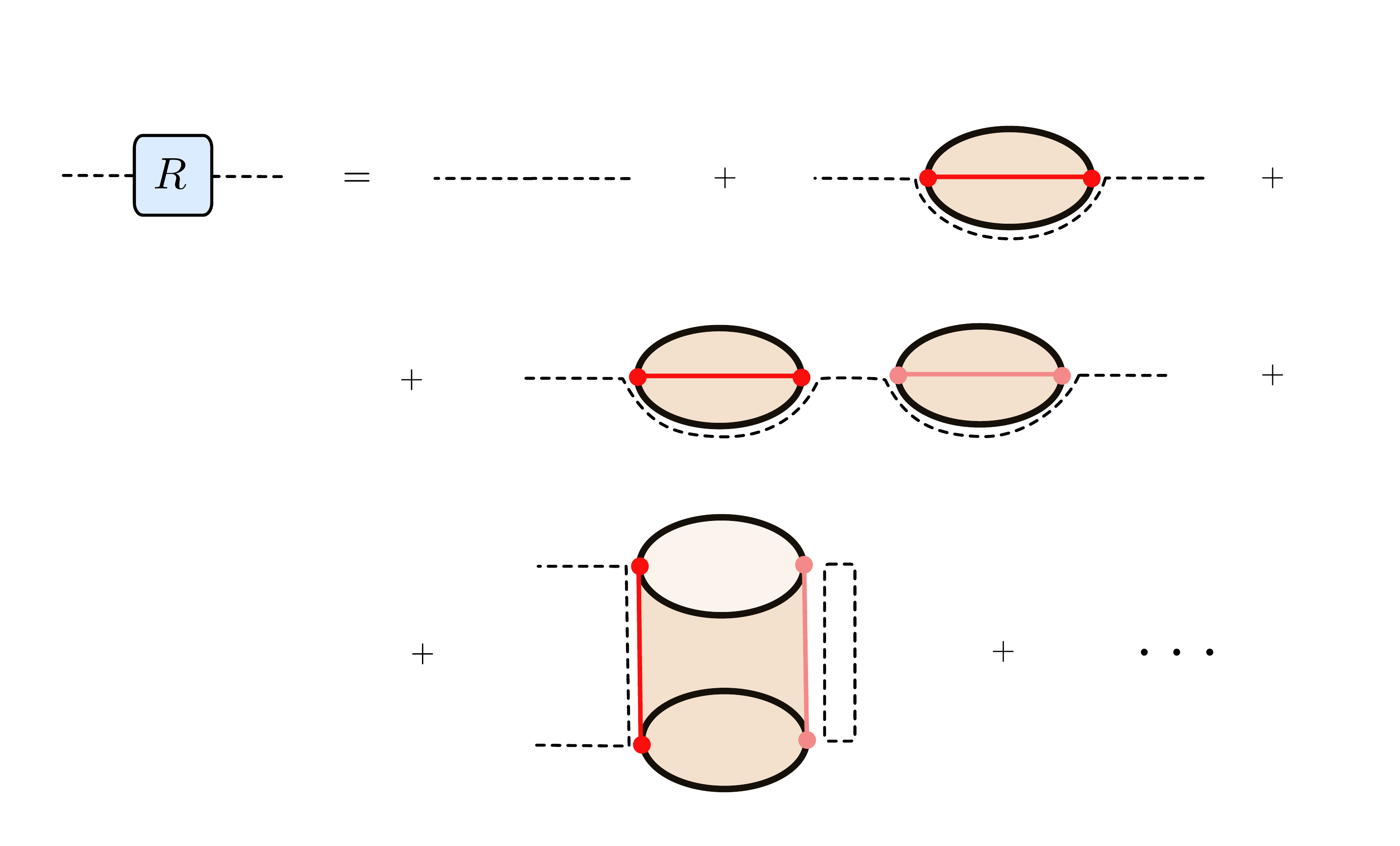}
		\caption{Expansion of the resolvent as computed in the leading semiclassical gravity approximation. The colored blob in the second term represents gravitational saddle point computing the first moment of the inner product. It corresponds to the second term in Fig.~\ref{fig:sd}. The third and fourth terms are the two gravitational saddlepoint contributions (disconnected and wormhole respectively) to the computationa of the 
        third term in Fig.~\ref{fig:sd}.}
		\label{fig:sdp}
	\end{figure}

Given this diagrammatic expansion, we can  compute the  resolvent matrix by using the gravitational path integral. As before, we denote a gravity computation in the leading approximation by an overline, and we arrive at
\be \label{resolav}
\overline{R_{ij}(\lambda)}\,=\,\frac{1}{\lambda}\,\delta_{ij}+\sum\limits_{n=1}^{\infty}\,\frac{1}{\lambda^{n+1}}\,\overline{(G^n)_{ij}} \;,
\ee
This expansion can also be depicted graphically (Fig.~\ref{fig:sdp}).  Interestingly,  semiclassical gravity produces two terms when it is applied to the second term in the resolvent expansion depicted in Fig.~\ref{fig:sd}; the first is disconneected, while the second  is a connected wormhole contribution. In the computation of this expansion in Fig.~\ref{fig:sdp}, we have only included diagrams that are ``planar'' in the sense explained in \cite{Penington:2019kki}, since non-planar diagrams will give subleading contributions in the limit of large entropy and large matrix $G$, which is the limit we are interested in.

The elements of $(G^n)_{ij}$ are products of the overlaps $\langle \Psi_i | \Psi_j \rangle$.  Above we showed from the gravitational path integral that these overlaps display a universal form for sufficiently high mass,\footnote{Notice that for any family of black hole microstates we choose, most states will have sufficiently large mass in this sense, since the mass can increase without bound. Else one can choose a family whose first state starts already at sufficient high mass.}
   \be\label{UniAmp}
    \overline{\bra{\Psi_{\mathbf{m}}} \ket{\Psi_{\mathbf{m}'}}...\bra{\Psi_{\mathbf{m}'...'}} \ket{\Psi_{\mathbf{m}}}} \, \simeq \, \dfrac{Z(n\beta)^{2}}{Z(\beta)^{2n}}\equiv\frac{Z_n}{Z_1^n}\;,
    \ee
where $n$ is the number of inner products on the left hand side, and where we recall that $Z(n\beta)^{2}\neq Z_n$, because there are cancellations between numerator and denominator. Notice the right hand side is universal and it does not depend on the masses of the states.

\begin{figure}[h]
		\centering
		\includegraphics[width = \textwidth]{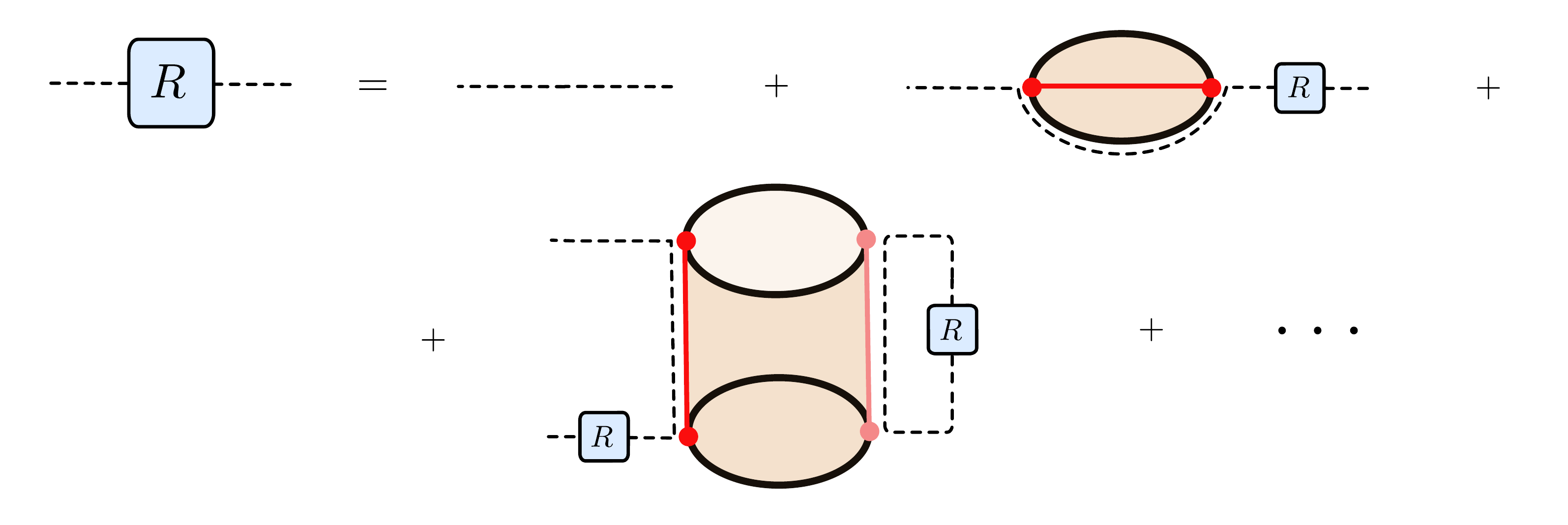}
		\caption{Schwinger-Dyson equation for the resolvent, in the leading semiclassical approximation.  This is a self-consistent reorganization of Fig.~\ref{fig:sdp}.}
		\label{fig:sdpii}
	\end{figure}

Using this, we can perform the sum in (\ref{resolav}) by observing that the expansion can be reorganized in a self-consistent way as depicted in Fig.~\ref{fig:sdpii}. This is analogous to the scenario developed in \cite{Penington:2019kki}.  The upshot is that we obtain the  Schwinger-Dyson equation
\be 
\overline{R_{ij}(\lambda)}=\frac{1}{\lambda}\,\delta_{ij}+\frac{1}{\lambda}\,\sum\limits_{n=1}^{\infty}\,\frac{Z_n}{Z_1^n}\,\overline{R(\lambda)}^{n-1}\,\overline{R_{ij}(\lambda)}\;,
\ee
where $R(\lambda)$ is the trace of the resolvent. Taking the trace of this equation we arrive at
\be \label{Req}
\lambda\,\overline{R(\lambda)}=\Omega+\,\sum\limits_{n=1}^{\infty}\,\frac{Z_n}{Z_1^n}\,\overline{R(\lambda)}^{n}\;.
\ee

In order to count black hole microstates for a given energy from these overlaps, i.e., the microcanonical degeneracy, we have to first account for the fact that any entry in the Gram matrix is actually a sum over microcanonical windows in the energy basis.\footnote{Notice that different energy bands are orthogonal to each other and so the inner products of states in the different bands just partition.}  This is because we construted fixed temperature, rather than fixed mass, microstates.  So, to proceed we must project our states via an inverse Laplace transform into the microcanonical window around energy $E$, 
\be \label{eq:Laplace}
Z(n\beta)^{2}=\left( \int dE \,z(E)\,e^{-n\beta\,E}\right)^2 \;,
\ee
where $z(E)$ is the inverse Laplace transform of $Z(n\beta)$. The square appearing in the previous and subsequent formulas just comes from the fact that we are considering eternal black holes, and the entropies/microcanonical degeneracies will be doubled since we have two copies of the same quantum gravity theory. 

As in \cite{Penington:2019kki}, we now define the functions
\be \label{eq:defS}
e^\textbf{S}\equiv\,\left(\,z(E)\,\Delta E\,\right)^2\;,\,\,\,\,\,\,\,\,\,\,\,\,\,\,\,\,\,\,\,\,\,\textbf{Z}_n\equiv\,\left(\,z(E)\,e^{-n\beta\,E}\,\Delta E\,\right)^2\;.
\ee
Inserting these expressions after projecting~(\ref{Req}) into the microcanonical band we obtain the simple result
\be 
\lambda\,\overline{R(\lambda)}=\Omega+\,e^\textbf{S}\sum\limits_{n=1}^{\infty}\,\left(\,\frac{\overline{R(\lambda)}}{e^\textbf{S}}\,\right)^n=\Omega+\,\frac{e^\textbf{S}\,\overline{R(\lambda)}}{e^\textbf{S}-\overline{R(\lambda)}}\;,
\ee
leading to a quadratic equation for the resolvent
\be 
\overline{R(\lambda)}^2+\left(\,\frac{e^\textbf{S}-\Omega}{\lambda}-e^\textbf{S}\,\right)\,\overline{R(\lambda)}+\dfrac{\Omega}{\lambda}\,e^\textbf{S}=0\;.
\ee
As we said above, the density of states follows from the discontinuity across the real axis of the trace of the resolvent of the Gram matrix (Eq.~\ref{eq:densitydisc}) as computed in semiclassical gravity. 
This finally gives
\be \label{denG}
\overline{D(\lambda)}=\frac{e^\textbf{S}}{2\pi\lambda}\sqrt{\,\left[\lambda-\left(1-\Omega^{1/2}\, e^{-\textbf{S}/2} \right)^2\,\right]\left[\,\left(1+\Omega^{1/2} \,e^{-\textbf{S}/2} \right)^2-\lambda \right]}+\delta(\lambda)\left(\Omega-e^{\mathbf{S}}\right)\theta(\Omega-e^{\mathbf{S}})\;,
\ee
where ${\bf S} = A/4G$.  Ultimately the value $A/4G$ arises from evaluation of the gravitational action of the wormhole contribution to the overlap moments \eqref{eq:Laplace} and \eqref{eq:defS}. This density of states has a continuous part and a singular part. The continuous part is supported on 
\be 
\left(1-\Omega^{1/2}\, e^{-\textbf{S}/2} \right)^2\,<\,\lambda\,<\,\left(1+\Omega^{1/2} \,e^{-\textbf{S}/2} \right)^2\;.
\ee
The eigenvalues accounted for by this part are all positive definite. The reason is that this part is continuous and the measure at  $\lambda=0$ is zero.\footnote{Notice that all eigenvalues are positive semidefinite, as follows from the positive semidefinite nature of the Gram matrix.} The singular part counts the number of zero eigenvalues. It only appears when $\Omega>e^{\textbf{S}}$ due to the Heaviside factor. Regardless, the number of positive eigenvalues, i.e., the rank of the Gram matrix, is accounted for by the continuous part of the distribution.

Thus, by separately integrating the continuous and singular parts of the eigenvalue distribution we find that:
\begin{itemize}
    \item For $\Omega<e^{\textbf{S}}$, where $\textbf{S}$ is the Bekenstein-Hawking entropy, the Gram matrix $G$ has no zero eigenvalues. The number of non-zero eigenvalues, i.e., the rank of $G$, equals the dimension spanned by the black hole microstates $\vert\Psi_p\rangle$ and is given by $\Omega$. These statements are true on average in the effective random matrix ensemble for $G$ that gravity provides.  
    \item For $\Omega>e^{\textbf{S}}$, the Gram matrix $G$ has $\Omega-e^{\textbf{S}}$  zero eigenvalues on average. The number of non-zero eigenvalues, i.e., the rank of $G$, equals the dimension spanned by the black hole microstates $\vert\Psi_p\rangle$, and is given by $e^{\textbf{S}}$.
    \item The black hole microstate degeneracy, equal to the number of possible orthogonal states in a given energy band is  $e^{\textbf{S}}=\left(\,z(E)\,\Delta E\,\right)^2$, equal to the exponential of the Bekenstein-Hawking entropy.
\end{itemize}
All of these are statements are valid on average in the effective ensemble provided by semiclassical gravity. But the variances associated with the densities of eigenvalues in Random Matrix Theory \cite{10.1093/oxfordhb/9780198744191.001.0001}  are suppressed by the dimension of the associated matrix. In this case they will be suppressed by factors of $e^{-\textbf{S}}$. It will be interesting to compute the effects of these subleading corrections to the Bekenstein-Hawking entropy.

Intuitively, if we keep adding potential microstates to a system, there is a point at which these states cannot be orthogonal anymore. In our case, this point is controlled by the universal statistics of the inner product displayed in Sec.~\ref{SecIV}, which is in turn controlled by the Bekenstein-Hawking entropy.  Thus  the solution to the problem of understanding the entropy of general black holes is not to construct a specific set of $e^{\textbf{S}}$  microstates. Indeed, there may be infinite numbers of such sets, even when they are constrained to be semiclassical and geometrical. The problem is really to show that any such choice gives rise to the same Hilbert space with the right Bekenstein-Hawking dimension.

We have proved the linear dependence of the semiclassical geometrical microstates that we examined. Still, we have not proved the completeness of the basis generated from these states. This basis we generate of course has the right dimension to explain the Bekenstein-Hawking entropy and  is complete in its span. But one could ask whether there are states in the black hole that cannot be expanded in this basis.  If that occurs, the true entropy of the black hole is greater than the Bekenstein-Hawking entropy.   We have derived the black hole degeneracy at leading order. However, there are subleading corrections to our results, and some of these could correspond to additional states that cannot be expanded in this way. These corrections will be non-perturbatively suppressed and we expect that adding them could modify the Hilbert space dimension at subleading order in the $G\rightarrow0$ expansion.  That said, we indeed expect most states to be expandable in the basis generated from our shell states. For example, we could take the Thermofield Double (TFD) state without any shell, along with the small excitations around this background. These states again have non-zero, exponentially suppressed, overlaps with all shell states that we have considered. This overlap can be computed by the same methods as in previous sections.  The dominant contribution will come from a wormhole saddle with one shell behind the horizon.  Thus adding the TFD family of states to the Gram matrix will not increase its rank. Similarly, other semiclassical black hole microstates can be understood as superpositions of our dust shell states, provided that the characteristic overlaps, given by wormhole configurations, have the correct magnitude.

We could have similarly considered microstates with  inhomogeneities,  microstates rotating with some angular momentum, or charged microstates. 
However, generically in statistical mechanics such states make sub-leading contributions to the entropy. For example, very much like the shells of dust we are considering, most states of the particles of a gas in a room are approximately homogeneously spread over the volume. Indeed, configurations with inhomogeneities are  suppressed in their phase space volume. The same is true if the system has, say, electric charge. For a given energy, there will be many more configurations of the system with equal numbers of positive and negative charges, and hence vanishing total charge.  Similarly, at a given energy there are also many more configurations with net vanishing rotation. These arguments can be made more precise in our context by considering an ensemble of shells in which the probability of having a dust particle in a certain position is constant along the sphere, and in which the probability of having a dust particle with positive or negative electric charge is the same. This ensemble now contains microstates with inhomogeneites and charge, but average quantities such as the entropy will still be dominated  by the spherically symmetric shells. Similar considerations also apply to dust shell configurations with net angular momentum. Note also that if a shell has Planck-sized inhomogeneities, it will be semiclassically described by the same symmetric shells that we have studied.

Finally, we notice that subleading-in-$G$ corrections to the black hole entropy can be easily incorporated using these methods. In particular, a class of corrections at next-to-leading order arises from the one-loop corrections to the overlaps of the thin shell microstates in the large mass limit. Another class of corrections arises from the wormhole variance to the computation of the rank of the Gram matrix. The former will depend on the particular gravitational theory in question. For theories including charged black holes in the near-extremal limit, it has been recently understood that these corrections will universally become important at very low temperatures, and will drive the density of states to zero for non-supersymmetric black holes \cite{Stanford:2017thb,Iliesiu:2020qvm}. For BPS black holes, methods similar to the ones presented here have appeared in \cite{Boruch:2023trc} to account for the extremal degeneracy. In \cite{AnaC} these quantum corrections to our construction for generic black holes will be described in detail.

\subsection{Evaporating black holes and the Page curve}\label{Sec:4_3}

Refs. \cite{Penington:2019kki,Almheiri:2019qdq} provide a proof of the semiclassical island formula \cite{Penington:2019npb,Almheiri:2019psf,Almheiri:2019hni} that  reproduces the Page curve \cite{Page:1993df,Page:1993wv}. One of these scenarios, often called the ``west coast model'' \cite{Penington:2019kki} allows detailed computation in the context of 2d JT gravity.\footnote{The field of two dimensional gravity has seen a tremendous rebirth in the last decade. See \cite{Sarosi:2017ykf,Mertens:2022irh} for two beautiful reviews.} Meanwhile, the ``east coast model'' \cite{Almheiri:2019qdq} presents a general argument, based on gravitational path integrals in general dimensions. Some aspects of the west coast model have been extended to three~\cite{Balasubramanian:2020hfs} and higher \cite{Chandra:2022fwi} dimensions as well.

Here we have reproduced the advantages of the west coast model within general relativity in general dimensions without including unknown degrees of freedom, thus giving a finer microscopic understanding of the east coast model.  In particular, the west coast model assumes the existence of End-Of-The-World (EOW) branes equipped with somewhat mysterious color degrees of freedom. Microscopically, this assumes the existence of $k$ interior black hole microstates $\vert\psi_i\rangle_{\textbf{B}}$ that are orthogonal to each other, or at least only have non-perturbatively small overlaps. At the same time, one assumes that the radiation of the black hole leaks into an external reservoir with basis of states given by $\lbrace \ket{i}_{\textbf{R}}\rbrace $.  The full state of the system, composed by the black hole and the radiation reservoir, can be Schmidt decomposed into the form
\be
\vert\Psi\rangle=\frac{1}{\sqrt{k}}\sum\limits_{i=1}^k\,\ket{\psi_i}_{\textbf{B}}\,\ket{i}_{\textbf{R}} 
\ee
The entanglement entropy of the radiation is computed from the reduced density matrix
\be 
\rho_{\textbf{R}}=\frac{1}{k}\,\sum\limits_{i,j=1}^k\,\ket{j}\bra{i}_{\textbf{R}}\,\braket{\psi_i}{\psi_j}=\frac{1}{k}\,\sum\limits_{i,j=1}^k\,\ket{j}\bra{i}_{\textbf{R}}\,G_{ij} = {G \over k}\;,
\ee
where $G_{ij}$ is the Gram matrix of the microstates with an EOW brane inside the black hole. The connection between this model of black hole evaporation and our approach to black hole microstate counting is now clear. At the end of the day, both computations depend on the eigenvalue statistics of the Gram matrix of interior microstates. To relate our results to the analysis of the Page curve we just need to make the following associations
\be 
k\rightarrow \Omega\, \,\,\,\,\,\,\,\,\,\,\,\,\,\,\,\,\rho_{\textbf{R}}=\frac{G}{k}=\frac{G}{\Omega}\,\,\,\,\,\,\,\,\,\,\,\,\,\,\,\,\,\lambda_{\rho_{\textbf{R}}}=\frac{\lambda_G}{\Omega}\,\,\,\,\,\,\,\,\,\,\,\,\,\,\,\,\,D(\lambda_{\rho_{\textbf{R}}})_{\rho_{\textbf{R}}}=\Omega\, D(\lambda_{G})\;.
\ee
where $\Omega$ is the dimension of the Gram matrix,  $\lambda_{\rho_{{\bf R}}}$ and $\lambda_G$ are eigenvalues of the the density matrix ${\rho_{{\bf R}}}$ and the Gram matrix $G$ respecitively, and $D$ are the eigenvalue densities.
This leads to the following conclusions:
\begin{itemize}
    \item Our approach extends the west coast model, in particular the derivation of the density of states and Page curve, to general relativity in general dimensions. Everything follows from the universal overlaps between black hole microstates derived in Sec.~\ref{SecIV}.
    \item Our approach clarifies the west coast model by considering physical microstates, such as dust shells with different masses, which do not require the inclusion of mysterious EOW branes with novel degrees of freedom.  In the west coast model the EOW brane states are taken to be naively orthogonal and then wormhole contributions determine a non-perturbative overlap.   In our case, the shells are classically orthogonal by also have a non-perturbative overlap.
    \item Our approach makes it clear that any basis, such as the EOW branes with color, cannot have infinitely many orthogonal members, and that the true dimension of the Hilbert space is given by the Bekenstein-Hawking degeneracy. 
\end{itemize}
These points have important implications. Concerning the extension of the west coast model to a full theory of gravity, the authors of \cite{Geng:2021hlu} argued that such resolutions of the Page curve puzzle might fail in theories with long range gravity, such as general relativity itself.\footnote{See however \cite{Balasubramanian:2021wgd} which avoids this argument by considering entanglement between disjoint gravitating universes.} 
Our results achieve such an extension to general relativity and help to demystify the nature of the degrees of freedom needed to achieve a resolution of the Page curve conundrum. The universality of gravitational dynamics implies that any complete, or indeed overcomplete, basis will do the job.

\section{Einstein-Rosen volume saturation and complexity}\label{SecVI}

The authors of \cite{Susskind:2014rva,Stanford:2014jda} conjectured  that the volume of  Einstein-Rosen (ER) bridges is related to the ``quantum complexity'' \cite{Aaronson:2016vto} of the underlying state. In classical gravity, these volumes grow linearly with asymptotic time, and the conjecture therefore predicts that  ``complexity'' will increase linearly with time as well.  However,  the complexity of any circuit is bounded by the dimension of the Hilbert space on which the circuit acts (Chapter 7 of \cite{Aaronson:2016vto} and discussion in \cite{Balasubramanian:2019wgd}).  Thus, the  conjecture makes a second prediction:  the volume of an ER bridge   must saturate in  quantum gravity at a value  exponential in the black hole entropy, i.e., at $\mathcal{O}(e^{S_{BH}})$.  Plainly,   no such saturation is observed semiclassically.   One possibility is that we simply cannot make sense of semiclassical physics for times exponentially long in the entropy because the states cease to be geometrical by some still undiscovered quantum effect \cite{Susskind:2020wwe}. Alternatively, the geometric volume may stop being related to complexity at an exponential time. 
As we will see below,  our results suggest that the geometric volume, whatever its relation to complexity, cannot be measured a linear operator in the theory since we will show that long wormholes can be written as a superposition of short wormholes.

\subsection{Interior geometry of the microstates} 
\label{SecIII}

We start by computing  volumes of the Einstein-Rosen bridges for the families of shell states described in Sec.~\ref{SecII}.  We will also characterize the geometry of the ``python's lunch"  within these bridges, i.e.,  regions of maximal transversal area.  These quantities are conjectured to  relate to the  complexity of the underlying quantum state \cite{Stanford:2014jda} and the complexity of interior reconstruction \cite{Brown:2019rox} respectively.

Using the expressions in Sec.~\ref{SecII}, the volume at time $t=0$  in between the horizons for a geometry with a shell of mass $m$ is 
\be\label{vER}
	V(\ket{\Psi_m})\, = \, 2V_\Omega\,\int_{r_+}^{R_*} \dfrac{r^{d-1}\text{d}r}{\sqrt{f_+(r)}}\;,
	\ee
where $R^*$ is the minimum radius to which the shell arrives.  Here $V_\Omega$ is the volume of the transverse sphere. According to \cite{Stanford:2014jda}, the relative complexity between this state and the TFD is
\be\label{eq:complexityER}
	\mathcal{C}(\ket{\Psi_m})\, = \,\dfrac{d-1}{8\pi G\ell}\,V(\ket{\Psi_m})=\,\dfrac{d-1}{8\pi G\ell}\,\left(\text{Vol}(\Sigma)-\text{Vol}(\Sigma_{0}) \right)\,\;,
	\ee
where $\Sigma$ is the reflection-symmetric Cauchy slice in the geometry associated to $\ket{\Psi_m}$, and $\Sigma_0$ is the corresponding Cauchy slice of the eternal Schwarzschild black hole. Since the spacetime is the same outside the black hole, the relative complexity is  associated with the stretching of the Einstein-Rosen bridge, given by~(\ref{vER}).

For $d=2$, we can compute the integral explicitly
	\be
	\mathcal{C}(\ket{\Psi_m})\, = \, \dfrac{(R_*^2-r_+^2)^{1/2}}{2G}\, = \, m \ell = n\,m_{\Delta}\,\ell=n\, \sqrt{\Delta (\Delta -d)} \;,
	\ee
from the form of $R_*$ in \eqref{2+1radius}. This gives a strikingly simple relation between  the Einsten-Rosen volume and the  mass/scaling dimension for this family of states, supporting  the proposal  in \cite{Magan:2018nmu}, verified by different means in \cite{Magan:2020iac,Caputa:2021sib}, relating quantum  complexity and scaling dimensions in conformal field theories.

In higher dimensions, the relation between volume and the mass of the shell is more involved. However, for large enough shell masses we obtain
     \be
     R_*^{d-1} \approx \dfrac{4\pi G\, m \,\ell}{(d-1)\,V_\Omega}. 
     \ee
Using this expression in \eqref{eq:complexityER} we get
     \be\label{eq:complexityER2}
	 \mathcal{C}(\ket{\Psi_m})\, \approx \, \dfrac{V_\Omega R_*^{d-1}}{4\pi G}\,= \, \dfrac{m\ell}{d-1}\;,
	 \ee
again arriving at a simple relation with the scaling dimension.

For small masses, on the other hand, the solution to $V_{\text{eff}}(R_*) =0$ in \eqref{eq:Veff} satisfies
    \be
     R_*  \approx  r_+ + \dfrac{\beta_+}{4\pi} \left(\dfrac{4\pi G m }{(d-1)V_\Omega r_+^{d-2}}\right)^2\;.
    \ee
In this regime, \eqref{eq:complexityER} gives 
     \be\label{eq:complexityER3}
	 \mathcal{C}(\ket{\Psi_m})\, \approx  \, \dfrac{\beta_+ r_+ m}{4\pi \ell }\;,
	 \ee
Although still linear in the mass of the perturbation, the slope now depends on parameters of the original black hole.

\subsection*{Characterization of the python}

\begin{figure}[h]
    \centering
    \includegraphics[width= \textwidth]{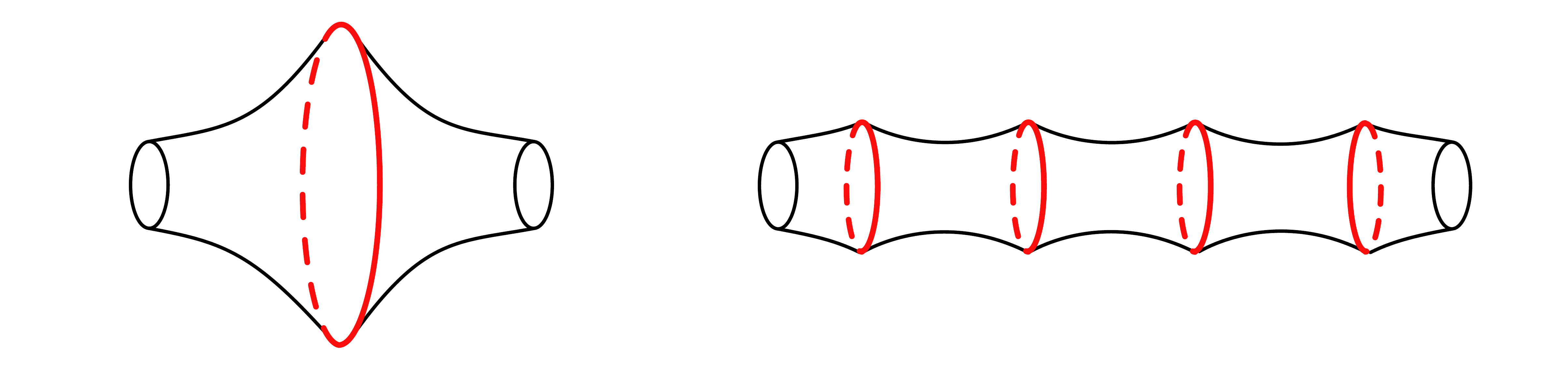}
    \caption{Geometry of the Einstein-Rosen bridge for the microstates. The single shell states with large $m$ contain large portions of a hyperbolic geometry in the interior, while the multi-shell states can contain large volume while keeping the spherical section relatively small.}
    \label{fig:ERbridges}
\end{figure}

The interior geometry of $\ket{\Psi_m}$ is a classical python's lunch, a geometry with a cross-sectional bulge at the position of the shell, $r=R_*$. For large $m \gg M$, this bulge has an area
\be
	S_{b} = \dfrac{\text{Area}_b}{4G} = \dfrac{V_\Omega R_*^{d-1}}{4G} \approx \dfrac{\pi}{d-1} m\ell \;.
\ee
Thus, most of the volume of the ER bridge comes from the region close to the bulge, since the geometry is locally hyperbolic.  To see this quantitatively, we can compute the proper length of the ER bridge
\be
	L = 2 \int_{r_+}^{R_*} \dfrac{\text{d}r}{\sqrt{f_+(r)}}	\approx 2 \ell \log \dfrac{R_*}{r_+} \approx 	 \dfrac{2 \ell}{d-1} \log m\ell\;,
\ee
so that
\begin{equation}
	S_b \approx \dfrac{\pi}{d-1}e^{\frac{d-1}{2\ell}L }\;,
	\ \ \ \ \ \ {\rm and} \ \ \ \ \ \ 
	\mathcal{C}(\ket{\Psi_m}) \approx \dfrac{1}{d-1}e^{\frac{d-1}{2\ell}L}\;,
\end{equation}
which is what we expect from the hyperbolic geometry of the black hole interior. Note that in order to have an exponentially large complexity in $\ell^{d-1}/G$, the length of the wormhole need only scale linearly  with $\ell^{d-1}/G$. So the single-shell pythons are relatively thick and short hyperbolic geometries. To build thin and long pythons, more similar to cylinders, we can concatenate multiple shells with reasonably low values of $m$. This is depicted in Fig~(\ref{fig:ERbridges}).

\subsection*{Multiple shells}

To construct long wormholes we consider microstates with k shells $\ket{\Psi_{\mathbf{m}}}$ for $\mathbf{m} = (m_1,...,m_k)$. The interior volume is additive
\be\label{vERm}
	\mathcal{C}(\ket{\Psi_{\mathbf{m}}})\, = \, \sum_{i=1}^k \dfrac{(d-1)V_\Omega}{4\pi G\ell} \int_{r_+}^{R^{i}_*} \dfrac{r^{d-1}\text{d}r}{\sqrt{f_+(r)}} \approx \sum_i \dfrac{m_i \ell}{d-1}\;.
\ee
For $d=2$, we again have
\be
	\mathcal{C}(\ket{\Psi_{\mathbf{m}}})\, = \, \sum_{i=1}^k m_i \ell  \;,
\ee
	The python will now contain multiple lunches at $R_*^i$ (see \cite{Susskind:1993if}), with associated entropy
\be
	S_{b_i}= \dfrac{\text{Area}_{b_i}}{4G} = \dfrac{V_\Omega (R^i_*)^{d-1}}{4G} \approx \dfrac{\pi}{d-1} m_i\ell \;.
\ee
and total length
\be
	L = \sum_{i} 2 \int_{r_+}^{R^i_*} \dfrac{\text{d}r}{\sqrt{f_+(r)}}	 \approx 	\sum_i \dfrac{2 \ell}{d-1} \log m_i\ell\;.
\ee
Thus, if we wish, we can build wormholes with high volumes and small bulges by considering many shells, each with $m_i \ell \sim \ell^{d-1}/G_N$.

\subsection*{Summary}

Since the mass, or equivalently the dimension, of the shell is unbounded from above, we conclude that we can construct well-controlled geometrical states with Einstein-Rosen bridges of any size. In particular there is no evident saturation of either the wormhole volume or the size of the python's lunch, both of which can be super-exponential in the black hole entropy already at $t=0$.  Thus, super-exponential wormholes need not be obstructed by quantum effects, as they can be classically well-defined.

\subsection{Saturation of Einstein-Rosen bridges}

We now restrict, without loss of generality, to the one-shell states $\ket{\Psi_p ^\beta}$ with masses $m_p=p\,m$, where $p=1,2,\cdots$ and $m\sim \mathcal{O}(\ell^d/G)$. We are now explicitly labeling the states with the inverse temperature $\beta$ of the black hole, with associated ADM energy $E$. 

In Sec.~\ref{SecV} we have shown that, when projected onto the  microcanonical window $[E,E+\Delta E]$ associated to the black hole, using the orthogonal projector $\Pi_E$, the unnormalized states
\be
\ket{\Psi^{E}_p} = \Pi_E\ket{\Psi^\beta_p}\;,
\ee
for $p=1,...,e^{\textbf{S}(E_i)}$ likely generate a basis of the microcanonical black hole Hilbert space 
\be    
\mathcal{H}_E = 
\textrm{Span}\,\left\lbrace   \vert\Psi_p^E \rangle : p=1,\cdots ,e^{\textbf{S}(E)}\right\rbrace\;.
\ee
All of these basis elements have shell masses, and hence wormhole sizes, at most exponential in entropy, and are thus ``short'' wormholes. The microcanonical Hilbert space dimension is set by the Bekenstein-Hawking entropy
\be 
e^{\textbf{S}(E)}\equiv\,\left(\,z(E)\,\Delta E\,\right)^2\;.
\ee

Now consider a one-shell state $\ket{\Psi^\beta_q}$ which has a super-exponential Einstein-Rosen bridge, $q\gg e^{\textbf{S}(E)}$. Evidently, its projection into the microcanonical window can be expanded in the previous basis built out of shorter wormholes
\be
q>e^{\textbf{S}(E_i)}\,\,\,\,\,\,\,\rightarrow\,\,\,\,\,\,\,\ket{\Psi^E_q}=\sum\limits_{p=1}^{e^{\textbf{S}(E)}}\,c_p \ket{\Psi^{E}_p}\;.
\ee
The wavefunction of the full state $\ket{\Psi^\beta_q}$ that we  prepared in previous section will however spread into different microcanonical windows, and if we we want to generate it with short wormholes, we will need to consider states $\ket{\Psi^{E_i}_p}$ associated to different energy windows $E_i$.  These states are constructed from projections of short wormhole states with different temperatures $\ket{\Psi^{\beta_i}_p}$ into   microcanonical windows, $\Pi_{E_i}\ket{\Psi^{\beta_i}_p}$, specified by the ADM masses $E_i$ of the corresponding large black holes. Therefore we can expand the state as
\be\label{eq:expansionlong}
\ket{\Psi_q^\beta}=\sum_i\sum\limits_{p=1}^{e^{\textbf{S}(E_i)}}\,c^{i}_p \ket{\Psi^{E_i}_p} \;.
\ee

Equivalently,  we can also write the state in a basis of fixed temperature states.  These are linearly independent when projected to the corresponding microcanonical windows, and thus they are also linearly independent in the infinite dimensional Hilbert space.   Making the change of basis we get
\be\label{eq:expansionlong2}
\ket{\Psi_q^\beta}= \sum_{i}\sum_p^{e^{\textbf{S}(E_i)}} \alpha_p^i \ket{\Psi_p^{\beta_i}}\;.
\ee
in terms of some new coefficients $\alpha_p^i$. In this last expression we have explicitly written the large wormhole as a linear superposition of {\it geometric} short wormhole states $\ket{\Psi_p^{\beta_i}}$ for different temperatures of the black hole.   The wavefunction of the long wormhole in the short wormhole basis, $\alpha_p^i$, will follow an approximately thermal distribution, so that the sum \eqref{eq:expansionlong2} is dominated by short wormholes with the temperature $\beta$. To see this, one can explicitly compute as $\alpha_p^i \approx \bra{\Psi_q^\beta}\ket{\Psi_p^{\beta_i}}$. The overlap is given in terms of a generalization of the two-boundary Euclidean wormhole of Sec.~\ref{SecIV}, for states with different temperatures. In the large mass limit repeating the calculation that led to (\ref{eq:overlap}) shows that the overlap is independent of $p$, and is given by the wormhole contribution 
\be\label{tfdov}
|\alpha_p^i| \sim \dfrac{Z(\beta + \beta_i)}{Z(\beta)Z(\beta_i)}\;.
\ee
For large temperatures $\beta_i \ll \beta$, the coefficients $\alpha_p^i$ are  suppressed by the free energy. For small temperatures $\beta_i \gg \beta$, the amplitude becomes constant, of value $Z(\beta)^{-1}$.

Thus, the wormhole state $\ket{\Psi_q^\beta}$ has most of its support around the microcanonical energy window, defined by the equation of state that relates the exterior temperature $\beta$ and the mass of the black hole $M_{BH}$, up to exponentially decaying Gibbs tails in the distribution.\footnote{To be precise, the $\alpha^i_p$ coefficients increases at low energies, but the number of microscopic configurations grows with energy, so most of the support will be at the temperature $\beta$, as in standard statistical mechanics.} We conclude that, when $q>e^{\textbf{S}(M_{BH})}$, i.e., for super-exponential values, the shell states can be written as complicated linear superpositions of at most exponentially large wormholes. This demonstrates a sense in which the volume of Einstein-Rosen bridges is indeed bounded, and relates this bounding quantum mechanically to the finiteness of black hole entropy.

\subsection{Concerning volume operators}

These considerations make it clear that the volume of the interior cannot be measured by a linear operator in quantum mechanics.   To sharpen this point, we can try to define a  ``volume operator'' in the naive Hilbert space of infinite dimension spanned by the family of thin-shell states, by assigning a geometric volume to each of them: \footnote{Notice that the geometric shell states are classically well-defined. They should be understood quantum mechanically as some sort of coherent state.  As such they cannot really be eigenvectors of the true volume operator.  Equivalently, we see this because the uncertainty in the dual  variable in canonical quantum gravity, the extrinsic curvature, is small. In the classical limit they should be thought of as minimum uncertainty states in both volume and curvature, although as discussed above, the volume of the classical solutions cannot be a linear operator of the quantum theory.}
\be\label{eq:volume1}
\hat{V}_{\text{naive}}\ket{\Psi_p}=V_p\ket{\Psi_p}\,\,\,\,\,\,\,\,\,\,\,\,\,\,\,p\in \lbrace 1,...,\infty\rbrace \;,
\ee
where the $V_p$ on the right side is the classically computed volume.
However, we have shown that the fundamental Hilbert space of the black hole is spanned by a finite number of these states, in fact by precisely $e^{\textbf{S}}$ of them. Therefore, it is clear that the observable $\hat{V}_{\text{naive}}$ in \eqref{eq:volume1} cannot exist in the fundamental description of the black hole, since there cannot be so many linearly independent eigenstates.  We could try to remedy this by defining a linear operator restricted to sub-exponential states
\be\label{eq:volume2}
\hat{V}\ket{\Psi_p}=V_p\ket{\Psi_p}\,\,\,\,\,\,\,\,\,\,\,\,\,\,\,p\in \lbrace 1,...,e^{\textbf{S}}\rbrace \;.
\ee
However,  if $q\geq e^{\textbf{S}} $ this linear operator obviously fails to provide the geometric volume of the state
\be
\bra{\Psi_q}\hat{V} \ket{\Psi_q} \approx \sum_{p=1}^{\mathbf{e^{\textbf{S}}}} |\alpha_p|^2 \,V_p  \neq V_q
\ee
This suggests that the geometric volume of the interior states of the black hole must be encoded in a non-linear or state-dependent way in the fundamental description. Such a semiclassical volume function could still exist in the Hilbert space of the black hole. In the next section we describe linear volume operators related to sensible notions of state complexity that are correspondingly bounded above.

\subsection{Toy model of interior growth and saturation of complexity}

We have shown that the super-exponential wormhole states can be written as superpositions of exponential ones. With this in mind, in this section we  will construct a toy model describing a time evolution between wormholes of different length. We will then define a notion of volume in these universes which is related to the ``spread complexity'' of the underlying state in a basis of volume eigenstates, recently introduced in \cite{Balasubramanian:2022tpr} (reviewed in Appendix~\ref{ApSpread}).  This notion of volume and the spread complexity will both saturate at values exponential in the entropy. \footnote{We remind the reader that, although this notion of volume is natural from the point of view of spread complexity, which involves a minimization over linear operators, it is still not clear whether the non-linear notion of semiclassical volume of our microstates, viewed as a function in Hilbert space, displays saturation.}

\begin{figure}[h]
	\centering
	\includegraphics[width=.6\textwidth]{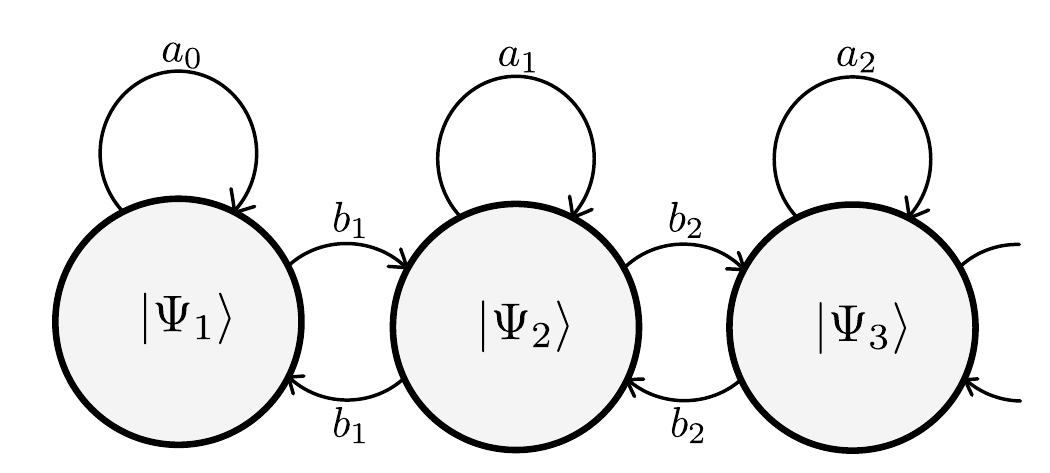}
	\caption{Simple Hamiltonian that effects transition between subsequent shell states. This Hamiltonian is simple since it acts on $\mathcal{O}(S)$ shell particles at a given time. }
	\label{HopingFig2}
\end{figure}

We start with the family of one-shell states used above. These are
\be    
\ket{\Psi_p}\;,\,\,\,\,\,\,\,\,\,\,\,\,\,\,\,\,\,m_p=p\,m\;,\,\,\,\,\,\,\,\,\,\,\,\,\,\,\,\,\,m\sim\mathcal{O}(\ell^{d-2}/G)\;.
\label{triH}
\ee
As proved previously, projecting these states to energy windows $E_i$  of size $\Delta E$, gives rise to finite dimensional Hilbert spaces
\be    
\mathcal{H}_i = 
\textrm{Span}\,\left\lbrace   \vert\Psi_p^i \rangle : p=1,\cdots ,e^{\textbf{S}(E_i)}\right\rbrace\;.
\ee
In this scenario,  we can create a toy model of black hole interior growth by defining, for each window of energy $E$, a ``simple'' hopping Hamiltonian that produces transitions between subsequent shell states of growing volume. This is shown schematically  in Fig.~\ref{HopingFig2}.  At late times the wavefunction will be completely spread out over the  sub-exponential wormhole microstates, but the amplitudes will keep evolving with time.  We showed that all the classically super-exponential wormholes are superpositions of these sub-exponential ones, which means that detailed differences between amplitudes in the superposition, including the phases, must be involved in determining the classical length.  The details of this will depend on the choice of the hopping parameters, a question  that we will leave for future work.

The Hamiltonian  takes the form\footnote{Here we are using the fact that the shell states can be chosen to form an approximately orthonormal basis.  To be absolutely precise one should slightly modify each shell state to make them exactly orthornormal, but we will not do that in this toy model.}
\be\label{trih}
H= \begin{pmatrix}
a_0 & b_1 &  & & & \\
b_1 & a_1 & b_2 & & &\\
& b_2 & a_2 & b_3 & & \\
& & \ddots & \ddots & \ddots & \\
& & & b_{e^{\textbf{S}(E)}-2} & a_{e^{\textbf{S}(E)}-2} & b_{e^{\textbf{S}(E)}-1}\\
&  & &  &b_{e^{\textbf{S}(E)}-1} & a_{e^{\textbf{S}(E)}-1}
\end{pmatrix}\;,
\ee
for some choice of $a$'s and $b$'s. This Hamiltonian is `simple' because it acts on small groups of the dust particles making up the shells at any given time, thereby locally changing the dust density and the wormhole volume. Starting from the state with shortest wormhole, namely $\ket{\Psi_1^E}$, time evolution with this Hamiltonian will make the expected wormhole size grow since the wavefunction will move along the ``1d-chain''.

We will study the spread complexity of this evolution. As explained in detail in  Appendix~\ref{ApSpread}, this is the minimum spread of the wave-function over all choices of basis of the Hilbert space. The solution to this minimization is the Krylov basis. For a given initial state $\vert \Psi\rangle$ and a Hamiltonian $H$, the Krylov basis is the the one that arises by the Gram-Schmidt procedure applied to  $H^n\vert \Psi\rangle$. The Krylov basis is simple to compute in our scenario, in which we start with $\ket{\Psi_1^E}$ and evolve with the tridigonal Hamiltonian $H$, beacause the Krylov basis is just the basis of the shell states themselves $\ket{\Psi_{p}^E}$. The wavefunction can be expanded in this basis
\be
|\Psi_1(t)\rangle= \sum_{p}\psi_p (t)\,|\Psi_{p}^E\rangle\;.
\label{eq:krylovexap}
\ee
In terms of the probabilities of each of the shell states as time evolves, spread complexity can be computed as
\be 
C_{\textrm{Spread}}(t) = C_\mathcal{K}(t) =\sum_{p} p \,\vert \psi_p(t)\vert^2\;.
\ee
This notion of complexity is just the expectation value of the position operator along the 1d-chain. Since the position in the chain is the mass of the shell in appropriate units, and the classical volume is proportional to the mass as well, the spread complexity and the expected Einstein-Rosen volume in the shell basis match. For example, we can use the precise analytical formulas in $d=2$. When acting on shell states we thus identify the position operator\footnote{In the context of spread complexity one typically defines the first state to have position, and therefore spread complexity, zero. Here we will define it to have initial position $1$. This is just a constant shift that accounts for the initial complexity of the shell operator. In any case this shift is irrelevant at late times.} in the chain with a linear volume operator as
\be     
\hat{p}\,\ket{\Psi_p}=p\,\ket{\Psi_p} \,\,\,\,\,\,\,\,\,\,\,\, \hat{\mathcal{C}}\equiv\dfrac{d-1}{8\pi G\ell}\,\hat{V}\ket{\Psi_p}=p\,m\,\ell\,\ket{\Psi_p}\;.
\ee
Now we  equate the spread complexity position operator to a ``holographic complexity operator'' as
\be \label{match} 
\hat{\mathcal{C}}=m\,\ell\,\hat{p}
\ee

Tridiagonal Hamiltonians such as~(\ref{trih}) were studied extensively in quantum chaotic models in \cite{Balasubramanian:2022tpr,Balasubramanian:2022dnj}.  Following that work, the spread complexity will increase for an exponentially long time, and saturate at an exponentially large value.  By construction in our toy model,  the average wormhole volume as measured by the linear volume operator defined above will also show this behavior.  Meanwhile, the state itself can continue to evolve with time, and the classical volume, no longer simply related to the volume operator we constructed, or to the complexity of time evoultion, could continue to increase.

\section{Summary and discussion}\label{SecVII}

The key insight in our approach is that to understand the microscopic origin of black hole entropy it is enough to construct any well-controlled and sufficiently large space of states. The rank of the matrix of overlaps between these states then reveals the Hilbert space dimension.  We have shown that there are many families of such microstates that are geometrical and under semiclassical control.  These states are highly atypical in the Hilbert space, and do not have the standard Schwarzschild interior. But nevertheless, they are sufficient to demonstrate that the Bekenstein-Hawking entropy can be explained as the dimension of an underlying quantum Hilbert space of the black hole.

Our results followed from the appearance of quantum wormholes that contribute to the  overlaps between apparently distinct classical states.  These are non-perturbative effects in quantum gravity. We also showed that all our results could be understood phenomenologically by assuming that the geometric configurations we construct are random phase superpositions of energy eigenstates.

An outstanding question that the field has been recently grappling with is, ``Why is the semiclassical gravitational path integral so clever?'' It knows how to compute the black hole entropy,\footnote{See \cite{Iliesiu:2022kny} for a recent application in the context of supersymmetric black holes in theories with a high degree of supersymmetry. It would be interesting to understand to what extent our techniques can be extended to such scenarios.} and moreover it appears to resolve some aspects of the information paradox, such as  the expected decay of the Page curve at late times in black hole evaporation.  At first sight this seems preposterous, because both questions would seem to require access to the complete space of microstates.
Here, we have shown that there are enough states under semiclassical control to give access to the dimension of the full Hilbert space.  This makes it possible for the semiclassical path integral to correctly answer questions that depend on the Hilbert space dimension, like the entropy of black holes or the decay of the Page curve.

Finally, our results do not explicitly use any details of string theory, AdS/CFT, or any other formulation of quantum gravity.  Indeed, the only assumptions we have used are that: (a) there is some ultraviolet completion, and (b) the semiclassical Euclidean path integral provides sensible information about the ultraviolet completion.  With this assumptions, our results provide an explanation for the entropy of black holes in any theory that has general relativity coupled to massive matter as a low-energy limit. In particular, this construction works in specific top-down models of quantum gravity, such as those appearing in string theory and AdS/CFT, since these theories contain the necessary ingredients to construct our microstates. Perhaps this explains the universality of the Bekenstein-Hawking entropy formula. We have worked in universes with a negative cosmological constant because there are additional tools in this case that allow us to put our state-construction methods on an entirely firm footing. But there is no obstacle to repeating the analysis in, for example, asymptotically flat spacetimes \cite{Balasubramanian:2022lnw}.

\section*{Acknowledgments}

We would like to thank Jos\'{e} Barb\'{o}n, Horacio Casini and Roberto Emparan for useful discussions.  VB and JM are supported in part by the Department of Energy through  DE-SC0013528 and  QuantISED DE-SC0020360, as well as by the Simons Foundation through the It From Qubit Collaboration (Grant No. 38559). 
AL and MS are supported in part by the Department of Energy through DE-SC0009986 and QuantISED DE-SC0020360. This preprint is assigned the code BRX-TH-6713.

\appendix

\section{Thin shell formalism}
\label{appendix:A}
	
In this appendix we provide a review of the thin-shell formalism. The objective is to glue two Euclidean Schwarzshild-AdS regions $X^\pm$ with local geometry
	\be\label{eq:appAgeometry}
	\text{d}s_\pm^2\, = \,  f_\pm(r)\,\text{d}\tau_\pm \,+\,\dfrac{\text{d}r^2}{f_\pm(r)}\,+\,r^2\,\text{d}	 \Omega_{d-1}^2\;.
	\ee
The gluing of 
the geometries
along the trajectory $\mathcal{W}$ of a codimension-one domain is performed using Israel's junction conditions. We denote by $(h^\pm_{ab},K_{ab}^\pm)$ the induced metrics $h^\pm_{ab}$ and extrinsic curvatures $K_{ab}^\pm$ in terms of the metric on each side $X^\pm$. These quantities are evaluated at $\mathcal{W}$. Denoting also $\Delta h_{ab} = h_{ab}^+ - h_{ab}^-$ and $\Delta K_{ab} = K_{ab}^+ - K_{ab}^-$, the junction conditions are simply
	\begin{gather}
	\Delta h_{ab} =0 \,,\label{junc1}\\[.4cm]
	\Delta K_{ab}- h_{ab} \Delta K = -8\pi G S_{ab}\,,\label{junc2}
	\end{gather}
where $\Delta K  = h^{ab}\Delta K_{ab}$,\footnote{Notice a relative minus sign in this expression with respect to the Lorentzian version. This arises since the hypersurface $\mathcal{W}$ is spacelike.} and $S_{ab}$ is the energy-momentum of the thin domain wall.  For a dust shell, we have $S_{ab} = -\sigma u_au_b$, where the minus sign in $S_{ab}$  comes from the analytic continuation $\sigma \rightarrow -\sigma$ in the Lorentzian fluid.
	
The junction conditions determine the motion for the shell,  namely $R=R(T)$ with $T$ the synchronous proper time of the shell. First, the angular parts of \eqref{junc2} impose the conservation of
	\be
	m = \sigma V_\Omega R^{d-1}\,
	\ee
along $\mathcal{W}$. This is the rest mass of the shell. Second, from the continuity of the metric \eqref{junc1} we arrive at
	\be\label{eq:appextrinsic}
	f_\pm \dot{\tau}_\pm = \sqrt{-\dot{R}^2 + f_\pm} \,,
	\ee	
where we use the notation $\dot{x} = dx/dT$ and the square root can have either sign, giving different particular trajectories.
	
The remaining component of \eqref{junc2} finally gives
	\be\label{eq:appjcext}
	\kappa_+ \sqrt{-\dot{R}^2+f_+(R)} - \kappa_- \sqrt{-\dot{R}^2+f_-(R)} = \dfrac{8\pi G m}{(d-1)V_\Omega R^{d-2}}\,,
	\ee
where $\kappa_\pm = \text{sign}(\dot{\tau}_\pm)$ is the sign of the extrinsic curvature. We now square this expression, getting an effective equation of motion for a non-relativistic particle of zero total energy
        \be\label{eq:appeom}
	\dot{R}^2 + V_{\text{eff}}(R) = 0\;,
	\ee
where the effective potential reads
	\be\label{eq:appeffpotential}
	V_{\text{eff}}(R) = -f_+(R) + \left(\dfrac{M_+-M_-}{m} - \dfrac{4\pi G m}{(d-1)V_\Omega R^{d-2}}\right)^2\;.
	\ee
In this non-relativistic dynamics describing the trajectory of the shell in the Euclidean geometry, the shell starts at the boundary $R=r_\infty$, and bounces at $R = R_*$ with $V(R_*)=0$. Notice that $R_*\geq r_\pm$ for the respective horizon radii. The time the shell takes to do this motion can be computed from \eqref{eq:appextrinsic}. It is given by the integral
	\be\label{eq:appetime}
	\Delta \tau_\pm \, = \, 2\,\int_{R_*}^\infty \dfrac{\text{d}R}{f_\pm}\,\sqrt{\dfrac{f_\pm + {V}_{\text{eff}}}{-{V}_{\text{eff}}}}\;.
	\ee

\subsection*{2+1 dimensions}
	
The  $d=2$ case can be worked out more explicitly. For the BTZ black hole, the event horizon radii are $r_\pm = \sqrt{8GM_\pm}$, the temperatures by $\beta_\pm     = 2\pi/r_\pm$, and the effective potential \eqref{eq:appeffpotential} reads
	\be\label{2+1potential2}
	{V}_{\text{eff}}(R)\, = \, -(r^2-R_*^2)\;,
	\ee  
where the turning point takes the explicit form
	\be\label{2+1radius2}
	R_*\, = \, \sqrt{r_+^2\,+\,\left(\dfrac{M_+-M_-}{m}\,-\,2Gm\right)^2}\;.
	\ee
The exact solution for the shell's motion is \eqref{eq:appeom} can be explicitly found
	\be\label{eq:app2+1sol2}
	R(T)\, = \, R_*\,\cosh T\;,
	\ee
where we have chosen the initial conditions forcing the shell to pass through $R_*$ at proper time $T=0$.
Finally, the Euclidean time elapsed by the shell \eqref{eq:appetime} during its motion can also be computed to give
	\be\label{etime2+1}
	\Delta \tau_\pm \, = \beta_\pm \,\dfrac{\arcsin(r_\pm/R_*)}{\pi}\;.
	\ee

\section{Higher moments and multi-boundary wormholes}
\label{appendix:B}

In this Appendix we provide details of the semiclassical computation of higher moments of the overlaps for one-shell states. The connected part of the $n^{\text{th}}$ moment is
\be
\overline{\bra{\Psi_{m_1}} \ket{\Psi_{m_2}}\bra{\Psi_{m_2}} \ket{\Psi_{m_3}}...\bra{\Psi_{m_n}} \ket{\Psi_{m_1}}}|_c =\dfrac{Z_n}{Z^{m_1}_1...Z^{m_n}_1}\,,
\ee
where 
\be
Z_n = e^{-I[X_n]}\;,
\ee
is the contribution from the gravitational action of the $n$-boundary wormhole $X_n$ that we will now construct.  The normalizations $Z^{m_i}_1$ are the ones computed in Sec. \ref{SecIV}. 

\begin{figure}[h]
		\centering
		\includegraphics[width=.55\textwidth]{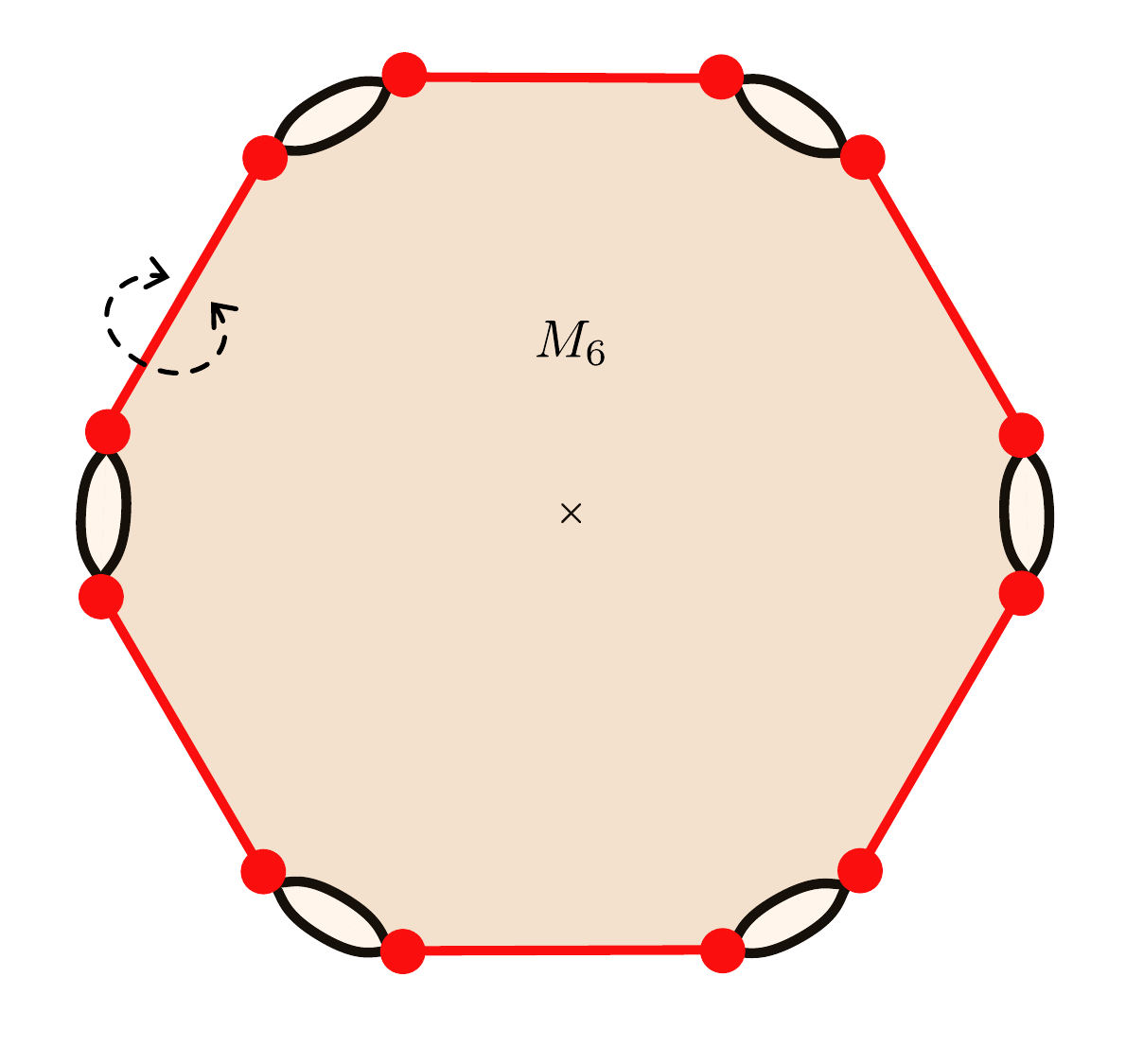}
		\caption{The wormhole $X_6$ consists of a pair of Euclidean black holes of the same mass, $M_6$, which are glued together along the trajectory of the $6$ thin shells (red lines).  The black holes are the boundaries of the geometry.  The second black hole is behind the figure in this perspective, as indicated by the dashed arrows.}
		\label{fig:nwh}
	\end{figure}

The wormhole $X_n$ with $n$ boundaries consists of a pair of Euclidean black holes of the same temperature $\beta_n$,\footnote{Here we assume that the original temperature is so small that the $n$-boundary saddle is still dominated by a large AdS black hole.} which are glued together along the trajectory of the $n$ thin shells. This is shown in Fig.~\ref{fig:nwh} for the sixth moment. The saddle-point equations read
    \be
    \beta_n = \sum_{i=1}^n \left(\tbeta_{m_i} +  \Delta \tau_{m_i}\right)\;.
    \label{eq:betanconstraintn}
    \ee
In this equation, the $n$ preparation temperatures $\tilde{\beta}_{m_i}$ have already been fixed by  Eq.~\ref{betapb}. Then in the wormhole solution the equation of motion (\ref{eq:shifttime}) fixes the elapsed Euclidean time for each shell as a function of the mass of the shell and the background black hole geometry parametrized by $\beta_n$.   The constraint (\ref{eq:betanconstraintn}) then fixes $\beta_n$.

Having found this solution, the action of the wormhole after counterterm subtraction can be derived by the same logic as in Sec. \ref{SecIV}, giving
\be\label{eq:whaction}
    I[X_n] = 2 \left(\sum_{i=1}^n \tbeta_{m_i}\right) F(\beta_n) + \sum_{i=1}^n I[X_{\text{shell}}^{m_i}]\;.
\ee
Including the contributions from the normalizations $Z_1$ and $Z_1'$ we obtain the semiclassical overlap squared
    \be\label{eq:Bfunctionn}
  \overline{\bra{\Psi_{m_1}} \ket{\Psi_{m_2}}\bra{\Psi_{m_2}} \ket{\Psi_{m_3}}...\bra{\Psi_{m_n}} \ket{\Psi_{m_1}}}|_c = e^{-2\,(\sum_i \tbeta_{m_i}) \,\Delta F - \sum_i \Delta I[X_{\text{shell}}^{m_i}]} \;,
\ee
where $\Delta F = F(\beta_n) - F(\beta)$ and $\Delta I[X^{m_i}_{\text{shell}}] = I[X^{m_i}_{\text{shell}}]|_{\beta_n} - I[X^{m_i}_{\text{shell}}]|_{\beta}$.

\subsection*{Large mass limit and universality}

In the limit $m_i\gg M$ for all of the shells, the preparation temperature coincides with the physical temperature, $\tbeta_{m_i} \approx \beta$. The equation \eqref{eq:betanconstraintn} gives $\beta_n \approx n\beta$ and the moment \eqref{eq:Bfunctionn} reduces to
\be\label{eq:Bfunctionnmass}
  \overline{\bra{\Psi_{m_1}} \ket{\Psi_{m_2}}\bra{\Psi_{m_2}} \ket{\Psi_{m_3}}...\bra{\Psi_{m_n}} \ket{\Psi_{m_1}}}|_c \approx \dfrac{Z(n\beta)^2}{Z(\beta)^{2n}}  \;.
\ee
This is the main result for the overlap that we use in Sec. \ref{SecV}. It is straightforward to show that \eqref{eq:Bfunctionnmass} holds also for higher moments of the overlaps between general multi-shell states in the large mass limit.

\section{Spread complexity}\label{ApSpread}
In this Appendix we briefly review of the notion of ``spread complexity'', recently introduced in \cite{Balasubramanian:2022tpr}. Consider a time-independent Hamiltonian $H$.  Time evolution of a initial state $\vert \psi \rangle$ is determined by the Schr\"{o}dinger equation
\be
i\partial_t\vert \psi (t)\rangle=H\vert \psi (t)\rangle \; .
\label{eq:se}
\ee
A natural notion of quantum complexity arises from quantification of the spread of  $\vert \psi(t) \rangle$  over the Hilbert space. To this end, we can define a cost function relative to a complete, orthonormal, ordered basis,  
$\mathcal{B} = 
\set{\vert B_n \rangle : n=0,1,2,\cdots}$ for the Hilbert space
\begin{equation}
C_\mathcal{B}(t) 
=\sum_n n\,\abs{\braket{\psi(t)}{B_n}}^2\equiv \sum_n n\,p_{\mathcal{B}}(n,t)  \, ,
\label{eq:cost}
\end{equation}
Unitarity of time evolution implies that the cost of a wavefunction increases if it spreads deeper into the basis. The question then arises as to which basis should we use to measure the spread. Inspired by other notions of complexity, such as Kolmogorov complexity \cite{doi:10.1080/00207166808803030}, we now define ``spread complexity'' as the minimum over a finite time interval of this cost function over all bases $\mathcal{B}$  %
\begin{equation}
    C(t) = \min_\mathcal{B}C_\mathcal{B}(t) \, .
    \label{eq:complexitydef}
\end{equation}
Ref. \cite{Balasubramanian:2022tpr} then shows that, under some assumptions, there is an essentially unique basis minimizing (\ref{eq:complexitydef}) across a finite time  domain. This basis follows the expansion
\be\label{exp2}
\vert \psi (t)\rangle=\sum^\infty_{n=0}\frac{(-it)^n}{n!}\vert \psi_n\rangle\;,
\ee
where we have defined
$
\vert \psi_n\rangle=H^n\vert \psi \rangle 
$. 
These states are neither orthogonal, nor normalized. They do however generate an ordered, orthonormal basis $\mathcal{K}=\set{\ket{K_n}: n=0,1,2,\cdots}$, by means of the Gram–Schmidt procedure with initial condition $|\psi(0) \rangle \equiv |K_0\rangle$. The basis $\mathcal{K}$ is called the Krylov basis in the recent literature.

To calculate this notion of spread complexity we must derive the Krylov basis ${\cal K}$. This is achieved via the Lanczos algorithm \cite{Lanczos1950AnIM,Viswanath1994TheRM}, which recursively applies the Gram–Schmidt procedure to $\vert \psi_n\rangle = H^n |\psi(0)\rangle$ to generate 
$\mathcal{K}=\set{\ket{K_n}: n=0,1,2,\cdots}$ in the following way: 
\be
|A_{n+1}\rangle=(H-a_{n})|K_n\rangle-b_n|K_{n-1}\rangle,\quad |K_n\rangle=b^{-1}_n|A_n\rangle\;,
\label{eq:Lrecursion}
\ee
where  $a_n$ and $b_n$ are dubbed the Lanczos coefficients
\be
a_n=\langle K_n|H|K_n\rangle,\qquad b_n=\langle A_n|A_n\rangle^{1/2}\;.
\ee
The initial conditions for this recursive generation are $b_0 \equiv 0$ and $|K_0\rangle=|\psi(0)\rangle$ being the initial state.

It is immediate that the Lanczos algorithm (\ref{eq:Lrecursion}) implies that
\be\label{Hact}
H|K_n\rangle=a_n|K_n\rangle+b_{n+1}|K_{n+1}\rangle+b_n|K_{n-1}\rangle \; .
\ee
Equivalently, the Hamiltonian becomes a tri-diagonal matrix in the Krylov basis. For finite-dimensional systems, like the ones we consider in this paper, this representation is known as the ``Hessenberg form'' of the Hamiltonian:
\be\label{triH}
H= \begin{pmatrix}
a_0 & b_1 &  & & & \\
b_1 & a_1 & b_2 & & &\\
& b_2 & a_2 & b_3 & & \\
& & \ddots & \ddots & \ddots & \\
& & & b_{N-2} & a_{N-2} & b_{N-1}\\
&  & &  &b_{N-1} & a_{N-1}
\end{pmatrix}\;.
\ee
This leads to the interesting observation that every quantum dynamical evolution can be framed in terms of an appropriate one-dimensional hopping Hamiltonian.

To compute the spread complexity, we must expand the state in the basis in which the spread is minimized:
\be
|\psi(t)\rangle=\sum_{n}\psi_n (t)|K_n\rangle \;.
\label{eq:krylovexap2}
\ee
Finally, given $\psi_n(t)$ we  apply the definition of complexity in (\ref{eq:cost}, \ref{eq:complexitydef}):
\be 
C(t) = C_\mathcal{K}(t) =
\sum_{n} n \,p_n(t)=\sum_{n} n \,\vert \psi_n(t)\vert^2\;.
\ee
This notion of spread complexity can be seen as a generalization of the notion of Krylov operator complexity, introduced in \cite{Parker:2018yvk}, to quantum states, now understood in terms of a minimization of the spread of the wave function over all choices of basis.

\bibliographystyle{utphys}
\bibliography{main}

\providecommand{\href}[2]{#2}\begingroup\raggedright\begin{thebibliography}{100}

\bibitem{PhysRevD.7.2333}
J.~D. Bekenstein, ``Black holes and entropy,''
  \href{http://dx.doi.org/10.1103/PhysRevD.7.2333}{{\em Phys. Rev. D}
  {\bfseries 7} (Apr, 1973) 2333--2346}.
  \url{https://link.aps.org/doi/10.1103/PhysRevD.7.2333}.

\bibitem{Hawking:1975vcx}
S.~W. Hawking, ``{Particle Creation by Black Holes},''
  \href{http://dx.doi.org/10.1007/BF02345020}{{\em Commun. Math. Phys.}
  {\bfseries 43} (1975) 199--220}. [Erratum: Commun.Math.Phys. 46, 206 (1976)].

\bibitem{Hawking1979}
S.~W. Hawking, {\em Euclidean Quantum Gravity},
  \href{http://dx.doi.org/10.1007/978-1-4613-2955-8_4}{pp.~145--173}.
\newblock Springer US, Boston, MA, 1979.
\newblock \url{https://doi.org/10.1007/978-1-4613-2955-8_4}.

\bibitem{Maldacena:1997re}
J.~M. Maldacena, ``{The Large N limit of superconformal field theories and
  supergravity},'' \href{http://dx.doi.org/10.1023/A:1026654312961}{{\em Adv.
  Theor. Math. Phys.} {\bfseries 2} (1998) 231--252},
  \href{http://arxiv.org/abs/hep-th/9711200}{{\ttfamily arXiv:hep-th/9711200}}.

\bibitem{Witten:1998qj}
E.~Witten, ``{Anti-de Sitter space and holography},''
  \href{http://dx.doi.org/10.4310/ATMP.1998.v2.n2.a2}{{\em Adv. Theor. Math.
  Phys.} {\bfseries 2} (1998) 253--291},
  \href{http://arxiv.org/abs/hep-th/9802150}{{\ttfamily arXiv:hep-th/9802150}}.

\bibitem{Gubser:1998bc}
S.~S. Gubser, I.~R. Klebanov, and A.~M. Polyakov, ``{Gauge theory correlators
  from noncritical string theory},''
  \href{http://dx.doi.org/10.1016/S0370-2693(98)00377-3}{{\em Phys. Lett. B}
  {\bfseries 428} (1998) 105--114},
  \href{http://arxiv.org/abs/hep-th/9802109}{{\ttfamily arXiv:hep-th/9802109}}.

\bibitem{Ryu:2006bv}
S.~Ryu and T.~Takayanagi, ``{Holographic derivation of entanglement entropy
  from AdS/CFT},'' \href{http://dx.doi.org/10.1103/PhysRevLett.96.181602}{{\em
  Phys. Rev. Lett.} {\bfseries 96} (2006) 181602},
  \href{http://arxiv.org/abs/hep-th/0603001}{{\ttfamily arXiv:hep-th/0603001}}.

\bibitem{Lewkowycz:2013nqa}
A.~Lewkowycz and J.~Maldacena, ``{Generalized gravitational entropy},''
  \href{http://dx.doi.org/10.1007/JHEP08(2013)090}{{\em JHEP} {\bfseries 08}
  (2013) 090}, \href{http://arxiv.org/abs/1304.4926}{{\ttfamily arXiv:1304.4926
  [hep-th]}}.

\bibitem{Strominger:1996sh}
A.~Strominger and C.~Vafa, ``{Microscopic origin of the Bekenstein-Hawking
  entropy},'' \href{http://dx.doi.org/10.1016/0370-2693(96)00345-0}{{\em Phys.
  Lett. B} {\bfseries 379} (1996) 99--104},
  \href{http://arxiv.org/abs/hep-th/9601029}{{\ttfamily arXiv:hep-th/9601029}}.

\bibitem{https://doi.org/10.48550/arxiv.2204.13113}
I.~Bena, E.~J. Martinec, S.~D. Mathur, and N.~P. Warner, ``Fuzzballs and
  microstate geometries: Black-hole structure in string theory,'' 2022.
\newblock \url{https://arxiv.org/abs/2204.13113}.

\bibitem{Bombelli:1986rw}
L.~Bombelli, R.~K. Koul, J.~Lee, and R.~D. Sorkin, ``{A Quantum Source of
  Entropy for Black Holes},''
  \href{http://dx.doi.org/10.1103/PhysRevD.34.373}{{\em Phys. Rev. D}
  {\bfseries 34} (1986) 373--383}.

\bibitem{Srednicki:1993im}
M.~Srednicki, ``{Entropy and area},''
  \href{http://dx.doi.org/10.1103/PhysRevLett.71.666}{{\em Phys. Rev. Lett.}
  {\bfseries 71} (1993) 666--669},
  \href{http://arxiv.org/abs/hep-th/9303048}{{\ttfamily arXiv:hep-th/9303048}}.

\bibitem{tHooft:1984kcu}
G.~'t~Hooft, ``{On the Quantum Structure of a Black Hole},''
  \href{http://dx.doi.org/10.1016/0550-3213(85)90418-3}{{\em Nucl. Phys. B}
  {\bfseries 256} (1985) 727--745}.

\bibitem{Jacobson:2012ek}
T.~Jacobson and A.~Satz, ``{Black hole entanglement entropy and the
  renormalization group},''
  \href{http://dx.doi.org/10.1103/PhysRevD.87.084047}{{\em Phys. Rev. D}
  {\bfseries 87} no.~8, (2013) 084047},
  \href{http://arxiv.org/abs/1212.6824}{{\ttfamily arXiv:1212.6824 [hep-th]}}.

\bibitem{Cooperman:2013iqr}
J.~H. Cooperman and M.~A. Luty, ``{Renormalization of Entanglement Entropy and
  the Gravitational Effective Action},''
  \href{http://dx.doi.org/10.1007/JHEP12(2014)045}{{\em JHEP} {\bfseries 12}
  (2014) 045}, \href{http://arxiv.org/abs/1302.1878}{{\ttfamily arXiv:1302.1878
  [hep-th]}}.

\bibitem{Chandrasekaran:2022eqq}
V.~Chandrasekaran, G.~Penington, and E.~Witten, ``{Large N algebras and
  generalized entropy},'' \href{http://arxiv.org/abs/2209.10454}{{\ttfamily
  arXiv:2209.10454 [hep-th]}}.

\bibitem{Susskind:2014rva}
L.~Susskind, ``{Computational Complexity and Black Hole Horizons},''
  \href{http://dx.doi.org/10.1002/prop.201500092}{{\em Fortsch. Phys.}
  {\bfseries 64} (2016) 24--43},
  \href{http://arxiv.org/abs/1403.5695}{{\ttfamily arXiv:1403.5695 [hep-th]}}.
  [Addendum: Fortsch.Phys. 64, 44--48 (2016)].

\bibitem{Stanford:2014jda}
D.~Stanford and L.~Susskind, ``{Complexity and Shock Wave Geometries},''
  \href{http://dx.doi.org/10.1103/PhysRevD.90.126007}{{\em Phys. Rev. D}
  {\bfseries 90} no.~12, (2014) 126007},
  \href{http://arxiv.org/abs/1406.2678}{{\ttfamily arXiv:1406.2678 [hep-th]}}.

\bibitem{Aaronson:2016vto}
S.~Aaronson, ``{The Complexity of Quantum States and Transformations: From
  Quantum Money to Black Holes},''
\newblock 7, 2016.
\newblock \href{http://arxiv.org/abs/1607.05256}{{\ttfamily arXiv:1607.05256
  [quant-ph]}}.

\bibitem{Balasubramanian:2019wgd}
V.~Balasubramanian, M.~Decross, A.~Kar, and O.~Parrikar, ``{Quantum Complexity
  of Time Evolution with Chaotic Hamiltonians},''
  \href{http://dx.doi.org/10.1007/JHEP01(2020)134}{{\em JHEP} {\bfseries 01}
  (2020) 134}, \href{http://arxiv.org/abs/1905.05765}{{\ttfamily
  arXiv:1905.05765 [hep-th]}}.

\bibitem{Iliesiu:2021ari}
L.~V. Iliesiu, M.~Mezei, and G.~S\'arosi, ``{The volume of the black hole
  interior at late times},''
  \href{http://dx.doi.org/10.1007/JHEP07(2022)073}{{\em JHEP} {\bfseries 07}
  (2022) 073}, \href{http://arxiv.org/abs/2107.06286}{{\ttfamily
  arXiv:2107.06286 [hep-th]}}.

\bibitem{Martin}
M.~Sasieta, ``{Wormholes from heavy operator statistics in AdS/CFT},''
  \href{http://dx.doi.org/10.1007/JHEP03(2023)158}{{\em JHEP} {\bfseries 03}
  (2023) 158}, \href{http://arxiv.org/abs/2211.11794}{{\ttfamily
  arXiv:2211.11794 [hep-th]}}.

\bibitem{Marolf:2020xie}
D.~Marolf and H.~Maxfield, ``{Transcending the ensemble: baby universes,
  spacetime wormholes, and the order and disorder of black hole information},''
  \href{http://dx.doi.org/10.1007/JHEP08(2020)044}{{\em JHEP} {\bfseries 08}
  (2020) 044}, \href{http://arxiv.org/abs/2002.08950}{{\ttfamily
  arXiv:2002.08950 [hep-th]}}.

\bibitem{Balasubramanian:2020jhl}
V.~Balasubramanian, A.~Kar, S.~F. Ross, and T.~Ugajin, ``{Spin structures and
  baby universes},'' \href{http://dx.doi.org/10.1007/JHEP09(2020)192}{{\em
  JHEP} {\bfseries 09} (2020) 192},
  \href{http://arxiv.org/abs/2007.04333}{{\ttfamily arXiv:2007.04333
  [hep-th]}}.

\bibitem{Balasubramanian:2022fiy}
V.~Balasubramanian, A.~Kar, Yue, C.~Li, and O.~Parrikar, ``{Quantum Error
  Correction in the Black Hole Interior},''
  \href{http://arxiv.org/abs/2203.01961}{{\ttfamily arXiv:2203.01961
  [hep-th]}}.

\bibitem{Akers:2022qdl}
C.~Akers, N.~Engelhardt, D.~Harlow, G.~Penington, and S.~Vardhan, ``{The black
  hole interior from non-isometric codes and complexity},''
  \href{http://arxiv.org/abs/2207.06536}{{\ttfamily arXiv:2207.06536
  [hep-th]}}.

\bibitem{Kar:2022qkf}
A.~Kar, ``{Non-Isometric Quantum Error Correction in Gravity},''
  \href{http://arxiv.org/abs/2210.13476}{{\ttfamily arXiv:2210.13476
  [hep-th]}}.

\bibitem{Chakravarty:2020wdm}
J.~Chakravarty, ``{Overcounting of interior excitations: A resolution to the
  bags of gold paradox in AdS},''
  \href{http://dx.doi.org/10.1007/JHEP02(2021)027}{{\em JHEP} {\bfseries 02}
  (2021) 027}, \href{http://arxiv.org/abs/2010.03575}{{\ttfamily
  arXiv:2010.03575 [hep-th]}}.

\bibitem{chao2017overlapping}
R.~Chao, B.~W. Reichardt, C.~Sutherland, and T.~Vidick, ``Overlapping qubits,''
  {\em arXiv preprint arXiv:1701.01062} (2017) .

\bibitem{Cotler:2016fpe}
J.~S. Cotler, G.~Gur-Ari, M.~Hanada, J.~Polchinski, P.~Saad, S.~H. Shenker,
  D.~Stanford, A.~Streicher, and M.~Tezuka, ``{Black Holes and Random
  Matrices},'' \href{http://dx.doi.org/10.1007/JHEP05(2017)118}{{\em JHEP}
  {\bfseries 05} (2017) 118}, \href{http://arxiv.org/abs/1611.04650}{{\ttfamily
  arXiv:1611.04650 [hep-th]}}. [Erratum: JHEP 09, 002 (2018)].

\bibitem{Saad:2018bqo}
P.~Saad, S.~H. Shenker, and D.~Stanford, ``{A semiclassical ramp in SYK and in
  gravity},'' \href{http://arxiv.org/abs/1806.06840}{{\ttfamily
  arXiv:1806.06840 [hep-th]}}.

\bibitem{Saad:2019lba}
P.~Saad, S.~H. Shenker, and D.~Stanford, ``{JT gravity as a matrix integral},''
  \href{http://arxiv.org/abs/1903.11115}{{\ttfamily arXiv:1903.11115
  [hep-th]}}.

\bibitem{Mertens:2022irh}
T.~G. Mertens and G.~J. Turiaci, ``{Solvable Models of Quantum Black Holes: A
  Review on Jackiw-Teitelboim Gravity},''
  \href{http://arxiv.org/abs/2210.10846}{{\ttfamily arXiv:2210.10846
  [hep-th]}}.

\bibitem{Wheeler}
J.~Wheeler, ``{Relativity, groups and topology},'' {\em edited by B. S. DeWitt
  and C. M. DeWitt. Gordon and Breach, New York, 1964} .

\bibitem{Almheiri:2020cfm}
A.~Almheiri, T.~Hartman, J.~Maldacena, E.~Shaghoulian, and A.~Tajdini, ``{The
  entropy of Hawking radiation},''
  \href{http://dx.doi.org/10.1103/RevModPhys.93.035002}{{\em Rev. Mod. Phys.}
  {\bfseries 93} no.~3, (2021) 035002},
  \href{http://arxiv.org/abs/2006.06872}{{\ttfamily arXiv:2006.06872
  [hep-th]}}.

\bibitem{Brown:2019rox}
A.~R. Brown, H.~Gharibyan, G.~Penington, and L.~Susskind, ``{The
  Python\textquoteright{}s Lunch: geometric obstructions to decoding Hawking
  radiation},'' \href{http://dx.doi.org/10.1007/JHEP08(2020)121}{{\em JHEP}
  {\bfseries 08} (2020) 121}, \href{http://arxiv.org/abs/1912.00228}{{\ttfamily
  arXiv:1912.00228 [hep-th]}}.

\bibitem{Balasubramanian:2022tpr}
V.~Balasubramanian, P.~Caputa, J.~M. Magan, and Q.~Wu, ``{Quantum chaos and the
  complexity of spread of states},''
  \href{http://dx.doi.org/10.1103/PhysRevD.106.046007}{{\em Phys. Rev. D}
  {\bfseries 106} no.~4, (2022) 046007},
  \href{http://arxiv.org/abs/2202.06957}{{\ttfamily arXiv:2202.06957
  [hep-th]}}.

\bibitem{Balasubramanian:2022dnj}
V.~Balasubramanian, J.~M. Magan, and Q.~Wu, ``{A Tale of Two Hungarians:
  Tridiagonalizing Random Matrices},''
  \href{http://arxiv.org/abs/2208.08452}{{\ttfamily arXiv:2208.08452
  [hep-th]}}.

\bibitem{Kim:2020cds}
I.~Kim, E.~Tang, and J.~Preskill, ``{The ghost in the radiation: Robust
  encodings of the black hole interior},''
  \href{http://dx.doi.org/10.1007/JHEP06(2020)031}{{\em JHEP} {\bfseries 06}
  (2020) 031}, \href{http://arxiv.org/abs/2003.05451}{{\ttfamily
  arXiv:2003.05451 [hep-th]}}.

\bibitem{Susskind:1993if}
L.~Susskind, L.~Thorlacius, and J.~Uglum, ``{The Stretched horizon and black
  hole complementarity},''
  \href{http://dx.doi.org/10.1103/PhysRevD.48.3743}{{\em Phys. Rev. D}
  {\bfseries 48} (1993) 3743--3761},
  \href{http://arxiv.org/abs/hep-th/9306069}{{\ttfamily arXiv:hep-th/9306069}}.

\bibitem{Frolov:1998wf}
V.~P. Frolov and I.~D. Novikov, eds.,
  \href{http://dx.doi.org/10.1007/978-94-011-5139-9}{{\em {Black hole physics:
  Basic concepts and new developments}}}.
\newblock 1998.

\bibitem{Hsu:2008yi}
S.~D.~H. Hsu and D.~Reeb, ``{Unitarity and the Hilbert space of quantum
  gravity},'' \href{http://dx.doi.org/10.1088/0264-9381/25/23/235007}{{\em
  Class. Quant. Grav.} {\bfseries 25} (2008) 235007},
  \href{http://arxiv.org/abs/0803.4212}{{\ttfamily arXiv:0803.4212 [hep-th]}}.

\bibitem{Fu:2019oyc}
Z.~Fu and D.~Marolf, ``{Bag-of-gold spacetimes, Euclidean wormholes, and
  inflation from domain walls in AdS/CFT},''
  \href{http://dx.doi.org/10.1007/JHEP11(2019)040}{{\em JHEP} {\bfseries 11}
  (2019) 040}, \href{http://arxiv.org/abs/1909.02505}{{\ttfamily
  arXiv:1909.02505 [hep-th]}}.

\bibitem{Kourkoulou:2017zaj}
I.~Kourkoulou and J.~Maldacena, ``{Pure states in the SYK model and
  nearly-$AdS_2$ gravity},'' \href{http://arxiv.org/abs/1707.02325}{{\ttfamily
  arXiv:1707.02325 [hep-th]}}.

\bibitem{Goel:2018ubv}
A.~Goel, H.~T. Lam, G.~J. Turiaci, and H.~Verlinde, ``{Expanding the Black Hole
  Interior: Partially Entangled Thermal States in SYK},''
  \href{http://dx.doi.org/10.1007/JHEP02(2019)156}{{\em JHEP} {\bfseries 02}
  (2019) 156}, \href{http://arxiv.org/abs/1807.03916}{{\ttfamily
  arXiv:1807.03916 [hep-th]}}.

\bibitem{Chandra:2022fwi}
J.~Chandra and T.~Hartman, ``{Coarse graining pure states in AdS/CFT},''
  \href{http://arxiv.org/abs/2206.03414}{{\ttfamily arXiv:2206.03414
  [hep-th]}}.

\bibitem{Balasubramanian:1998de}
V.~Balasubramanian, P.~Kraus, A.~E. Lawrence, and S.~P. Trivedi, ``{Holographic
  probes of anti-de Sitter space-times},''
  \href{http://dx.doi.org/10.1103/PhysRevD.59.104021}{{\em Phys. Rev. D}
  {\bfseries 59} (1999) 104021},
  \href{http://arxiv.org/abs/hep-th/9808017}{{\ttfamily arXiv:hep-th/9808017}}.

\bibitem{Banks:1998dd}
T.~Banks, M.~R. Douglas, G.~T. Horowitz, and E.~J. Martinec, ``{AdS dynamics
  from conformal field theory},''
  \href{http://arxiv.org/abs/hep-th/9808016}{{\ttfamily arXiv:hep-th/9808016}}.

\bibitem{Balasubramanian:1999ri}
V.~Balasubramanian, S.~B. Giddings, and A.~E. Lawrence, ``{What do CFTs tell us
  about Anti-de Sitter space-times?},''
  \href{http://dx.doi.org/10.1088/1126-6708/1999/03/001}{{\em JHEP} {\bfseries
  03} (1999) 001}, \href{http://arxiv.org/abs/hep-th/9902052}{{\ttfamily
  arXiv:hep-th/9902052}}.

\bibitem{Hamilton:2006az}
A.~Hamilton, D.~N. Kabat, G.~Lifschytz, and D.~A. Lowe, ``{Holographic
  representation of local bulk operators},''
  \href{http://dx.doi.org/10.1103/PhysRevD.74.066009}{{\em Phys. Rev. D}
  {\bfseries 74} (2006) 066009},
  \href{http://arxiv.org/abs/hep-th/0606141}{{\ttfamily arXiv:hep-th/0606141}}.

\bibitem{Balasubramanian:1999re}
V.~Balasubramanian and P.~Kraus, ``{A Stress tensor for Anti-de Sitter
  gravity},'' \href{http://dx.doi.org/10.1007/s002200050764}{{\em Commun. Math.
  Phys.} {\bfseries 208} (1999) 413--428},
  \href{http://arxiv.org/abs/hep-th/9902121}{{\ttfamily arXiv:hep-th/9902121}}.

\bibitem{Henningson:1998gx}
M.~Henningson and K.~Skenderis, ``{The Holographic Weyl anomaly},''
  \href{http://dx.doi.org/10.1088/1126-6708/1998/07/023}{{\em JHEP} {\bfseries
  07} (1998) 023}, \href{http://arxiv.org/abs/hep-th/9806087}{{\ttfamily
  arXiv:hep-th/9806087}}.

\bibitem{Skenderis:2002wp}
K.~Skenderis, ``{Lecture notes on holographic renormalization},''
  \href{http://dx.doi.org/10.1088/0264-9381/19/22/306}{{\em Class. Quant.
  Grav.} {\bfseries 19} (2002) 5849--5876},
  \href{http://arxiv.org/abs/hep-th/0209067}{{\ttfamily arXiv:hep-th/0209067}}.

\bibitem{Israel:1966rt}
W.~Israel, ``{Singular hypersurfaces and thin shells in general relativity},''
  \href{http://dx.doi.org/10.1007/BF02710419}{{\em Nuovo Cim. B} {\bfseries
  44S10} (1966) 1}. [Erratum: Nuovo Cim.B 48, 463 (1967)].

\bibitem{Maldacena:2001kr}
J.~M. Maldacena, ``{Eternal black holes in anti-de Sitter},''
  \href{http://dx.doi.org/10.1088/1126-6708/2003/04/021}{{\em JHEP} {\bfseries
  04} (2003) 021}, \href{http://arxiv.org/abs/hep-th/0106112}{{\ttfamily
  arXiv:hep-th/0106112}}.

\bibitem{Emparan:1999pm}
R.~Emparan, C.~V. Johnson, and R.~C. Myers, ``{Surface terms as counterterms in
  the AdS / CFT correspondence},''
  \href{http://dx.doi.org/10.1103/PhysRevD.60.104001}{{\em Phys. Rev. D}
  {\bfseries 60} (1999) 104001},
  \href{http://arxiv.org/abs/hep-th/9903238}{{\ttfamily arXiv:hep-th/9903238}}.

\bibitem{Penington:2019kki}
G.~Penington, S.~H. Shenker, D.~Stanford, and Z.~Yang, ``{Replica wormholes and
  the black hole interior},''
  \href{http://dx.doi.org/10.1007/JHEP03(2022)205}{{\em JHEP} {\bfseries 03}
  (2022) 205}, \href{http://arxiv.org/abs/1911.11977}{{\ttfamily
  arXiv:1911.11977 [hep-th]}}.

\bibitem{Stanford:2020wkf}
D.~Stanford, ``{More quantum noise from wormholes},''
  \href{http://arxiv.org/abs/2008.08570}{{\ttfamily arXiv:2008.08570
  [hep-th]}}.

\bibitem{PhysRevLett.125.021601}
J.~Pollack, M.~Rozali, J.~Sully, and D.~Wakeham, ``Eigenstate thermalization
  and disorder averaging in gravity,''
  \href{http://dx.doi.org/10.1103/PhysRevLett.125.021601}{{\em Phys. Rev.
  Lett.} {\bfseries 125} (Jul, 2020) 021601}.
  \url{https://link.aps.org/doi/10.1103/PhysRevLett.125.021601}.

\bibitem{Maloney:2020nni}
A.~Maloney and E.~Witten, ``{Averaging over Narain moduli space},''
  \href{http://dx.doi.org/10.1007/JHEP10(2020)187}{{\em JHEP} {\bfseries 10}
  (2020) 187}, \href{http://arxiv.org/abs/2006.04855}{{\ttfamily
  arXiv:2006.04855 [hep-th]}}.

\bibitem{Cotler:2020ugk}
J.~Cotler and K.~Jensen, ``{AdS$_{3}$ gravity and random CFT},''
  \href{http://dx.doi.org/10.1007/JHEP04(2021)033}{{\em JHEP} {\bfseries 04}
  (2021) 033}, \href{http://arxiv.org/abs/2006.08648}{{\ttfamily
  arXiv:2006.08648 [hep-th]}}.

\bibitem{Altland:2020ccq}
A.~Altland and J.~Sonner, ``{Late time physics of holographic quantum chaos},''
  \href{http://dx.doi.org/10.21468/SciPostPhys.11.2.034}{{\em SciPost Phys.}
  {\bfseries 11} (2021) 034}, \href{http://arxiv.org/abs/2008.02271}{{\ttfamily
  arXiv:2008.02271 [hep-th]}}.

\bibitem{Altland:2021rqn}
A.~Altland, D.~Bagrets, P.~Nayak, J.~Sonner, and M.~Vielma, ``{From operator
  statistics to wormholes},''
  \href{http://dx.doi.org/10.1103/PhysRevResearch.3.033259}{{\em Phys. Rev.
  Res.} {\bfseries 3} no.~3, (2021) 033259},
  \href{http://arxiv.org/abs/2105.12129}{{\ttfamily arXiv:2105.12129
  [hep-th]}}.

\bibitem{Chandra:2022bqq}
J.~Chandra, S.~Collier, T.~Hartman, and A.~Maloney, ``{Semiclassical 3D gravity
  as an average of large-c CFTs},''
  \href{http://arxiv.org/abs/2203.06511}{{\ttfamily arXiv:2203.06511
  [hep-th]}}.

\bibitem{Betzios:2021fnm}
P.~Betzios, E.~Kiritsis, and O.~Papadoulaki, ``{Interacting systems and
  wormholes},'' \href{http://dx.doi.org/10.1007/JHEP02(2022)126}{{\em JHEP}
  {\bfseries 02} (2022) 126}, \href{http://arxiv.org/abs/2110.14655}{{\ttfamily
  arXiv:2110.14655 [hep-th]}}.

\bibitem{PhysRevA.43.2046}
J.~M. Deutsch, ``Quantum statistical mechanics in a closed system,''
  \href{http://dx.doi.org/10.1103/PhysRevA.43.2046}{{\em Phys. Rev. A}
  {\bfseries 43} (Feb, 1991) 2046--2049}.
  \url{https://link.aps.org/doi/10.1103/PhysRevA.43.2046}.

\bibitem{Srednicki_1994}
M.~Srednicki, ``Chaos and quantum thermalization,''
  \href{http://dx.doi.org/10.1103/physreve.50.888}{{\em Physical Review E}
  {\bfseries 50} no.~2, (Aug, 1994) 888--901}.

\bibitem{Belin:2020hea}
A.~Belin and J.~de~Boer, ``{Random statistics of OPE coefficients and Euclidean
  wormholes},'' \href{http://dx.doi.org/10.1088/1361-6382/ac1082}{{\em Class.
  Quant. Grav.} {\bfseries 38} no.~16, (2021) 164001},
  \href{http://arxiv.org/abs/2006.05499}{{\ttfamily arXiv:2006.05499
  [hep-th]}}.

\bibitem{Jafferis:2022uhu}
D.~L. Jafferis, D.~K. Kolchmeyer, B.~Mukhametzhanov, and J.~Sonner, ``{Matrix
  models for eigenstate thermalization},''
  \href{http://arxiv.org/abs/2209.02130}{{\ttfamily arXiv:2209.02130
  [hep-th]}}.

\bibitem{Maldacena:2015waa}
J.~Maldacena, S.~H. Shenker, and D.~Stanford, ``{A bound on chaos},''
  \href{http://dx.doi.org/10.1007/JHEP08(2016)106}{{\em JHEP} {\bfseries 08}
  (2016) 106}, \href{http://arxiv.org/abs/1503.01409}{{\ttfamily
  arXiv:1503.01409 [hep-th]}}.

\bibitem{Balasubramanian:2005kk}
V.~Balasubramanian, V.~Jejjala, and J.~Simon, ``{The Library of Babel},''
  \href{http://dx.doi.org/10.1142/S0218271805007826}{{\em Int. J. Mod. Phys. D}
  {\bfseries 14} (2005) 2181--2186},
  \href{http://arxiv.org/abs/hep-th/0505123}{{\ttfamily arXiv:hep-th/0505123}}.

\bibitem{Balasubramanian:2005mg}
V.~Balasubramanian, J.~de~Boer, V.~Jejjala, and J.~Simon, ``{The Library of
  Babel: On the origin of gravitational thermodynamics},''
  \href{http://dx.doi.org/10.1088/1126-6708/2005/12/006}{{\em JHEP} {\bfseries
  12} (2005) 006}, \href{http://arxiv.org/abs/hep-th/0508023}{{\ttfamily
  arXiv:hep-th/0508023}}.

\bibitem{Saad:2019pqd}
P.~Saad, ``{Late Time Correlation Functions, Baby Universes, and ETH in JT
  Gravity},'' \href{http://arxiv.org/abs/1910.10311}{{\ttfamily
  arXiv:1910.10311 [hep-th]}}.

\bibitem{Pollack:2020gfa}
J.~Pollack, M.~Rozali, J.~Sully, and D.~Wakeham, ``{Eigenstate Thermalization
  and Disorder Averaging in Gravity},''
  \href{http://dx.doi.org/10.1103/PhysRevLett.125.021601}{{\em Phys. Rev.
  Lett.} {\bfseries 125} no.~2, (2020) 021601},
  \href{http://arxiv.org/abs/2002.02971}{{\ttfamily arXiv:2002.02971
  [hep-th]}}.

\bibitem{Schlenker:2022dyo}
J.-M. Schlenker and E.~Witten, ``{No ensemble averaging below the black hole
  threshold},'' \href{http://dx.doi.org/10.1007/JHEP07(2022)143}{{\em JHEP}
  {\bfseries 07} (2022) 143}, \href{http://arxiv.org/abs/2202.01372}{{\ttfamily
  arXiv:2202.01372 [hep-th]}}.

\bibitem{Guhr:1997ve}
T.~Guhr, A.~Muller-Groeling, and H.~A. Weidenmuller, ``{Random matrix theories
  in quantum physics: Common concepts},''
  \href{http://dx.doi.org/10.1016/S0370-1573(97)00088-4}{{\em Phys. Rept.}
  {\bfseries 299} (1998) 189--425},
  \href{http://arxiv.org/abs/cond-mat/9707301}{{\ttfamily
  arXiv:cond-mat/9707301}}.

\bibitem{Papadodimas:2015xma}
K.~Papadodimas and S.~Raju, ``{Local Operators in the Eternal Black Hole},''
  \href{http://dx.doi.org/10.1103/PhysRevLett.115.211601}{{\em Phys. Rev.
  Lett.} {\bfseries 115} no.~21, (2015) 211601},
  \href{http://arxiv.org/abs/1502.06692}{{\ttfamily arXiv:1502.06692
  [hep-th]}}.

\bibitem{Verlinde:2021jwu}
H.~Verlinde, ``{Deconstructing the Wormhole: Factorization, Entanglement and
  Decoherence},'' \href{http://arxiv.org/abs/2105.02142}{{\ttfamily
  arXiv:2105.02142 [hep-th]}}.

\bibitem{delCampo:2017bzr}
A.~del Campo, J.~Molina-Vilaplana, and J.~Sonner, ``{Scrambling the spectral
  form factor: unitarity constraints and exact results},''
  \href{http://dx.doi.org/10.1103/PhysRevD.95.126008}{{\em Phys. Rev. D}
  {\bfseries 95} no.~12, (2017) 126008},
  \href{http://arxiv.org/abs/1702.04350}{{\ttfamily arXiv:1702.04350
  [hep-th]}}.

\bibitem{Stanford:2022fdt}
D.~Stanford and Z.~Yang, ``{Firewalls from wormholes},''
  \href{http://arxiv.org/abs/2208.01625}{{\ttfamily arXiv:2208.01625
  [hep-th]}}.

\bibitem{Saad:2022kfe}
P.~Saad, D.~Stanford, Z.~Yang, and S.~Yao, ``{A convergent genus expansion for
  the plateau},'' \href{http://arxiv.org/abs/2210.11565}{{\ttfamily
  arXiv:2210.11565 [hep-th]}}.

\bibitem{Blommaert:2022lbh}
A.~Blommaert, J.~Kruthoff, and S.~Yao, ``{An integrable road to a perturbative
  plateau},'' \href{http://arxiv.org/abs/2208.13795}{{\ttfamily
  arXiv:2208.13795 [hep-th]}}.

\bibitem{tHooft:1993dmi}
G.~'t~Hooft, ``{Dimensional reduction in quantum gravity},'' {\em Conf. Proc.
  C} {\bfseries 930308} (1993) 284--296,
  \href{http://arxiv.org/abs/gr-qc/9310026}{{\ttfamily arXiv:gr-qc/9310026}}.

\bibitem{Susskind:1994vu}
L.~Susskind, ``{The World as a hologram},''
  \href{http://dx.doi.org/10.1063/1.531249}{{\em J. Math. Phys.} {\bfseries 36}
  (1995) 6377--6396}, \href{http://arxiv.org/abs/hep-th/9409089}{{\ttfamily
  arXiv:hep-th/9409089}}.

\bibitem{Bousso:1999xy}
R.~Bousso, ``{A Covariant entropy conjecture},''
  \href{http://dx.doi.org/10.1088/1126-6708/1999/07/004}{{\em JHEP} {\bfseries
  07} (1999) 004}, \href{http://arxiv.org/abs/hep-th/9905177}{{\ttfamily
  arXiv:hep-th/9905177}}.

\bibitem{Hsin:2020mfa}
P.-S. Hsin, L.~V. Iliesiu, and Z.~Yang, ``{A violation of global symmetries
  from replica wormholes and the fate of black hole remnants},''
  \href{http://dx.doi.org/10.1088/1361-6382/ac2134}{{\em Class. Quant. Grav.}
  {\bfseries 38} no.~19, (2021) 194004},
  \href{http://arxiv.org/abs/2011.09444}{{\ttfamily arXiv:2011.09444
  [hep-th]}}.

\bibitem{CVITANOVIC198149}
P.~Cvitanović, ``Planar perturbation expansion,''
  \href{http://dx.doi.org/https://doi.org/10.1016/0370-2693(81)90801-7}{{\em
  Physics Letters B} {\bfseries 99} no.~1, (1981) 49--52}.
  \url{https://www.sciencedirect.com/science/article/pii/0370269381908017}.

\bibitem{https://doi.org/10.48550/arxiv.0911.0087}
R.~Speicher, ``Free probability theory,'' 2009.
\newblock \url{https://arxiv.org/abs/0911.0087}.

\bibitem{10.1093/oxfordhb/9780198744191.001.0001}
G.~Akemann, J.~Baik, and P.~Di~Francesco, {\em {The Oxford Handbook of Random
  Matrix Theory}}.
\newblock Oxford University Press, 09, 2015.

\bibitem{Stanford:2017thb}
D.~Stanford and E.~Witten, ``{Fermionic Localization of the Schwarzian
  Theory},'' \href{http://dx.doi.org/10.1007/JHEP10(2017)008}{{\em JHEP}
  {\bfseries 10} (2017) 008}, \href{http://arxiv.org/abs/1703.04612}{{\ttfamily
  arXiv:1703.04612 [hep-th]}}.

\bibitem{Iliesiu:2020qvm}
L.~V. Iliesiu and G.~J. Turiaci, ``{The statistical mechanics of near-extremal
  black holes},'' \href{http://dx.doi.org/10.1007/JHEP05(2021)145}{{\em JHEP}
  {\bfseries 05} (2021) 145}, \href{http://arxiv.org/abs/2003.02860}{{\ttfamily
  arXiv:2003.02860 [hep-th]}}.

\bibitem{Boruch:2023trc}
J.~Boruch, L.~V. Iliesiu, and C.~Yan, ``{Constructing all BPS black hole
  microstates from the gravitational path integral},''
  \href{http://arxiv.org/abs/2307.13051}{{\ttfamily arXiv:2307.13051
  [hep-th]}}.

\bibitem{AnaC}
{A. Climent, R. Emparan, J. Magan, M. Sasieta, A. V. Lopez}, ``{To appear},''.

\bibitem{Almheiri:2019qdq}
A.~Almheiri, T.~Hartman, J.~Maldacena, E.~Shaghoulian, and A.~Tajdini,
  ``{Replica Wormholes and the Entropy of Hawking Radiation},''
  \href{http://dx.doi.org/10.1007/JHEP05(2020)013}{{\em JHEP} {\bfseries 05}
  (2020) 013}, \href{http://arxiv.org/abs/1911.12333}{{\ttfamily
  arXiv:1911.12333 [hep-th]}}.

\bibitem{Penington:2019npb}
G.~Penington, ``{Entanglement Wedge Reconstruction and the Information
  Paradox},'' \href{http://dx.doi.org/10.1007/JHEP09(2020)002}{{\em JHEP}
  {\bfseries 09} (2020) 002}, \href{http://arxiv.org/abs/1905.08255}{{\ttfamily
  arXiv:1905.08255 [hep-th]}}.

\bibitem{Almheiri:2019psf}
A.~Almheiri, N.~Engelhardt, D.~Marolf, and H.~Maxfield, ``{The entropy of bulk
  quantum fields and the entanglement wedge of an evaporating black hole},''
  \href{http://dx.doi.org/10.1007/JHEP12(2019)063}{{\em JHEP} {\bfseries 12}
  (2019) 063}, \href{http://arxiv.org/abs/1905.08762}{{\ttfamily
  arXiv:1905.08762 [hep-th]}}.

\bibitem{Almheiri:2019hni}
A.~Almheiri, R.~Mahajan, J.~Maldacena, and Y.~Zhao, ``{The Page curve of
  Hawking radiation from semiclassical geometry},''
  \href{http://dx.doi.org/10.1007/JHEP03(2020)149}{{\em JHEP} {\bfseries 03}
  (2020) 149}, \href{http://arxiv.org/abs/1908.10996}{{\ttfamily
  arXiv:1908.10996 [hep-th]}}.

\bibitem{Page:1993df}
D.~N. Page, ``{Average entropy of a subsystem},''
  \href{http://dx.doi.org/10.1103/PhysRevLett.71.1291}{{\em Phys. Rev. Lett.}
  {\bfseries 71} (1993) 1291--1294},
  \href{http://arxiv.org/abs/gr-qc/9305007}{{\ttfamily arXiv:gr-qc/9305007}}.

\bibitem{Page:1993wv}
D.~N. Page, ``{Information in black hole radiation},''
  \href{http://dx.doi.org/10.1103/PhysRevLett.71.3743}{{\em Phys. Rev. Lett.}
  {\bfseries 71} (1993) 3743--3746},
  \href{http://arxiv.org/abs/hep-th/9306083}{{\ttfamily arXiv:hep-th/9306083}}.

\bibitem{Sarosi:2017ykf}
G.~S\'arosi, ``{AdS$_{2}$ holography and the SYK model},''
  \href{http://dx.doi.org/10.22323/1.323.0001}{{\em PoS} {\bfseries Modave2017}
  (2018) 001}, \href{http://arxiv.org/abs/1711.08482}{{\ttfamily
  arXiv:1711.08482 [hep-th]}}.

\bibitem{Balasubramanian:2020hfs}
V.~Balasubramanian, A.~Kar, O.~Parrikar, G.~S\'arosi, and T.~Ugajin,
  ``{Geometric secret sharing in a model of Hawking radiation},''
  \href{http://dx.doi.org/10.1007/JHEP01(2021)177}{{\em JHEP} {\bfseries 01}
  (2021) 177}, \href{http://arxiv.org/abs/2003.05448}{{\ttfamily
  arXiv:2003.05448 [hep-th]}}.

\bibitem{Geng:2021hlu}
H.~Geng, A.~Karch, C.~Perez-Pardavila, S.~Raju, L.~Randall, M.~Riojas, and
  S.~Shashi, ``{Inconsistency of islands in theories with long-range
  gravity},'' \href{http://dx.doi.org/10.1007/JHEP01(2022)182}{{\em JHEP}
  {\bfseries 01} (2022) 182}, \href{http://arxiv.org/abs/2107.03390}{{\ttfamily
  arXiv:2107.03390 [hep-th]}}.

\bibitem{Balasubramanian:2021wgd}
V.~Balasubramanian, A.~Kar, and T.~Ugajin, ``{Entanglement between two
  gravitating universes},''
  \href{http://dx.doi.org/10.1088/1361-6382/ac3c8b}{{\em Class. Quant. Grav.}
  {\bfseries 39} no.~17, (2022) 174001},
  \href{http://arxiv.org/abs/2104.13383}{{\ttfamily arXiv:2104.13383
  [hep-th]}}.

\bibitem{Susskind:2020wwe}
L.~Susskind, ``{Black Holes at Exp-time},''
  \href{http://arxiv.org/abs/2006.01280}{{\ttfamily arXiv:2006.01280
  [hep-th]}}.

\bibitem{Magan:2018nmu}
J.~M. Mag\'an, ``{Black holes, complexity and quantum chaos},''
  \href{http://dx.doi.org/10.1007/JHEP09(2018)043}{{\em JHEP} {\bfseries 09}
  (2018) 043}, \href{http://arxiv.org/abs/1805.05839}{{\ttfamily
  arXiv:1805.05839 [hep-th]}}.

\bibitem{Magan:2020iac}
J.~M. Mag\'an and J.~Sim\'on, ``{On operator growth and emergent Poincar\'e
  symmetries},'' \href{http://dx.doi.org/10.1007/JHEP05(2020)071}{{\em JHEP}
  {\bfseries 05} (2020) 071}, \href{http://arxiv.org/abs/2002.03865}{{\ttfamily
  arXiv:2002.03865 [hep-th]}}.

\bibitem{Caputa:2021sib}
P.~Caputa, J.~M. Magan, and D.~Patramanis, ``{Geometry of Krylov complexity},''
  \href{http://dx.doi.org/10.1103/PhysRevResearch.4.013041}{{\em Phys. Rev.
  Res.} {\bfseries 4} no.~1, (2022) 013041},
  \href{http://arxiv.org/abs/2109.03824}{{\ttfamily arXiv:2109.03824
  [hep-th]}}.

\bibitem{Iliesiu:2022kny}
L.~V. Iliesiu, S.~Murthy, and G.~J. Turiaci, ``{Black hole microstate counting
  from the gravitational path integral},''
  \href{http://arxiv.org/abs/2209.13602}{{\ttfamily arXiv:2209.13602
  [hep-th]}}.

\bibitem{Balasubramanian:2022lnw}
V.~Balasubramanian, A.~Lawrence, J.~M. Magan, and M.~Sasieta, ``{Microscopic
  origin of the entropy of astrophysical black holes},''
  \href{http://arxiv.org/abs/2212.08623}{{\ttfamily arXiv:2212.08623
  [hep-th]}}.

\bibitem{doi:10.1080/00207166808803030}
A.~N. Kolmogorov, ``Three approaches to the quantitative definition of
  information,'' {\em International Journal of Computer Mathematics} {\bfseries
  2} no.~1-4, (1968) 157--168.

\bibitem{Lanczos1950AnIM}
C.~Lanczos, ``An iteration method for the solution of the eigenvalue problem of
  linear differential and integral operators,'' {\em Journal of research of the
  National Bureau of Standards} {\bfseries 45} (1950) 255--282.

\bibitem{Viswanath1994TheRM}
V.~S. Viswanath and G.~M{\"u}ller, ``The recursion method : application to
  many-body dynamics,''
\newblock 1994.

\bibitem{Parker:2018yvk}
D.~E. Parker, X.~Cao, A.~Avdoshkin, T.~Scaffidi, and E.~Altman, ``{A Universal
  Operator Growth Hypothesis},''
  \href{http://dx.doi.org/10.1103/PhysRevX.9.041017}{{\em Phys. Rev. X}
  {\bfseries 9} no.~4, (2019) 041017},
  \href{http://arxiv.org/abs/1812.08657}{{\ttfamily arXiv:1812.08657
  [cond-mat.stat-mech]}}.

\end{thebibliography}\endgroup

\end{document}